\documentclass[oneside,english,british]{amsart}
\usepackage[T1]{fontenc}
\usepackage[latin9]{inputenc}
\usepackage{float}
\usepackage{bm}
\usepackage{amsthm}
\usepackage{amstext}
\usepackage{amssymb}
\usepackage{graphicx}
\usepackage{esint}
\usepackage[authoryear]{natbib}

\makeatletter

\floatstyle{ruled}
\newfloat{algorithm}{tbp}{loa}
\providecommand{\algorithmname}{Algorithm}
\floatname{algorithm}{\protect\algorithmname}

\numberwithin{equation}{section}
\numberwithin{figure}{section}
\theoremstyle{plain}
\newtheorem{thm}{\protect\theoremname}
\usepackage[usenames,dvipsnames]{color}

\newcommand{\eqdef}{\stackrel{\Delta}{=}}
\newcommand{\bt}{\bm{\theta}}
\newcommand{\R}{\mathbb R }

\newcommand{\Pb}{\mathbb P }
\def\1{1\!{\rm l}}  
\def\N{\mbox{I}\kern-.20em \mbox{N}}

\def\T{\bm{T}}
\def\tr{\mbox{tr}}
\def\be{\beta}
\def\txid{\bar{\bm{\theta}}_d}
\def\bd{\bar{\theta}_{d,k}}

\usepackage{babel}
\addto\captionsbritish{\renewcommand{\theoremname}{Theorem}}
  \addto\captionsenglish{\renewcommand{\theoremname}{Theorem}}
\addto\captionsenglish{\renewcommand{\algorithmname}{Algorithm}}
\providecommand{\theoremname}{Theorem}

\usepackage{babel}
\addto\captionsbritish{\renewcommand{\theoremname}{Theorem}}
  \addto\captionsenglish{\renewcommand{\theoremname}{Theorem}}
\addto\captionsenglish{\renewcommand{\algorithmname}{Algorithm}}
\providecommand{\theoremname}{Theorem}

\usepackage{babel}
\addto\captionsbritish{\renewcommand{\theoremname}{Theorem}}
  \addto\captionsenglish{\renewcommand{\theoremname}{Theorem}}
\addto\captionsenglish{\renewcommand{\algorithmname}{Algorithm}}
\providecommand{\theoremname}{Theorem}

\makeatother

\usepackage{babel}
  \addto\captionsbritish{\renewcommand{\theoremname}{Theorem}}
  \addto\captionsenglish{\renewcommand{\theoremname}{Theorem}}
\addto\captionsenglish{\renewcommand{\algorithmname}{Algorithm}}
\providecommand{\theoremname}{Theorem}

\begin{document}

\title{Computational aspects of Bayesian spectral density estimation}

\author{N. Chopin (CREST-ENSAE)}

\author{J. Rousseau (CREST-ENSAE and Universit\'e Paris Dauphine) }

\author{B. Liseo (Universita di Roma)}
\begin{abstract}
Gaussian time-series models are often specified through their spectral
density. Such models present several computational challenges, in particular
because of the non-sparse nature of the covariance matrix. We derive
a fast approximation of the likelihood for such models. We propose to sample 
from the approximate posterior (that is, the prior times the approximate 
likelihood), and then to recover the exact posterior through importance 
sampling. We show that the
variance of the importance sampling weights vanishes as the sample
size goes to infinity. We explain why the approximate posterior may typically multi-modal,
and we derive a Sequential Monte Carlo sampler based on an annealing
sequence in order to sample from that target distribution. Performance
of the overall approach is evaluated on simulated and real datasets. 
In addition, for one real world dataset, we provide some numerical evidence 
that a Bayesian approach to semi-parametric estimation of spectral density 
may provide more reasonable results than its Frequentist counter-parts. 
\end{abstract}


\maketitle

\section{Introduction}

Several models in the time series literature are defined through their
spectral density. Assuming Gaussianity, one observes a vector $\bm{x}$
of length $n$, from the Gaussian distribution 
\[
\bm{x}|\mu,f\sim N\left(\mu\bm{1},\bm{T}(f)\right)
\]
where $\bm{1}=(1,\ldots,1)'$, and $\bm{T}(f)$ is the $n\times n$
Toeplitz matrix associated to spectral density $f$, with entries
$\bm{T}(f)(l,m)=\gamma_{f}(l-m)$, 
\begin{equation}
\gamma_{f}(l)=\int_{-\pi}^{\pi}f(\lambda)e^{il\lambda}\, d\lambda,\quad l=-(n-1),\ldots,(n-1).\label{eq:FourierIntegral}
\end{equation}
The models vary with respect to the specification of $f$. For instance,
the FEXP parametrisation \citep{Hurvich2002,MoulinesSoulier2003}
assumes that 
\begin{equation}
f(\lambda)=\frac{1}{2\pi}\left|1-e^{-i\lambda}\right|^{-2d}g(\lambda),\quad g(\lambda)=\exp\left\{ \sum_{j=0}^{k}\xi_{j}\cos(j\lambda)\right\} .\label{eq:FEXPmodel}
\end{equation}
This specification conveniently separates the long-range behaviour,
as determined by parameter $d\in[0,1/2)$, and the short-memory part
$g$. By taking $k$ large enough, $g$ may be arbitrarily close to
any function in a certain regularity class. \citet{RousseauChopinLiseo2012}
show that, for a well chosen prior with respect to $(k,d,\xi_{1},\ldots)$,
the corresponding posterior is consistent for both $d$ and $f$,
under semi-parametric settings; that is, assuming that the true spectral
density belongs to a certain infinite-dimensional class of functions.

This FEXP parametrisation will be our running example. We note however
that many other parametrisations of $f$ are possible, which can also
be tackled by the methodology developed in this paper. One may take
$d=0$, and obtain a non-parametric procedure for estimating the spectral
density, under a short-memory assumption. One may replace $g$ by
another type of expansion, a spline regression, and so on. Finally,
one may also consider a purely parametric model, such as an ARFIMA$(p,q)$
model, given by: 
\[
f(\lambda)=\frac{1}{2\pi}\left|1-e^{-i\lambda}\right|^{-2d}\left|\frac{1+\sum_{j=1}^{q}\xi_{j}e^{-ij\lambda}}{1-\sum_{j=1}^{p}\phi_{j}e^{-ij\lambda}}\right|^{2}.
\]
ARMA models may also be defined through their spectral density, but
well-known specialised methods exist for such models, hence they fall
outside the scope of this paper.

Whatever the specification of $f$, parametric or semiparametric,
several computational difficulties arise regarding Bayesian inference
for such models. First, the likelihood of the data 
\begin{equation}
p(\bm{x}|\mu,f)=\left(2\pi\right)^{-n/2}\left|\bm{T}(f)\right|^{-1/2}\exp\left\{ -\frac{1}{2}\left(\bm{x}-\mu\bm{1}\right)^{T}\bm{T}(f)^{-1}\left(\bm{x}-\mu\bm{1}\right)\right\} \label{eq:standard_lik}
\end{equation}
involves a determinant and a quadratic form which are expensive to
compute, i.e. the cost is $O(n^{3})$ if one uses off-the-shelf methods.
Second, the entries of matrix $\bm{T}(f)$ itself, that is, the Fourier
integrals \eqref{eq:FourierIntegral} cannot be computed reliably
using standard quadrature methods, because of the many oscillations
of the integrand for large values of $k$. Third, Gibbs sampling is
generally not feasible for posterior distributions associated to this
likelihood. One usually resorts to the Metropolis-Hastings sampler
\citep[see e.g. ][Chap. 7]{RobCas}, which is difficult to tune in
order to obtain reasonable performance. This is quite problematic
in this context: since the likelihood is expensive, performing several
pilot runs in order to progressively tune the sampler may be a long
and tedious process for the user. This problem is compounded if $f$
is specified through a trans-dimensional prior; for instance, in the
FEXP model above, if $k$ is random. Then one needs to implement an
algorithm for trans-dimensional sampling spaces, such as Green's algorithm
\citep{Green:reversible,RichGreen}, also known as the reversible-jump
sampler, which is harder to tune yet.

Perhaps because of the above difficulties, literature on Bayesian analysis of the spectral density of a possibly long memory stationary process is not vast; 
see \citet{pai1998bayesian,pai2001bayesian,ravishanker1997bayesian} for approaches in the parametric (ARFIMA) case, and \citet{petris,Liseo2001,Choudhuri2004,holan2009bayesian} for semi-parametric approaches. 
These approaches often rely on some approximation of the likelihood; commonly the Whittle approximation is used, although that approximation is not reliable at low frequencies in  long range settings \citep{Robinson1995a}.
Some of these papers establish the consistency of the resulting pseudo-posterior, but under quite strict hypotheses: results of \citet{Liseo2001} hold 
under the assumption that the true model is a fractional Gaussian noise model, while those of \citet{Choudhuri2004} hold under short memory. Even under these assumptions, there is obviously some interest for doing exact Bayesian inference, rather than doing some approximation, the error of which is hard to assess for finite sample sizes.  

Current Frequentist approaches to  estimating of the long memory $d$ parameter
are not necessarily entirely satisfactory either, albeit for different reasons. A common semi-parametric approach in the Frequentist literature is to discard the higher frequencies in the periodogram, but this leads potentially to an important loss of information; see \citet{MoulinesSoulier2003} for a discussion and an alternative approach. Parametric (e.g. ARFIMA) Frequentist approaches on the other hand are known to provide unstable and inconsistent results under model misspecification. We give at the end of the paper a real data example where our Bayesian approach provides quite more satisfactory results than those provided by standard Frequentist procedures.

This paper proposes a novel approach for addressing the above problems in a unified manner, and is organised as follows. Section \ref{sec:Exact} discusses the exact computation the likelihood. It is seen that the cost of this
operation is $O(n^{3})$, but the constant in $O(n^{3})$ is typically
small, hence on modern hardware it remains possible to perform a reasonable
number of such evaluations provided $n$ is not too large.
Section \ref{sec:Approx} proposes a fast approximation of the likelihood,
the cost of which is essentially $O(n)$. This approximation scheme
motivates the following approach. In a first step, we perform Monte
Carlo simulation of the approximate posterior, that is, the prior
times the approximate likelihood. The cost of this step is $O(n)$ but,
typically, with a large constant in front of $n$, due to the intensive
nature of Monte Carlo algorithms. In a second step, which is necessary only if $n$
is not too large, we correct for the above approximation by doing importance
sampling on a reasonable number of simulated samples; the cost of
this second step is $O(n^{3})$; however this time the multiplying constant
in front of $n^{3}$ is small. In practice the
cost of the second step is negligible with respect to the
first step, at least for $n\ll 10^{4}$. (To put things into perspective, 
many real datasets discussed in the literature are such that $n<10^3$, 
and the largest real dataset we could find had $n=4000$, see end of Section 5
for a discussion.)  
 On the other hand, if
$n$ is large, we show that the importance sampling step becomes superfluous,
because the variance of the importance sampling weights converge to
$0$ as $n\rightarrow+\infty$, under appropriate assumptions.
Section \ref{sec:Monte-Carlo-sampling} discusses a Sequential Monte
Carlo algorithm (SMC) for sampling from the approximate posterior;
see \citet{DelDouJas:SMC} for a general introduction to SMC. We shall
see that the following advantages of SMC (relative to for instance
MCMC) are particularly useful in this specific context. First, as
explained above, one would like to apply the importance sampling step
to a sample as small as possible, because the exact likelihood is
expensive to compute. We shall see that our SMC sampler typically
generates particles that are close to IID samples from its target
distribution (in the sense that the Monte Carlo variance computed over
repeated SMC runs for certain test functions seems close to the variance 
that one would obtain from IID particles; this point will be discussed
more in detail in the paper). 
Second, as also mentioned above, one has little prior
information on the structure of the (approximate or exact) posterior.
It is easy however to make SMC adaptive, by learning iteratively features
of the target from the sample of simulated particles. Third, we observe
in certain settings that semi-parametric models such as the FEXP model,
may generate multi-modal posteriors. The SMC sampler we propose is
based on tempering ideas, and thus more able to escape from minor
local modes.

Section \ref{sec:Monte-Carlo-sampling} illustrates the proposed approach
on simulated and real data.

We shall use the following notations: vectors and matrices are always
in bold face, e.g. $\bm{x}$ and $\bm{\Sigma}$, the determinant,
transpose, and trace of $\bm{\Sigma}$ are denoted respectively $\left|\bm{\Sigma}\right|$,
$\bm{\Sigma}^{T}$ and $\tr(\bm{\Sigma})$.

\section{Exact computation of the likelihood\label{sec:Exact}}

\subsection{Marginalisation}

As a preliminary, we note that $f$ is often parametrised in such
a way that $f=\sigma^{2}\bar{f}_{\bm{\bm{\theta}}}$, where $\sigma$
is a scale parameter, and $\bm{\theta}$ is the vector of all remaining
parameters (except $\mu$). For instance, in the FEXP case, we may
set $\sigma^{2}=\exp\left(\xi_{0}\right)$, $\bm{\theta}=(k,\bm{\theta}_{k})$,
$\bm{\theta}_{k}=(\mathrm{logit}(2d),\xi_{1},\ldots,\xi_{k})$, and
\begin{equation}
\bar{f}_{\bm{\theta}}(\lambda)=\frac{1}{2\pi}\left|1-e^{-i\lambda}\right|^{-2d}\bar{g}_{\bm{\theta}}(\lambda),\quad\bar{g}_{\bm{\theta}}(\lambda)=\exp\left\{ \sum_{j=1}^{k}\xi_{j}\cos(j\lambda)\right\} .\label{eq:fexp_normalised}
\end{equation}
(The function $\mathrm{logit}$ is defined as $\mathrm{logit}(x)=\log(x)-\log(1-x)$,
and is used here to facilitate the construction of proposal distributions,
see Section \ref{sec:Monte-Carlo-sampling}.) It is then possible
to marginalise out both $\mu$ and $\sigma^{2}$ from the
likelihood, provided these two parameters are assigned the following
standard $g-$prior distribution, independently from $\bm{\theta}$:
$1/\sigma^{2}\sim\mathrm{Gamma}(a,b)$, $\mu|\sigma^{2}\sim N(m_{\mu},\sigma^{2}/g_{\mu})$.
The marginal likelihood reads: 
\begin{eqnarray}
p\left(\bm{x}|\bm{\theta}\right) & = & \int p(\bm{x}|\mu,\sigma^{2},\bm{\theta})p(\mu,\sigma^{2})\, d\mu d\sigma^{2}\nonumber \\
 & \propto & \left|\bm{T}(\overline{f}_{\bm{\theta}})+\frac{1}{g_{\mu}}\bm{E}\right|^{-1/2}\times\nonumber \\
 &  & \left\{ b+\frac{1}{2}\left(\bm{x}-m_{\mu}\bm{1}\right)^{T}\left(\bm{T}(\overline{f}_{\bm{\theta}})+\frac{1}{g_{\mu}}\bm{E}\right)^{-1}\left(\bm{x}-m_{\mu}\bm{1}\right)\right\} ^{-a-n/2},\label{eq:marg_lik}
\end{eqnarray}
where $\bm{E}$ is the $n\times n$ matrix filled with ones.
Marginalising out parameters usually improves the performance of any
sampling algorithm. We note however that the approach developed in
this paper would work with little modification for the unmarginalised
likelihood $p(\bm{x}|\mu,\sigma^{2},\bm{\theta})$. These two likelihood
functions both suffer from the same computational difficulties, which are
described in the next section.

\subsection{Computational difficulties associated to likelihood evaluation}

We review in this section the specific difficulties that arise when
evaluating either the standard likelihood function \eqref{eq:standard_lik},
or the marginal version \eqref{eq:marg_lik}. First, both likelihood
functions include some quadratic form $\bm{y}\bm{\Sigma}^{-1}\bm{y}$
involving the inverse of a $n\times n$ symmetric matrix $\bm{\Sigma};$
in \eqref{eq:standard_lik}, $\bm{\Sigma}=\bm{T}(f)$, and in \eqref{eq:marg_lik},
$\bm{\Sigma=}\bm{T}(\overline{f}_{\bm{\theta}})+\frac{1}{g_{\mu}}\bm{E}$.
Second, both functions involve the determinant of the same matrix
$\bm{\Sigma}$. Third, in both cases, evaluating the entries of $\bm{\Sigma}$
requires computing simultaneously $n$ Fourier integrals, see \eqref{eq:FourierIntegral}.

Our solution to the third difficulty is described in the two next
Sections. Regarding the two first points, the most direct solution
is to compute the Cholesky lower triangle of $\bm{\Sigma}$, $\bm{\Sigma=CC^{T}}$.
Then, one obtains the determinant by taking the square of the product
of the diagonal elements of $\bm{C}$, and one computes $\bm{y}^{T}\bm{\Sigma}^{-1}\bm{y}=\left(\bm{C}^{-1}\bm{y}\right)^{T}\bm{C}^{-1}\bm{y}$
as the norm of the solution (in $\bm{z}$) of the linear system $\bm{C}\bm{z}=\bm{y}$,
which is quickly obtained by back-substitution. The Cholesky decomposition
is a $O(n^{3})$ operation.

For the sake of completeness, we mention briefly faster, but specialised,
algorithms for solving directly the system $\bm{\Sigma}\bm{z}=\bm{y}$,
in order to compute $\bm{y}^{T}\bm{\Sigma}^{-1}\bm{y}$. Given that
$\bm{\Sigma}$ is Toeplitz, one may use Levinson's algorithm \citep[p. 96]{levinson1949wiener,NumericalRecipes3rdEd},
which is $O(n^{2})$. Alternatively, \citet{Chen2006} have developed
a variant of the conjugate gradient method, based on a a particular
preconditioned matrix. Their algorithm requires $O(\log^{3/2}n)$
iterations, each involving a FFT transform over $2n$ points. The
overall cost is therefore $O(n\log^{5/2}n)$.

Unfortunately, these alternative approaches do not provide an evaluation
of the determinant as a by-product. \citet{Chen2006}, \citet{holan2009bayesian}
approximate the determinant by using a particular asymptotic approximation,
which we describe later, but obviously this approach is not entirely
satisfactory when used within a non-asymptotic, and in particular,
a Bayesian, approach.

Perhaps more importantly, we note that the constant in front of the
$O(n^{3})$ cost of the Choleksy decomposition is typically very small,
due to very efficient implementation in most scientific software.
To give an order of magnitude, for $n=10^{3}$, 1000 of such operations
takes about one minute on the first author's computer. This observation
underpins the strategy laid out in the introduction: to run some Monte
Carlo algorithm so as to sample from an approximation of the posterior,
then, provided $n$ is not too large, to correct for the approximation
using importance sampling on a moderate (possibly sub-sampled from
the first step) Monte Carlo sample.

\subsection{Computing the Fourier coefficients\label{sub:Fourier-coefficients}}

In this section and the following, we consider the problem of evaluating
simultaneously the $n$ Fourier integrals defined in \eqref{eq:FourierIntegral}.
As noted in the introduction, using standard quadrature would work
very poorly, because the integrand in \eqref{eq:FourierIntegral}
strongly oscillates when $n$ gets large. The solution we describe
here seems well known in the numerical mathematics literature \citep[e.g. ][Chap. 13, Sect. 9]{NumericalRecipes3rdEd},
yet, to the best of our knowledge, it has not been used before in
the time series literature. Instead, previous approaches \citep[e.g. ][]{Chen2006,holan2009bayesian}
rely on more specific algorithms such as the splitting algorithm of
\citet{bertelli2002note}. The approach described here has the same
computational cost as such alternative approaches, that is, that of
a FFT (Fast Fourier Transform), i.e. $O(n\log n$). However, we find
our approach slightly more convenient, for the following reasons:
(a) this is a generic approach, which requires only pointwise evaluation
of the spectral density, whereas the splitting algorithm requires
exact expressions for the moving average coefficients (of the short
memory part) which are specific to the considered class of spectral
densities; for instance, it is unclear how one could use these methods
if $f$ would be specified through splines; (b) it is characterised
by only one level of approximation (see below), whereas the splitting
algorithm expresses first the moving averages coefficients as an infinite
sum, which must be truncated, then plug these coefficients into another
infinite sum, which must be truncated again, so assessing the numerical
error is slightly more delicate; and (c) in long memory settings,
the terms of these infinite sums are supposed to decay slowly, hence
one may need to truncate to a large number of terms.

Let $M$ be a power of two, $M=2^K$, such that $M\geq2n$. We explain first
how to compute efficiently and simultaneously the Fourier integrals 
\[
\gamma_{g}(l)=\int_{-\pi}^{\pi}g(\lambda)e^{il\lambda}\, d\lambda
\]
for a given bounded function $g$, and $0\leq l\leq M/2$. (If $n<M/2$,
simply discard the extra values.) If the spectral density $f$ itself
is bounded, that is, if it corresponds to a short memory process,
then the following method may be used directly by taking $g=f$. If
$f$ corresponds to a long-memory process, then $f$ diverges at $0$,
and a minor modification is required to use the following method,
see next section.

The idea is to replace $g$ by a linear interpolation $\widetilde{g}$:
\[
\tilde{g}(\lambda)=\sum_{j=0}^{M-1}g_{j}\psi(\frac{\lambda-\lambda_{j}}{\Delta})+g_{0}\varphi_{0}(\frac{\lambda-\lambda_{0}}{\Delta})+g_{M}\varphi_{M}(\frac{\lambda-\lambda_{M}}{\Delta})
\]
where $\Delta=2\pi/M$, $\lambda_{j}=-\pi+j\Delta$, $g_{j}=\tilde{g}(\lambda_{j})$
$j=0,\ldots,M$, and $\psi$ is the linear interpolation kernel, i.e.
$\psi(\lambda)=\left(1-|\lambda|\right)^{+}$; $\varphi_{0}$ and
$\varphi_{M}$ are boundary corrections, the expression of which may
be skipped for the rest of the discussion. Note $\tilde{g}$ is defined
on the entire real-line, and is zero outside $[-\pi,\pi]$. Applying
the operator $\int\left(\cdot\right)e^{il\lambda}\, d\lambda$ yields:

\begin{eqnarray*}
\int_{-\pi}^{\pi}\tilde{g}(\lambda)e^{il\lambda}\, d\lambda & = & \int_{-\infty}^{+\infty}\tilde{g}(\lambda)e^{il\lambda}\, d\lambda\\
 & = & \sum_{j=0}^{M-1}g_{j}\int_{-\infty}^{+\infty}\psi(\frac{\lambda-\lambda_{j}}{\Delta})e^{il\lambda}\, d\lambda\\
 &  & +g_{0}\int_{-\infty}^{+\infty}\varphi_{0}(\frac{\lambda-\lambda_{0}}{\Delta})e^{il\lambda}\, d\lambda+g_{M}\int_{-\infty}^{+\infty}\varphi_{M}(\frac{\lambda-\lambda_{M}}{\Delta})e^{il\lambda}\, d\lambda\\
 & = & \Delta(-1)^{l}\left\{ W(l\Delta)\sum_{j=0}^{M-1}g_{j}e^{ijl\frac{2\pi}{M}}+g_{0}\alpha_{0}(l\Delta)+g_{M}\alpha_{M}(l\Delta)\right\} 
\end{eqnarray*}
where $W(\lambda)=\int_{-\infty}^{\infty}\psi(u)e^{i\lambda u}\, du$,
$\alpha_{0}(\lambda)=\int_{-\infty}^{\infty}\varphi_{0}(u)e^{i\lambda u}\, du$,
$\alpha_{M}(\lambda)=\int_{-\infty}^{\infty}\varphi_{M}(u-M)e^{i\lambda u}\, du$.
The first sum may be computed using a FFT, in $O(M\log(M))$ time.
All the other functions admit close-form expressions, given by \citet[Chap. 13, Sect. 9]{NumericalRecipes3rdEd}:
\[
W(\lambda)=\frac{2(1-\cos\lambda)}{\lambda^{2}},\quad\alpha_{0}=\alpha_{M}=-\frac{W}{2}.
\]

The scheme above may be adapted so as to rely on a cubic (rather than
linear) interpolation. This point is interesting in settings where
the spectral density is parametrised in terms of cubic splines. Then,
the method described here becomes exact. Otherwise, the accuracy of
the method is determined by the size of the grid, $M+1$. We follow
\citet{NumericalRecipes3rdEd} and takes $M$ to be the smallest power
of two such that $M\geq2n$, and we observe in practice that it gives
very accurate results (in the sense that larger value of $M$ give
essentially the same values).

\subsection{Computing the Fourier coefficients when $f$ diverges at $0$}

In this Section, we explain how to adapt the method above for computing
Fourier integrals, in the situations where $f$ is a long-range dependent
spectral density, 
\[
f(\lambda)=\frac{1}{2\pi}\left|1-e^{-i\lambda}\right|^{-2d}g(\lambda)
\]
where $g$ is a bounded function, and $0<d<1/2$. In this case, $f$
diverges at $0$, hence it may not be well approximated by a piecewise
linear function. To address this problem, we simply decompose $f$
in two terms: 
\[
2\pi f(\lambda)=\left|1-e^{-i\lambda}\right|^{-2d}g(0)+\left|1-e^{-i\lambda}\right|^{-2d}\left\{ g(\lambda)-g(0)\right\} ,
\]
and compute each Fourier integral as a sum of two Fourier integrals
corresponding to each terms. The Fourier integrals of the first term
correspond to the autocovariance function of a fractionally integrated
noise, which admits a close-form expression \citep[Chap. 13]{brockwell2009time}:
\[
\frac{1}{2\pi}\int_{-\pi}^{\pi}\left|1-e^{-i\lambda}\right|^{-2d}e^{il\lambda}\, d\lambda=\begin{cases}
\frac{\Gamma(l+d)\Gamma(1-d)}{\Gamma(l-d+1)\Gamma(d)} & \mbox{ if }l\geq1,\\
\frac{\Gamma(1-2d)}{\Gamma(1-d)^{2}} & \mbox{ if }l=0.
\end{cases}
\]

The Fourier integrals of the second term may be computed directly
using the method of the previous section: assuming $g$ is differentiable
at $0$, the second term vanishes at $0$, since $\left|1-e^{-i\lambda}\right|^{-2d}\left\{ g(\lambda)-g(0)\right\} \sim g'(0)\lambda^{1-2d}$,
with $1-2d\geq0$.

\section{Approximated likelihood\label{sec:Approx}}

\subsection{Principle\label{sub:Principle}}

For convenience, we consider only the standard likelihood function
\eqref{eq:standard_lik} in this section, and furthermore we assume
that $\mu=0$, i.e. a model with zero mean. The main idea of our approximation
scheme is to replace $\bm{T}(f)^{-1}$ by $\bm{T}(1/4\pi^{2}f)$ in
the quadratic form, yielding 
\[
\tilde{p}(\bm{x}|f)=\left(2\pi\right)^{-n/2}\left|\bm{T}(f)\right|^{-1/2}\exp\left\{ -\frac{1}{2}\bm{x}^{T}\bm{T}(\frac{1}{4\pi^{2}f})\bm{x}\right\} .
\]
It is easy to see that the quadratic form within the exponential may
now be computed in $O(n\log n)$ time. We return to this point and
other implementation aspects in the next section.

The idea of approximating $\bm{T}(f)^{-1}$ by $\bm{T}(1/4\pi^{2}f)$
is related to the asymptotic theory of Toeplitz matrices. It is in
fact a common technical tool in the asymptotic theory of long-memory
processes \citep{Dahlhaus1989,RousseauChopinLiseo2012}, but to the
best of our knowledge it has not been used for computational reasons
before.

Now assume that $f$ is parametrised in some way, $f=f_{\bm{\theta}}$,
and denote $p(\bm{x}|f)=p(\bm{x}|\bm{\theta})$, $\tilde{p}(\bm{x}|f)=\tilde{p}(\bm{x}|\bm{\theta})$.
As explained in the introduction, our strategy boils down to sample
from the approximate posterior $\pi_{n}(\bt|\bm{x})\propto p(\bt)\tilde{p}(\bm{x}|\bt)$,
then to perform importance sampling from the approximate posterior
to the true posterior, that is, to assign some weight to any simulation
from the approximated posterior, with a weight function defined as:
\[
w_{\mathrm{Corr}}(\bm{\theta})\stackrel{\Delta}{=}\frac{p(\bm{x}|\bm{\theta})}{\tilde{p}(\bm{x}|\bm{\theta})}=\exp\left\{ -\frac{1}{2}\bm{x}^{T}\left[\bm{T}(f_{\bm{\theta}})^{-1}-\bm{T}(\frac{1}{4\pi^{2}f})\right]\bm{x}\right\} .
\]
The following theorem justifies this strategy. 
\begin{thm}
Consider the FEXP model, as defined by \eqref{eq:FEXPmodel}, and
let $\pi_{n}$ denote the approximated posterior distribution defined
as $\pi_{n}(\bm{\theta})\propto p(\bm{\theta})\tilde{p}(\bm{x}|\bm{\theta})$,
where $p(\bm{\theta})$ is some prior density with respect to parameter
$\bm{\theta}$. Then, under certain conditions (listed in the Appendix)
on the prior distribution $p(\bm{\theta})$ and the true distribution
of $\bm{x}$, with associated spectral density $f_{o}$, one has 
\[
\mathbb{E}^{\pi_{n}}\left[w_{\textrm{Corr}}(\bm{\theta})\right]=w_{\textrm{Corr}}^{0}
\left\{
1+o_{P}(1)\right\};
\]
\[
\mathbb{E}^{\pi_{n}}\left[w_{\textrm{Corr}}(\bm{\theta})^{2}
\right ]=\left(w_{\textrm{Corr}}^{0}\right)^{2}\left\{ 1+o_{P}(1)\right\}, 
\]
where $w_{\textrm{Corr}}^{0}=p(\bm{x}|f_{o})/\tilde{p}(\bm{x}|f_{o})$. 
\end{thm}
The proof and the technical conditions on the prior and the true spectral
density are given in the Appendix. Because this theorem relies heavily
on technical results of \citet{RousseauChopinLiseo2012}, it is restricted
to the semi-parametric FEXP model presented in the introduction, but
with zero mean. We believe it could be extended to other classes of
models with some extra effort. Note that the true spectral density is not
assumed to belong to a fixed-dimension FEXP parametric model (for some fixed $k$, in the notations used in  \eqref{FEXPmodel}), but, with greater generality, to a certain Sobolev class of infinite dimension; see the beginning of the Appendix for a more precise statement. 

In practical terms, this theorem says that the variance of the weights
goes to zero as $n$ goes to infinity, or, in other words, that the
importance weights become nearly constant as $n$ goes to infinity.

\subsection{Practical implementation\label{sub:Practical-implementation-approx}}

In this section, we work out a practical implementation of the approximation
scheme proposed in the previous section. We now turn our attention
to the marginalised likelihood defined in \eqref{eq:marg_lik}.

First, we simplify the quadratic form by ignoring the uncertainty
with respect to $\mu$, i.e. by taking $m_{\mu}=\bar{\bm{x}}$, $g_{\mu}=+\infty$,
so that the likelihood simplifies to 
\[
\tilde{p}(\bm{x}|\bm{\theta})\propto\left|\bm{T}(\overline{f}_{\bm{\theta}})\right|^{-1/2}\left\{ b+\frac{1}{2}\widetilde{\bm{x}}^{T}\bm{T}(\overline{f}_{\bm{\theta}})^{-1}\widetilde{\bm{x}}\right\} ^{-a+n/2},\quad\widetilde{\bm{x}}=\bm{x}-\bar{x}\bm{1}.
\]
We observe that this particular approximation is quite accurate in
practice, which relates to the fact, in long-memory scenarios, $\overline{x}$
is the standard estimator of $\mu$, and typically converges faster
than other features (e.g. $d$) of the model.

Second, as explained in the previous section, we replace the inverse
of $\bm{T}(\bar{f}_{\bm{\theta}})$ by $\bm{T}(4\pi^{2}/\bar{f}_{\bm{\theta}})$
which leads to the following approximation of the quadratic form:

\begin{equation}
\widetilde{\bm{x}}^{T}\bm{T}(\overline{f}_{\bm{\theta}})^{-1}\widetilde{\bm{x}}\approx\widetilde{\bm{x}}\bm{T}(\frac{4\pi^{2}}{\overline{f}_{\bm{\theta}}})\widetilde{\bm{x}}=\sum_{j=0}^{n-1}c_{j}(\tilde{\bm{x}})\gamma_{h}(j)\label{eq:approx_quad}
\end{equation}
where the coefficients $c_{k}(\tilde{\bm{x}})$ may be pre-computed,
once and for all, from the data (denoting $\tilde{x}_{i}=x_{i}-\bar{x}$
the components of $\tilde{\bm{x}}$): 
\[
c_{j}(\tilde{\bm{x}})=\left\{ \begin{array}{cc}
\sum_{i=1}^{n}\tilde{x}_{i}^{2} & \mbox{if }j=0\\
2\sum_{i=1}^{n-k}\tilde{x}_{i}\tilde{x}_{i+j} & \mbox{if }j=1,\ldots,n-1
\end{array}\right.
\]
and the $\gamma_{h}(j)$' are the coefficients of the Toeplitz matrix
$\bm{T}(4\pi^{2}/\overline{f}_{\bm{\theta}})$, that is, the Fourier
integrals corresponding to function $h=1/\left(4\pi^{2}\bar{f}\right)$.
Using the method described in Section \ref{sub:Fourier-coefficients},
one obtains a $O(n\log n)$ overall cost for evaluating the quadratic
form.

Third, we approximate the determinant using a $O(1)$ asymptotic approximation,
as explained in the next section.

\subsection{Determinant approximation\label{sub:Approximating-the-determinant}}

Approximations of determinants of the form $\left|\bm{T}(\bar{f}_{\bm{\theta}})\right|$
typically rely on asymptotic expansions. For instance, Whittle's approximation
consists in replacing $\log\left|\bm{T}(\bar{f}_{\bm{\theta}})\right|/n$
by its limit, 
\[
\frac{1}{n}\log\left|\bm{T}(\bar{f}_{\bm{\theta}})\right|\rightarrow\int_{-\pi}^{\pi}\log\left\{ 2\pi\bar{f}_{\bm{\theta}}(\lambda)\right\} \, d\lambda.
\]
\citet{Chen2006} use more refined asymptotic results on Toeplitz
matrices \citep[e.g. ][p. 177]{Bot_sil_99:book} to obtain a more
accurate approximation. Using their approach, one obtains in the FEXP
case 
\[
\log\left|\bm{T}(\bar{f})\right|\approx D_{n}(\overline{f}_{\bm{\theta}})\stackrel{\Delta}{=}d^{2}\log n+\frac{1}{4}\sum_{j=1}^{k}j\xi_{j}^{2}+d\sum_{j=1}^{k}j\xi_{j}+\log\frac{G(1-d)^{2}}{G(1-2d)}.
\]
where $G$ is Barnes' function \citep{barnes:function}. This approach
is easily adapted to other time series models, such as ARFIMA; we
refer to \citet{Chen2006} for more details and possible extensions
to other classes of models.

\subsection{Further approximation }

The cost of the approximation described in the previous section is
that of a FFT operation, that is $O(n\log n$). This cost may be further
reduced by remarking that the approximation of the quadratic form
may be rewritten as \citep[see e.g. ][Chap. 4]{palma2007long}: 
\[
\widetilde{\bm{x}}^{T}\bm{T}(\overline{f}_{\bm{\theta}})^{-1}\widetilde{\bm{x}}\approx\widetilde{\bm{x}}'\bm{T}(\frac{1}{4\pi^{2}\overline{f}_{\bm{\theta}}})\widetilde{\bm{x}}=\frac{n}{2\pi}\int_{-\pi}^{\pi}\frac{I(\lambda)}{\overline{f}_{\bm{\theta}}(\lambda)}\, d\lambda,\quad I(\lambda)=\left|\sum_{j=1}^{n}\tilde{x}_{j}e^{ij\lambda}\right|^{2},
\]
that is, $I(\lambda)$ is the periodogram of the (centred) dataset
$\tilde{\bm{x}}$.

It is relatively easier to evaluate $I(\lambda)$ at the Fourier frequencies
$\lambda_{j}=2\pi j/n$, which suggests a further approximation, where
this integral is replaced by a Riemann sum computed over the $\lambda_{j}$:
\[
\frac{n}{2\pi}\int_{-\pi}^{\pi}\frac{I(\lambda)}{\overline{f}_{\bm{\theta}}(\lambda)}\, d\lambda\approx\sum_{j=1}^{n}\frac{I(\lambda_{j})}{\overline{f}_{\bm{\theta}}(\lambda_{j})}.
\]
Computing simultaneously the $I(\lambda_{j})$'s requires performing
a FFT transform. The cost is $O(n\log n)$, but this needs to be done
only once, for a given dataset. Then the approximate likelihood may
be evaluated for many different values of $\bm{\theta}$, at a $O(n)$
cost. Since typically about $10^{4}-10^{6}$ such evaluations are
performed when Monte Carlo sampling from the approximate posterior,
one may for practical purposes ignore the pre-computation time, and
consider this further approximation as a $O(n)$ operation.

To conclude, our final approximation takes the following form:

\begin{equation}
\tilde{p}(\bm{x}|\bm{\theta})\propto D_{n}(\bm{\theta})^{-1/2}\left\{ b+\frac{1}{2}\sum_{j=1}^{n}\frac{I(\lambda_{j})}{\overline{f}_{\bm{\theta}}(\lambda_{j})}\right\} ^{-a-n/2}.\label{eq:further_approx}
\end{equation}
This is very close in spirit to Whittle's approximation of the likelihood,
which is based on the idea that the $I(\lambda_{j})$'s are nearly
independent, with variance $\overline{f}_{\bm{\theta}}(\lambda_{j})$;
see again e.g. \citet[Chap. 4]{palma2007long}. Rigorously speaking,
we have not been able to establish that this further approximation
is valid in the sense defined in Section \ref{sub:Principle}, that
is, that the variance of an importance sampling step from the approximated
to the true likelihood converges to zero. We merely observe empirically
that this approximate likelihood is nearly indistinguishable numerically
to the more principled approximation developed in the previous Sections,
that is: 
\[
\tilde{p}(\bm{x}|\bm{\theta})\propto D_{n}(\bm{\theta})^{-1/2}\left\{ b+\frac{1}{2}\widetilde{\bm{x}}'\bm{T}(\frac{1}{4\pi^{2}\overline{f}_{\bm{\theta}}})\widetilde{\bm{x}}\right\} ^{-a+n/2}.
\]
On the other hand, we also observe that the speed improvement brought
by this further approximation is rather modest, which is in line with
the respective theoretical costs of $O(n)$ and $O(n\log n)$. From
now on, we do not distinguish between these two approximations. We
only note that an additional advantage of \eqref{eq:further_approx}
is that its gradient is easy to compute, which would make it possible
to use Langevin-type MCMC moves.

\section{Monte Carlo sampling\label{sec:Monte-Carlo-sampling}}

\subsection{Background\label{sub:MCMC}}

This section discusses Monte Carlo sampling from the approximate posterior,
that is, 
\[
\tilde{\pi}_{n}(\bm{\theta})\propto p(\bm{\theta})\tilde{p}(\bm{x}|\bm{\theta})
\]
where $p(\bm{\theta})$ is some prior density defined with respect
to parameter $\bm{\theta}$, and $\tilde{p}(\bm{x}|\bm{\theta)}$
is the approximate likelihood defined in Section \ref{sec:Approx}.
Although this discussion is, as before, not specific to the FEXP model,
we shall assume for notational simplicity that the considered model
is parametrised as follows: 
\[
\bm{\theta}=(k,\bm{\theta}_{k})\in\cup_{i\in\mathbb{N}}\left\{ i\right\} \times\mathbb{R}^{i+1}.
\]
For instance, in the FEXP case, we have seen that $\bm{\theta}_{k}=\left(\mathrm{logit}(2d),\xi_{1},\ldots,\xi_{k}\right)$.
We shall also $ $assume a nested structure for the $\bm{\theta}_{k}'s$,
that is, $\bm{\theta}_{0}=(\theta_{0})$, $\bm{\theta}_{k+1}^{T}=\left(\bm{\theta}_{k}^{T},\theta_{k+1}\right)^{T}$.
Again, in the FEXP case, $\theta_{0}=\mathrm{logit}(2d)$, $\theta_{j}=\xi_{j}$
for $j\geq1$. Finally, we denote $p_{k}(\bm{\theta}_{k})=p(\bm{\theta}_{k}|k)$,
the prior of $\bm{\theta}_{k}$, conditional of $k$, $p_{k+1}(\theta_{k+1}|\bm{\theta_{k}})$,
the prior of component $\theta_{k+1}$, conditional on the first $k$
components of $\bm{\theta}_{k+1}$ being equal to those of vector
$\bm{\theta}_{k}$, and $\tilde{p}(\bm{x}|k,\bm{\theta}_{k})$ the
approximate likelihood $\tilde{p}(\bm{x}|\bm{\theta})$ for $\bm{\theta}=(k,\bm{\theta}_{k})$.

A common approach to this problem would be to alternate the steps
described below as Algorithms \ref{alg:Gaussian-random-walk} and
\ref{alg:Birth-and-death}, that is a random walk Metropolis step
with respect to $\bm{\theta}_{k}$, conditional on $k$, and a birth-and-death
step, that is, a particular instance of \citet{Green:reversible}'s
reversible jump sampler, which attempts at either incrementing (birth)
or decrementing (death) $k$. In case a birth is proposed, the current
vector $\bm{\theta}_{k}$ is completed with a draw from the conditional
prior $p_{k+1}(\theta_{k+1}|\bm{\theta}_{k})$. In case a death step
is proposed, component $\theta_{k}$ is deleted from the current vector
$\bm{\theta}_{k}$.

\begin{algorithm}
\caption{\label{alg:Gaussian-random-walk}Gaussian random walk Metropolis step
(conditional on $k$)}

\begin{raggedright} Input: $\bm{\theta}=(k,\bm{\theta}_{k})$

\end{raggedright}

\begin{raggedright} Output: $\bm{\theta}'=(k,\bm{\theta}_{k}^{'})$

\end{raggedright}

\begin{raggedright} 1. Sample $\bm{\theta}_{k}^{\star}\sim N(\bm{\theta}_{k},\bm{\Sigma}_{k})$.

\end{raggedright}

\raggedright{}2. With probability $1\wedge r$, take $\bm{\theta}_{k}^{'}=\bm{\theta}_{k}^{\star}$,
otherwise $\bm{\theta}_{k}^{'}=\bm{\theta}_{k}$, with 
\[
r=\frac{p_{k}(\bm{\theta}_{k}^{\star})\tilde{p}(\bm{x}|k,\bm{\theta}_{k}^{\star})}{p_{k}(\bm{\theta}_{k})\tilde{p}(\bm{x}|k,\bm{\theta}_{k})}.
\]
\end{algorithm}

\begin{algorithm}
\caption{\label{alg:Birth-and-death}Birth and death step}

\begin{raggedright} Input: $\bm{\theta}=(k,\bm{\theta}_{k})$

\end{raggedright}

\begin{raggedright} Output: $\bm{\theta}'=(k',\bm{\theta}'_{k'})$

\end{raggedright}

\begin{raggedright} Constants: $\rho_{k\rightarrow k+1}=\rho_{k\rightarrow k-1}=1/2$
for $k\geq1$, $\rho_{0\rightarrow1}=1$, $\rho_{0\rightarrow-1}=0$.

\end{raggedright}

\begin{raggedright} 1. Let $\bm{\theta}^{\star}=(k^{\star},\bm{\theta}_{k^{\star}}^{\star})$,
where, with probability $\rho_{k\rightarrow k+1}$, $k^{\star}=k+1$,
$\bm{\theta}_{k^{\star}}^{\star}=(\bm{\theta}_{k},\theta_{k+1})$,
and $\theta_{k+1}\sim p_{k+1}(\xi_{k+1}|\bm{\theta}_{k})$ (birth
step); otherwise, with probability $\rho_{k\rightarrow k-1}=1-\rho_{k\rightarrow k+1}$,
$k^{\star}=k-1$, $\bm{\theta}_{k^{\star}}^{\star}=(\theta_{0},\ldots,\theta_{k-1})$
(death step).

\end{raggedright}

\raggedright{}2. Set $\bm{\theta}'=\bm{\theta}^{\star}$ with probability
$1\wedge r$, otherwise $\bm{\theta}'=\bm{\theta}$, with 
\[
r=\frac{\rho_{k^{\star}\rightarrow k}p(k^{\star})\tilde{p}(\bm{x}|\bm{\theta}^{\star})}{\rho_{k\rightarrow k^{\star}}p(k)\tilde{p}(\bm{x}|\bm{\theta})}.
\]
\end{algorithm}

Of course, many variants of this basic MCMC strategy could be considered.
However, such variants are unlikely to directly address the following
two limitations of standard MCMC: (a) calibration; and (b) local behaviour.
Calibration refers to the difficulty of choosing certain tuning parameters,
such as that the scales $\bm{\Sigma}_{k}$ in the random walk step.
It is well known that the choice of such tuning parameters may have
a critical impact on the performance of the algorithm. We shall see
in our numerical study, Section \eqref{sec:numerics}, that tuning
manually the $\bm{\Sigma}_{k}$'s is rather tedious. A more principled
approach would be to use some form of adaptive MCMC sampling, see
\citet{andrieu2008tutorial} for a review, where one uses past samples
to iteratively adapt the tuning parameters to the target distribution.
However, adaptive MCMC is less straightforward to use in trans-dimensional
settings (as an infinite number of tuning parameters $\bm{\Sigma}_{k}$
must be learnt), and also does not address the second difficulty,
which we now comment upon.

The phrase ``local behaviour'' refers to the fact that MCMC samplers
have difficulties exploring multi-modal posteriors, as they tend to
get trapped in the attraction of meaningless modes. One may wonder
if multi-modality is an issue for the classes of models considered
in this paper.

\begin{figure}
\includegraphics[scale=0.3]{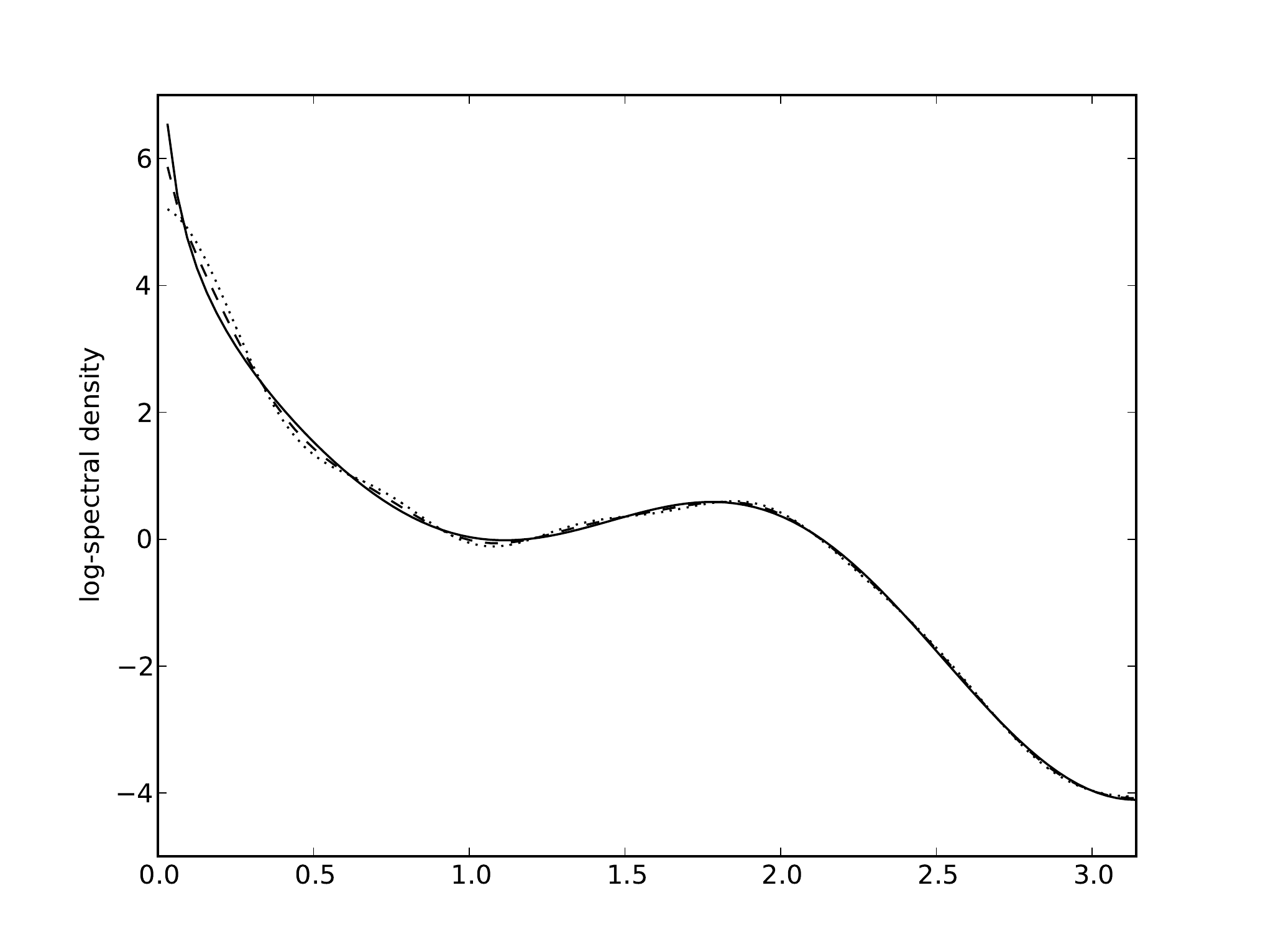}\caption{\label{fig:multi}Graphical illustration of a source of multimodality
for the FEXP model: FEXP log spectral densities, with $k=3$, $d=0.4$,
$(\xi_{1},\xi_{2},\xi_{3})=(1,-1,1)$ (solid line), compared to $k=10$,
$d=0$ (dashed line), and $d=0.2$ (dotted line), and coefficients
$\xi_{j}$ adjusted by least-squares estimation. }
\end{figure}

In our simulations, we observed that the FEXP model may indeed generate
multimodal posteriors, and we propose the following explanation. Figure
\ref{fig:multi} plots a spectral density of FEXP type, see \eqref{eq:fexp_normalised},
with $d=0.4$, $k=3$, and $(\xi_{1},\xi_{2},\xi_{3})=(1,-1,1)$.
Overlaid are two other FEXP spectral densities, but with $k=10$ in
both cases, $d=0.2$ (dotted line), $d=0$ (dashed line), and coefficients
$\xi_{j}$ fit by least-squares on the Fourier frequencies $\pi j/100,$
$j=1,\ldots,100$. These three densities are hard to distinguish except
maybe in the close vicinity of $0$. This indicates that, unless the
sample size is very large, the posterior distribution may be multi-modal,
with modes that may correspond to values of $d$ which are very far
from the true value, while $k$ is set to a larger value so as to
introduce additional components to compensate the bias in $d$.

We also refer the readers to Section \ref{sub:Performance-of-MCMC}
(and in particular Fig. \ref{fig:arfima_mcmc})
for numerical evidence of this type of multimodality in one example,
and its impact on the performance of MCMC.

\subsection{Sequential Monte Carlo}

Algorithm \ref{alg:Generic-SMC-sampler} describes a generic SMC sampler.
Compared to the more general framework of \citet{DelDouJas:SMC},
this algorithm is specialised to the case where the sequence of distributions
$(\eta_{t})$ is defined on a common sampling space, $\bm{\Theta}$,
and also to a specific choice of the backward kernel, see \citet{DelDouJas:SMC}
for more details. The algorithm depends on the specification of a
sequence of distributions $(\eta_{t})$ and a sequence of Markovian
kernels $(K_{t})$. The former must be such that it is easy to sample
from $\eta_{0}$, and that $\eta_{T}=\tilde{\pi}_{n}$, the target
distribution (the approximate posterior, as defined in the previous
section). The latter must be such that $K_{t}$ leaves invariant $\eta_{t}$,
and is typically a MCMC kernel.

\begin{algorithm}
\caption{\label{alg:Generic-SMC-sampler}Generic SMC sampler}

\begin{raggedright} All operations are for $j=1,\ldots,N$.

\end{raggedright} 
\begin{enumerate}
\item Sample $\bm{\theta}^{j}\sim\eta_{0}(\bm{\theta})$. 
\item For $t=1,\ldots,T$, do:

\begin{enumerate}
\item Reweighing step: Compute and normalise weights 
\[
w_{t}(\bm{\theta}^{j})\propto\frac{\eta_{t}(\bm{\theta}^{j})}{\eta_{t-1}(\bm{\theta}^{j})},\quad W_{t}^{j}=\frac{w_{t}(\bm{\theta}^{j})}{\sum_{l=1}^{N}w_{t}(\bm{\theta}^{l})}.
\]

\item Resampling step: sample $\hat{\bm{\theta}}^{j}$ from the multinomial
distribution that assigns probability $W_{t}^{l}$ to value $\bm{\theta}_{t}^{l}$,
$l=1,\ldots,N$. 
\item Move step: regenerate $\bm{\theta}^{j}$ as 
\[
\bm{\theta}^{j}\sim K_{t}(\hat{\bm{\theta}}^{j},d\bm{\theta})
\]
where $K_{t}$ is a Markovian kernel that leaves $\eta_{t}$ invariant. \end{enumerate}
\end{enumerate}
\end{algorithm}

For $(\eta_{t})$, we take a particular annealing sequence \citep{Neal:AIS},
that is, a geometric bridge between the prior and the approximate
posterior 
\begin{equation}
\eta_{t}(\bm{\theta})\propto p(\bm{\theta})\left\{ \tilde{p}(\bm{x}|\bm{\theta})\right\} ^{\gamma_{t}}\label{eq:tempering_sequence}
\end{equation}
with $\gamma_{0}=0<\ldots<\gamma_{T}=1$. The weight function is then:
\[
w_{t}(\bm{\theta})=\tilde{p}(\bm{x}|\bm{\theta})^{\alpha_{t}},\quad\alpha_{t}=(\gamma_{t}-\gamma_{t-1}).
\]

We shall briefly discuss in the conclusion an alternative choice for $(\eta_{t})$, based on the IBIS strategy of \cite{Chopin:IBIS}, which may be useful in sequential estimation scenarios. 

We follow \citet{jasrainference}, \citet{SchaferChopin}, and adjust
dynamically the annealing coefficients $\gamma_{t}$ by solving at
iteration $t$, with respect to variable $\alpha_{t}$, the equation
\[
\frac{\left(\sum_{j=1}^{N}w_{t}(\bm{\theta}^{j})\right)^{2}}{\sum_{j=1}^{N}w_{t}(\bm{\theta}^{j})^{2}}=cN
\]
for some fixed $c\in(0,1)$; in our simulations, we took the default
value $c=1/2$, and we used Brent's root-find algorithm \citep[Section 9.3]{NumericalRecipes3rdEd}
to solve numerically this equation. The left hand side is the effective
sample size of \citet{Kong1994}; it takes values in $[1,N]$, and
is a convenient measure of the weight degeneracy.

For $(K_{t})$, we use $M$ steps of the MCMC sampler described in
the previous section, that is, we repeat $M$ times the following
sequence: Algorithm \ref{alg:Gaussian-random-walk} (random walk Metropolis),
then Algorithm \ref{alg:Birth-and-death} (birth-and-death). A big
advantage of the SMC framework is that we can use the current particle
system to calibrate these MCMC steps. Specifically, before each move
step we set $\bm{\Sigma}_{l}=\tau_{l}\bm{S}_{l}$, $\tau_{l}=2.38^{2}/(l+1)$,
where $\bm{S}_{l}$ is the covariance matrix of those resampled particles
$\hat{\bm{\theta}}^{(j)}=\left(\hat{k}^{(j)},\bm{\theta}_{\hat{k}^{(j)}}^{(j)}\right)$
such that $\hat{k}^{(j)}=l$; in case this set is empty, we take instead
$\bm{S}_{l}=\bm{I}_{l}$. This particular choice is motivated by the
theory on optimal scaling of Hastings-Metropolis kernel, as developed
in \citet{RobertsRosenthal:OptimalScalingMH}.

The only tuning parameters of this algorithm are the number of particles
$N$, and the number of MCMC steps performed at each move step, $M$.
Regarding the choice of $M$, we observed the following interesting phenomenon.
Increasing $M$ from 1 to some small integer (say in the range $5-20$) often led to a dramatic improvement (as compared to increasing $N$, relative to the same CPU cost). In particular, one observes cases where particles are close to IID 
(in the sense that the empirical variance of certain particle estimates over repeated runs is close to the variance of the corresponding estimate for IID 
particles). At this point, increasing further $M$ seems to bring no improvement. 
This point is illustrated in our numerical study, see Section \ref{sec:numerics}.
This phenomenon seems related to a recent result (obtained in a slightly different context) by \citet{dubarry2011particle}, who establish a central limit theorem, with 
asymptotic variance equal to the variance under the target distribution (hence
the same as for IID particles), by taking $M=\log(N)$. This formal result 
seems to suggest that one does not need to take $M$ very large to obtain
nearly IID particles. At any rate, this point surely deserves more investigation
in other contexts.

Again, one may propose many variants of this SMC sampler. For instance,
one could also use adaptive reversible jump steps, where the jump
probabilities $\rho_{k\rightarrow k+1}$, $\rho_{k\rightarrow k-1}$
(currently set to $1/2$) could be optimised using the particle sample,
or even design more elaborate proposals based on a family of linear
transforms, as in \citet{GreenAdaptRJ}, where the corresponding matrices
may also be learnt from the particle sample. (We did some experiments
with this strategy, but did not obtain significantly better results
in the examples we looked at.) However, our main focus here is to
develop a black box algorithm, which requires as little input from
the user as possible, while giving reasonable performance, even in
the presence of multimodality. Our numerical experiments seem to indicate
this is the case, see Section \ref{sec:numerics}.

\section{Numerical experiments\label{sec:numerics}}

\subsection{Settings\label{sub:Settings}}

The first part of this numerical study focus on the following simulated
example. We sampled a long time series from an ARFIMA$(1,d,1)$ model,
with $d=0.45$, and AR (resp. MA) coefficient $-0.9$ (resp. $-0.2$),
and applied the FEXP model described in the Introduction. We find
this example to be challenging, because the spectral density of the
simulated process has a particular shape which is not easily approximated
by a FEXP spectral density, at least for small values of $k$; see
e.g. Fig. \ref{fig:Consistency} and additional comments around this
Figure. (From a modelling perspective, the presence of an autoregressive
root close to one makes it difficult to determine whether the persistence
in the data is really characteristic of long range behaviour.) Thus,
and as described more properly in \citet{RousseauChopinLiseo2012},
the FEXP model described in the introduction must be understood as
a semi-parametric procedure, which makes it possible to consistently
estimate the true spectral density (and related quantities, e.g. the
long-range coefficient $d$), provided this spectral density belongs
to some infinite-dimensional class of functions (of a certain regularity).
In particular, one expects the posterior of $k$ to shift towards
infinity as $n$ goes to infinity, which adds to the computational
difficulty. The second part of the study considers a real dataset,
see Section \ref{sub:Real-data-study}.

Except in the consistency study (Section \ref{sub:Consistency-study}),
the dataset consists of the $n=10^{4}$ first values of the simulated
time-series. In the consistency study, different sample sizes are
compared, by again taking the $n$ first points for different values
of $n$. Except in the prior sensitivity section (Section \ref{sub:prior-sensitivity-analysis}),
we always take the following prior: independently, $d\sim\mathrm{Uniform}[0,1/2]$,
$k\sim\mathrm{Geometric}(1/5)$ (with support $\left\{ 0,1,2,\ldots\right\} $),
$\xi_{j}\sim N(0,100j^{-2\beta})$, $\mu|\sigma\sim N(0,\sigma^{2}/g_{\mu})$,
$1/\sigma^{2}\sim\mathrm{Gamma}(a,b)$, with hyper-parameters $\beta=1$,
$a=b=0.5$, $g_{\mu}=0.1$. Note that this prior falls slightly outside
the class of prior distributions that ensures consistency, according
to \citet{RousseauChopinLiseo2012}. In fact, we found it interesting
to see whether the conditions to ensure consistency of \citet{RousseauChopinLiseo2012}
may be too strong; our numerical study seems to indicate it may indeed
be the case.

All the simulations presented in this Section were performed on a
standard 3GHz desktop computer, without resort to any form of parallelisation,
and implemented in the Python language (using the NumPy and SciPy
libraries); the programs are available on the first author's web-page.

\subsection{Performance of SMC\label{sub:Performance-of-SMC}}

This section is concerned with the performance of the SMC sampler,
regarding sampling the approximate posterior. Note that the final
correction step (i.e. the importance sampling step from the approximate
to the true posterior) is therefore not performed; see Section \ref{sub:Consistency-study}
for an evaluation of the performance of the correction step.

\begin{figure}
\begin{centering}
\includegraphics[scale=0.3]{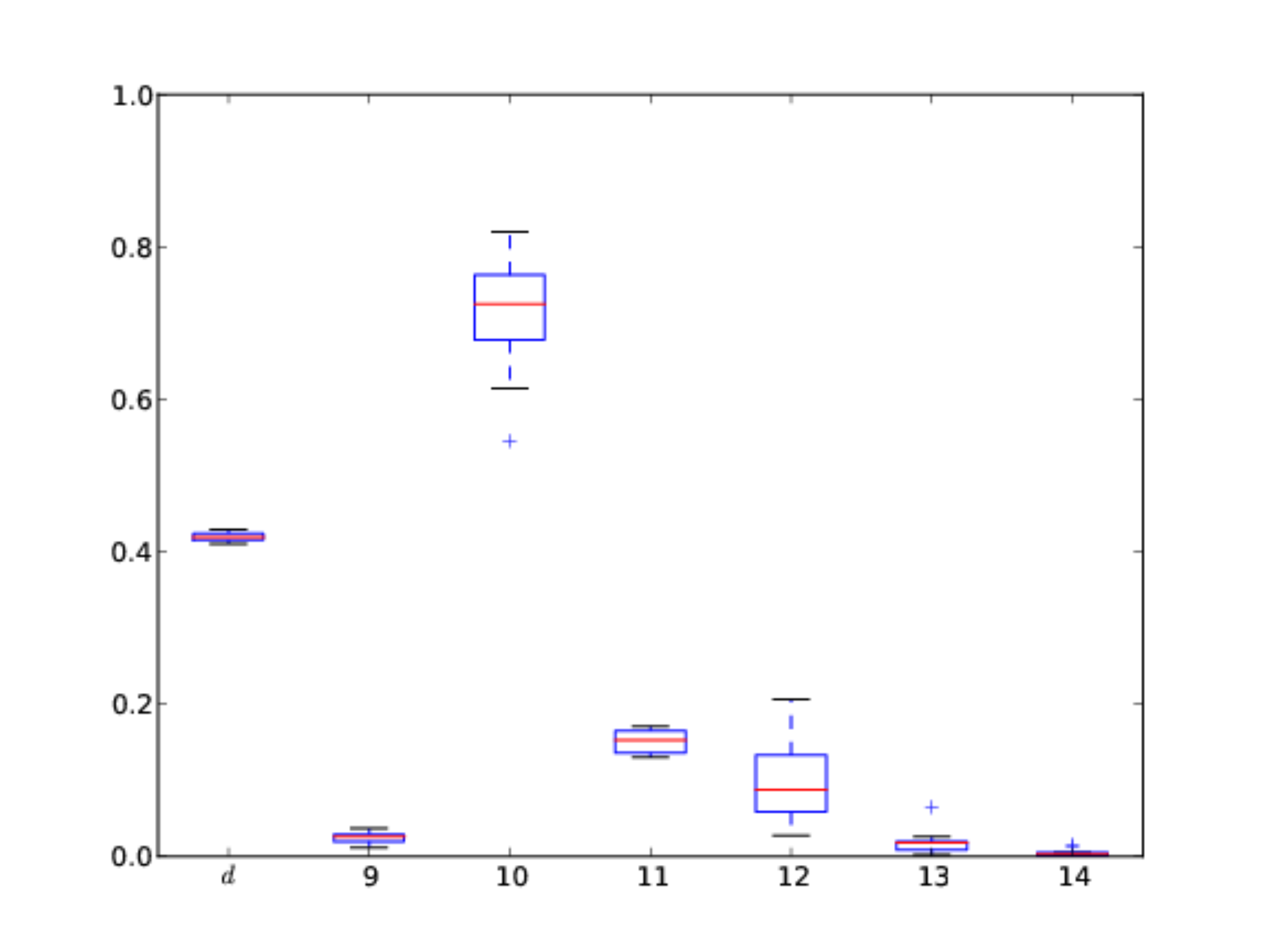}\includegraphics[scale=0.3]{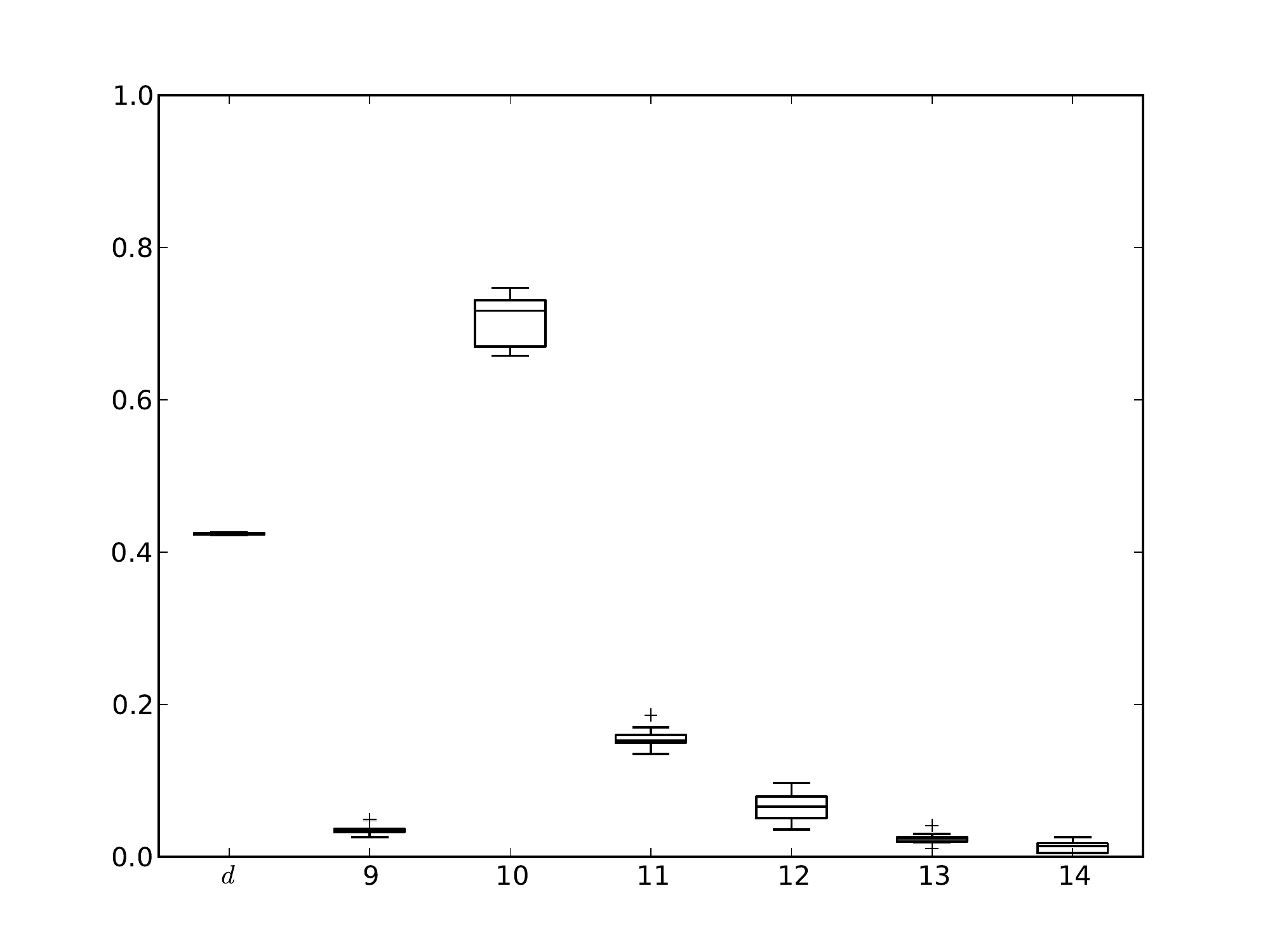} 
\par\end{centering}

\begin{centering}
\includegraphics[scale=0.3]{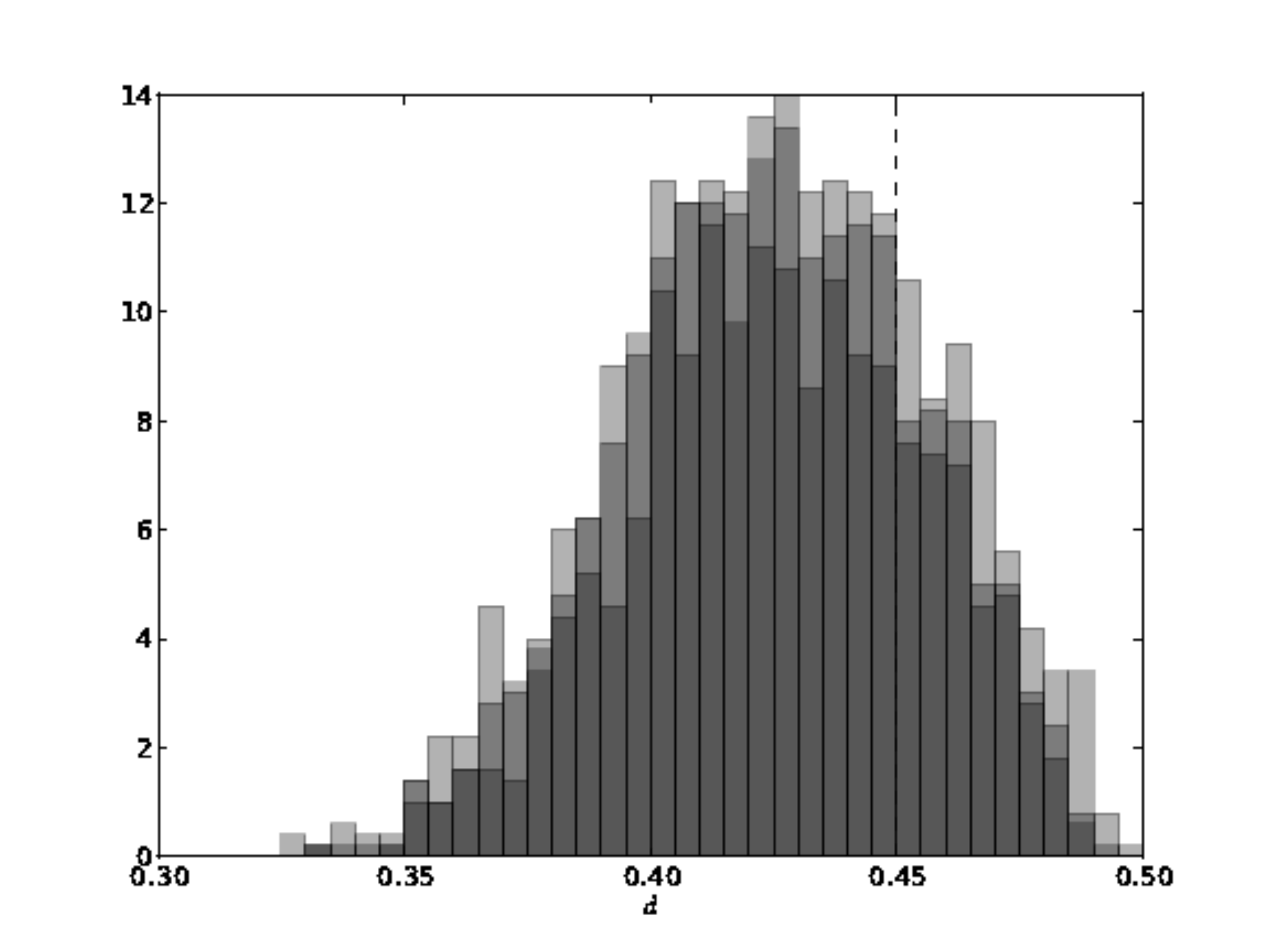}\includegraphics[scale=0.3]{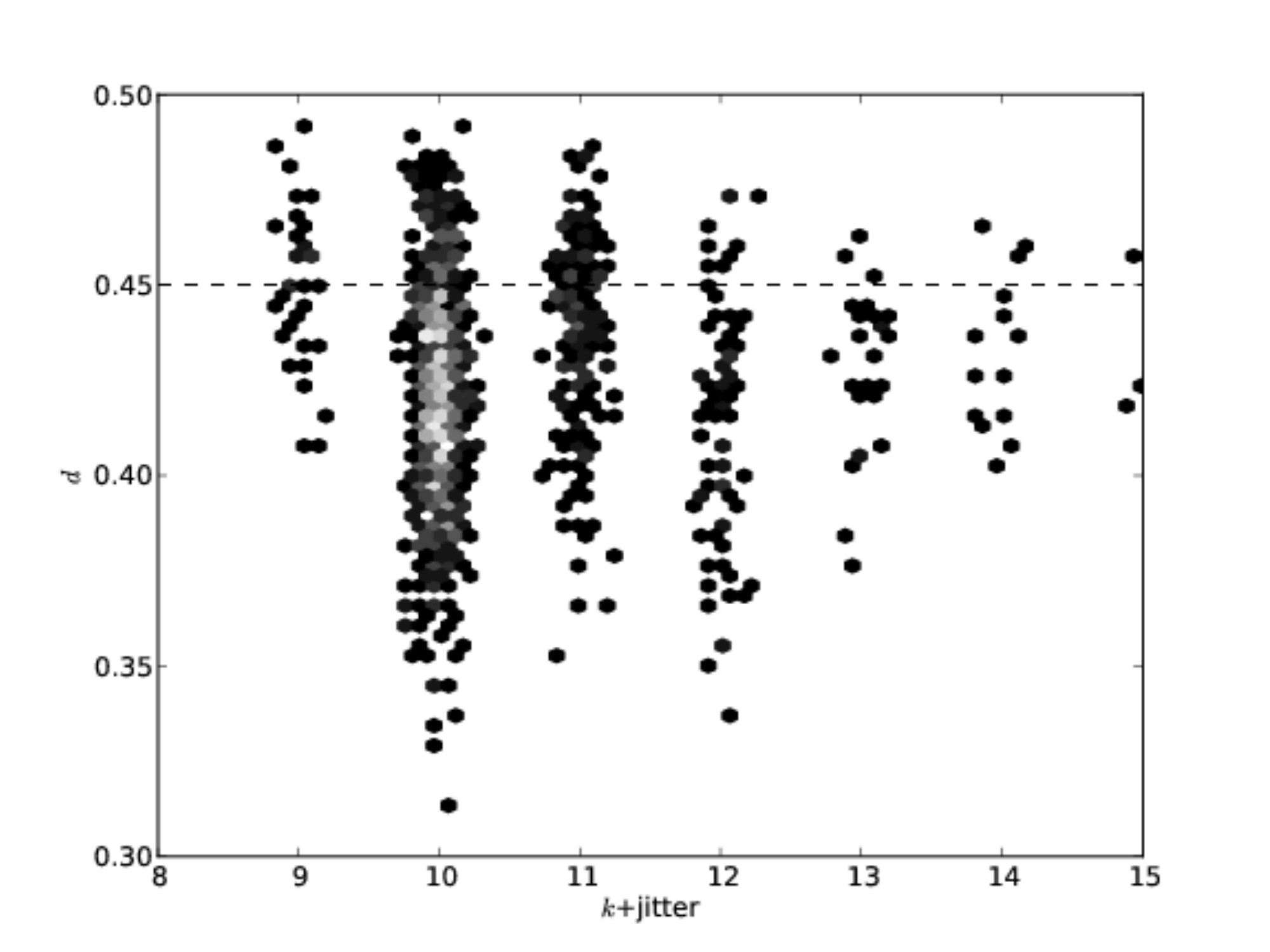} 
\par\end{centering}

\centering{}\caption{\selectlanguage{english}%
\label{fig:arfima}\foreignlanguage{british}{Top row: box-plots, over
$10$ repeated runs, of the SMC estimates of the posterior expectation
of $d$ and the posterior probabilities of $k=9,\ldots,14$ (Left:
$(N,M)=(10^{3},5)$; Right: $(N,M)=(10^{3},20)$); Bottom Left: marginal
posterior of $d$, as estimated by superimposed (with transparency
effects) weighted histograms obtained by 3 first SMC runs; Bottom
Right: hexagonal binning plot of $k$ versus $d$, from the first
SMC run; a $N(0,0.1^{2})$ jitter is added to $k$, and colour is
proportional to the sum of the weights of the particles falling in
each hexagon.}\selectlanguage{british}%
}
\end{figure}

\begin{figure}
\centering{}\includegraphics[scale=0.3]{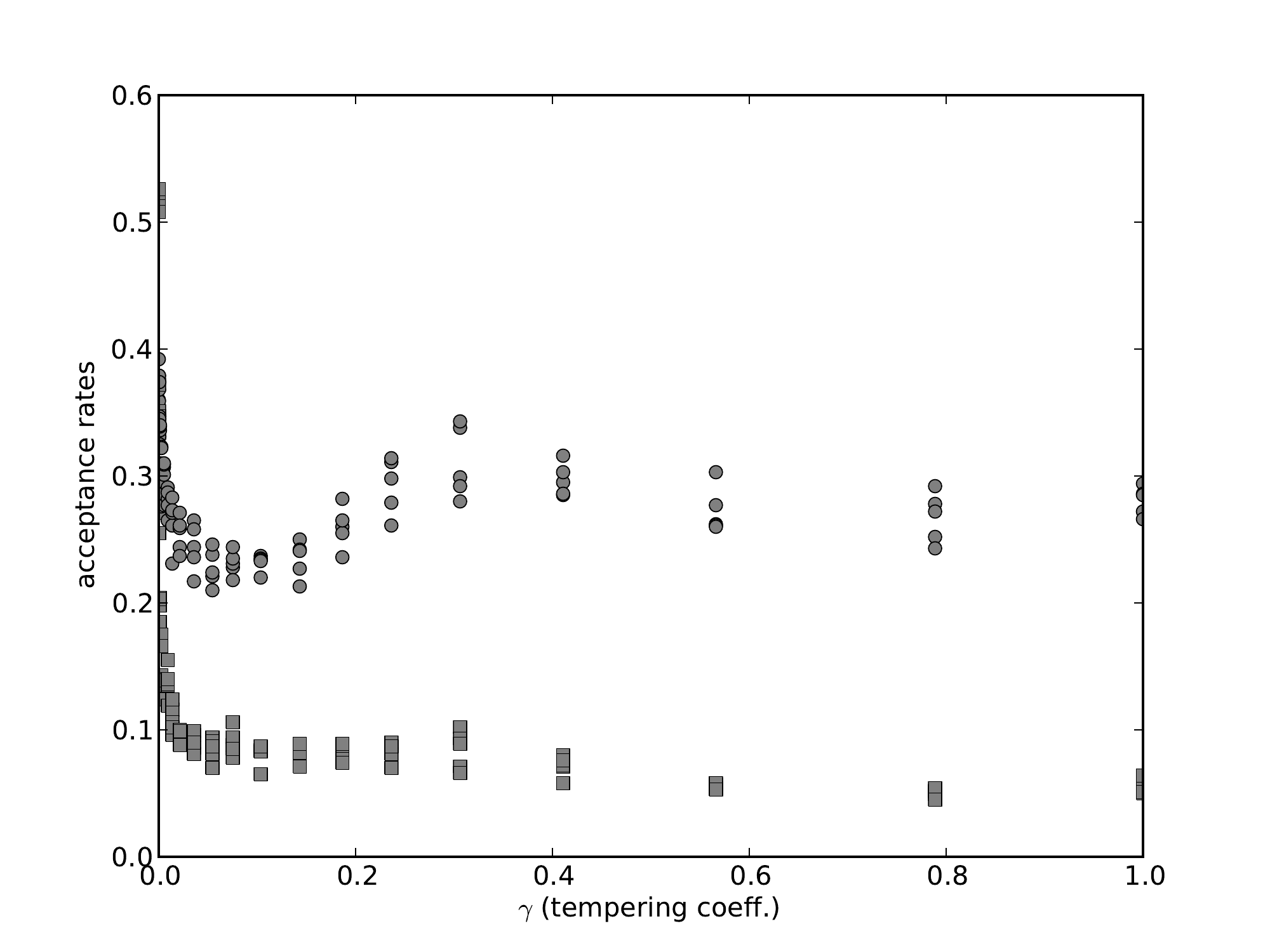}\includegraphics[scale=0.3]{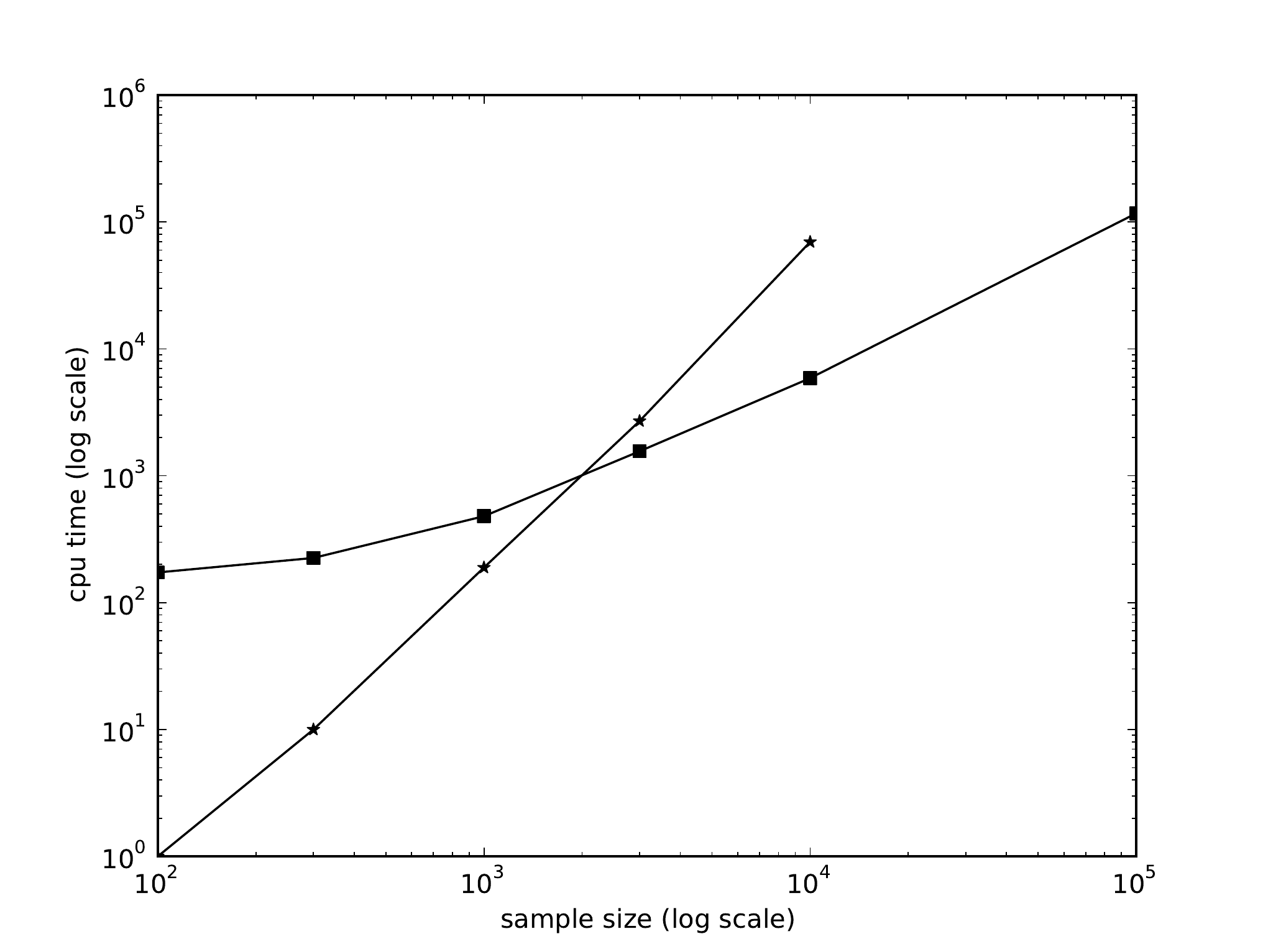}\caption{\label{fig:performanceSMC}Left: acceptance rates of the random walk
step (circles) and the birth and death step (squares) versus the tempering
coefficient $\gamma_{t}$, which progresses from $0$ to $1$ during
one run of the algorithm ($N=1000$, $M=5$). Right: CPU times in seconds for
the SMC sampler targeting the approximate posterior (squares) and
the correction step (stars) versus the sample size of the data. }
\end{figure}

We run the SMC sampler 10 times, for $N=1000$ particles, and $M=5$
MCMC steps; CPU time is 20 minutes per run. Variability over these
ten runs is small for posterior expectations of the $\xi_{j}$'s (not
shown) and $d$ (see top left plot of Figure \ref{fig:arfima}), but
is a bit larger for the posterior distribution of $k$ (see same plot).
The left plot of Figure \ref{fig:performanceSMC} reveals that the
acceptance rate of the birth and death step is is often below $10\%$,
which suggests $M=5$ steps may not be enough to successfully move
the particles with respect to component $k$. In that plot, the $x-$axis
gives the annealing coefficient; that is $\gamma_{t}$ at iteration
$t$, which goes from $\gamma_{0}=0$ to $\gamma_{T}=1$, 
see
\eqref{eq:tempering_sequence}. Note that this axis is a non-linear
function of elapsed CPU time, as the sequence $\gamma_{t}$ grows
slowly at the beginning, and progressively accelerates, whereas the
cost per iteration $t$ should be roughly constant.

We run again the SMC sampler with $N=1000$ particles, but this time
with $M=20$ steps; CPU time is then $1$ hour 20 minutes. We obtain
very satisfactory results; see the box-plots in Figure \ref{fig:arfima}.
In particular, regarding the posterior expectation $\hat{d}$ (with
respect to the approximated posterior) of component $d$, we observe
that the empirical variance of the estimates of $\hat{d}$ obtained
from the ten runs is very close to $\mathrm{Var}^{\tilde{\pi}_{n}}(d)$,
that is, the variance would be obtained from $N=1000$ independent
and identically distributed simulations from the posterior. Figure
\ref{fig:arfima} also reports superimposed weighted histograms obtained
from the second set of runs, and a particle approximation of the bivariate
posterior distribution of $(k,d)$. (Some noise was added to $k$
for the sake of visualisation.)

\subsection{Performance of MCMC\label{sub:Performance-of-MCMC}}

This section compares the performance of our SMC sampler with standard
MCMC. As the previous section, this comparison is in terms of sampling
from the approximate posterior, and no correction step is performed
at the end of the algorithm. The MCMC sampler we use is the one described
in Section \ref{sub:MCMC}.

We simulate several MCMC chains with different tuning parameters,
but always of size $N=1.6\times10^{5}$, which gives a running time
per chain of about 50 minutes (as compared to 20 minutes for the first
sequence of SMC runs).

We start by 10 runs, with $\bm{\Sigma}_{k}$, the scale of the random
walk in the $\bm{\theta}_{k}$-space, set to $\tau\bm{I}_{k+1}$ ($\tau$
times the identity matrix of rank $k+1$); after some pilot runs,
we take $\tau=0.015$ so as to obtain an acceptance rate for the random
walk step close to $25\%$.

\begin{figure}
\includegraphics[scale=0.3]{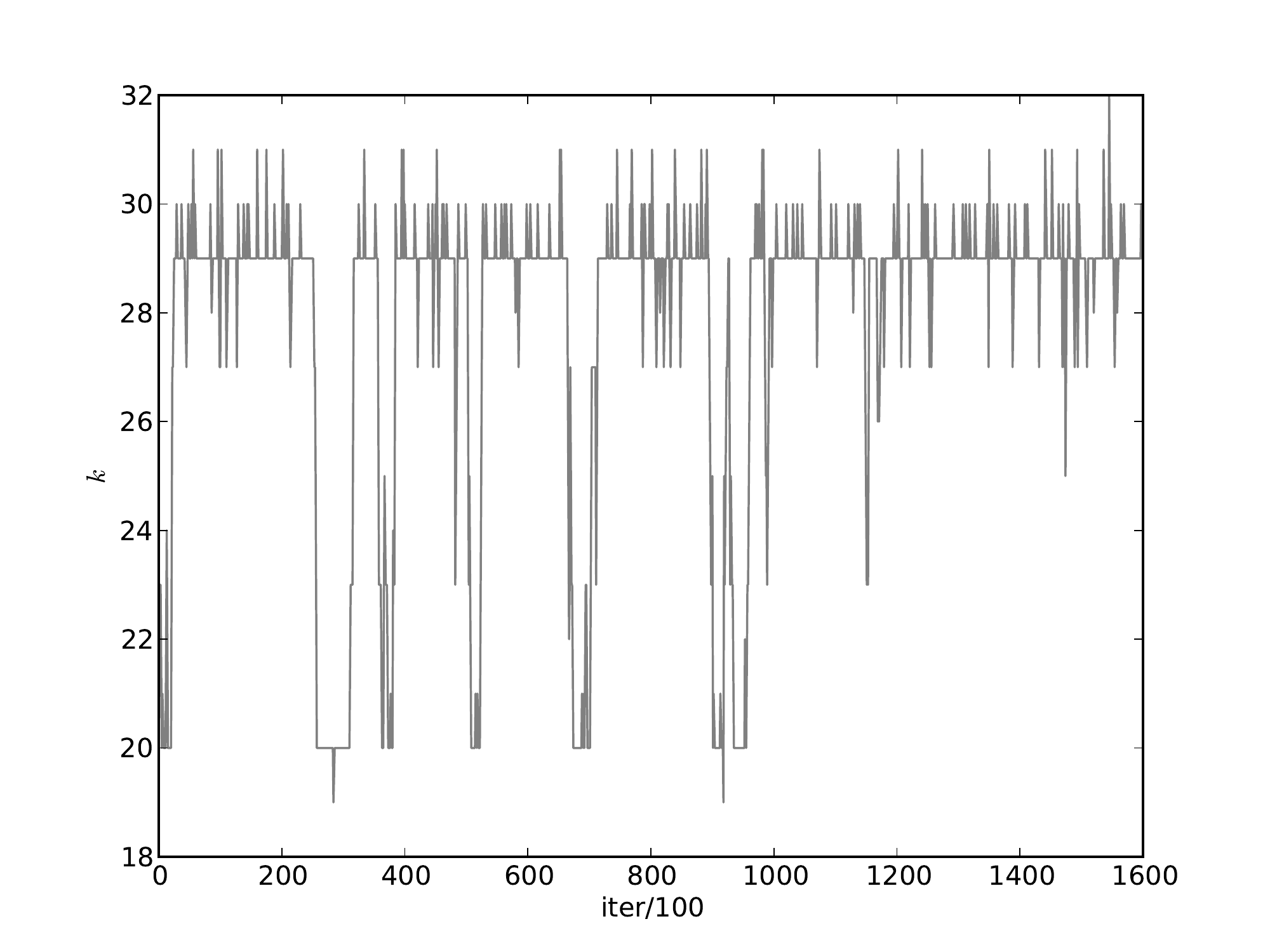}\includegraphics[scale=0.3]{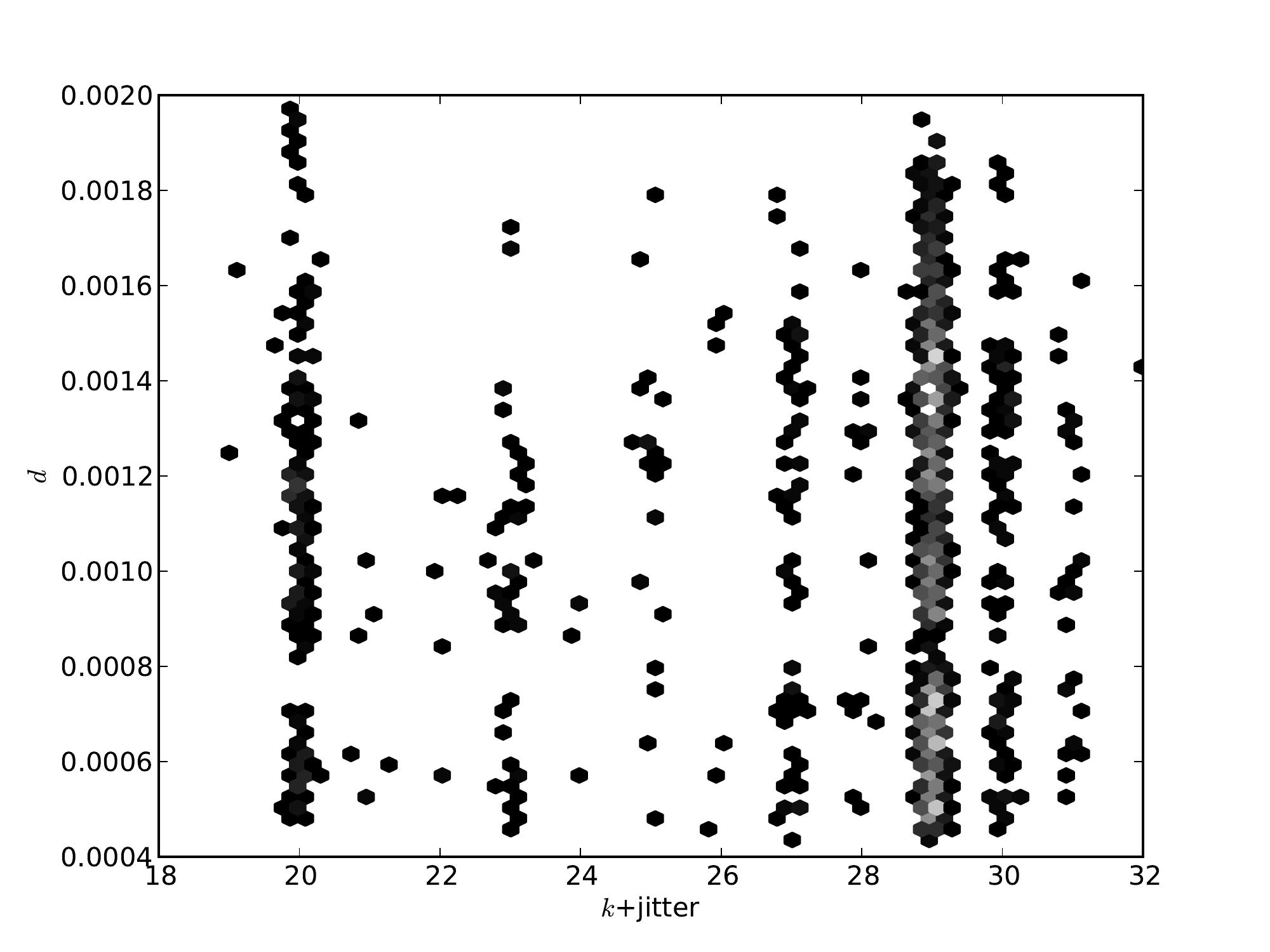}

\includegraphics[scale=0.3]{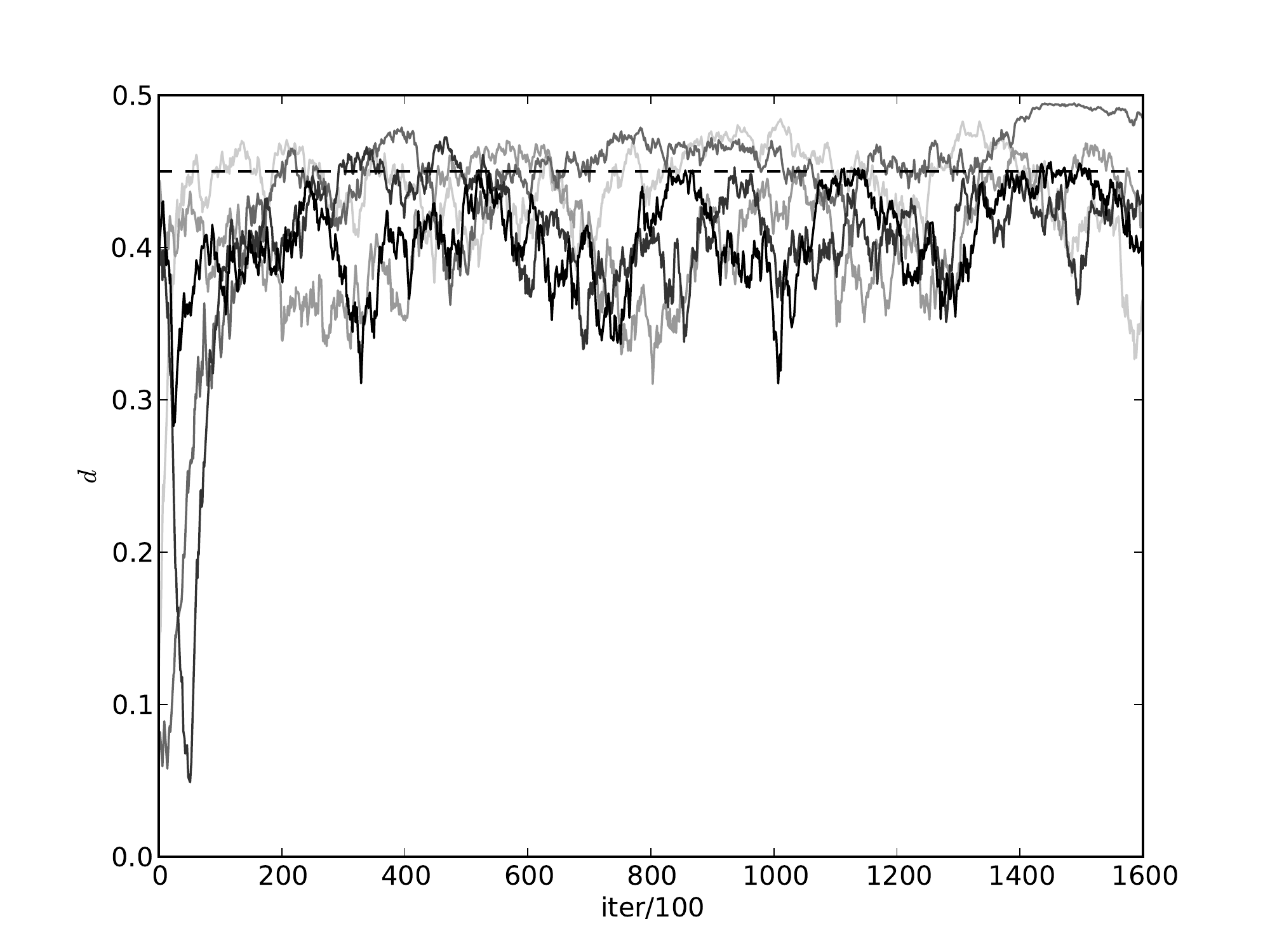}

\includegraphics[scale=0.3]{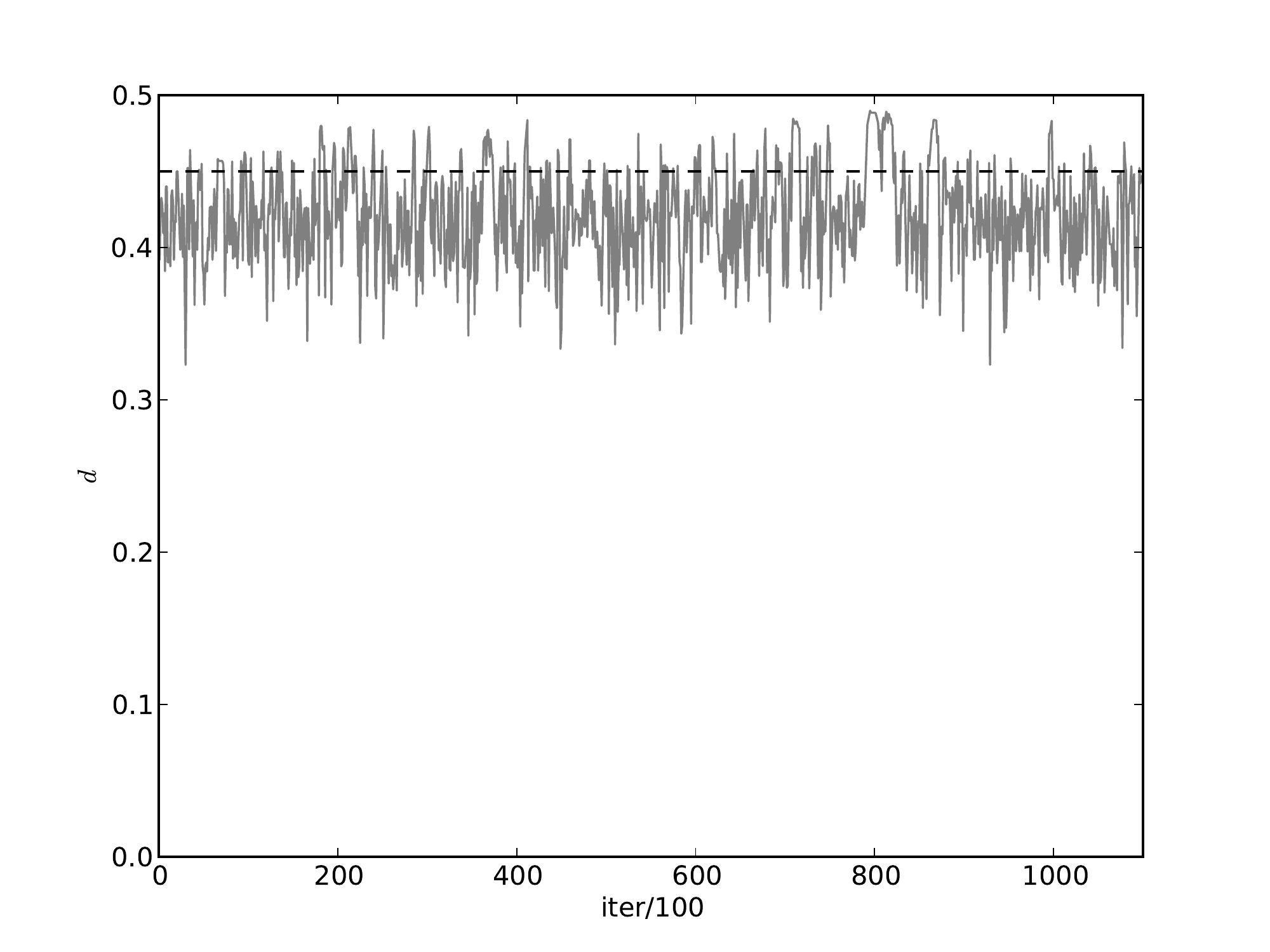}\includegraphics[scale=0.3]{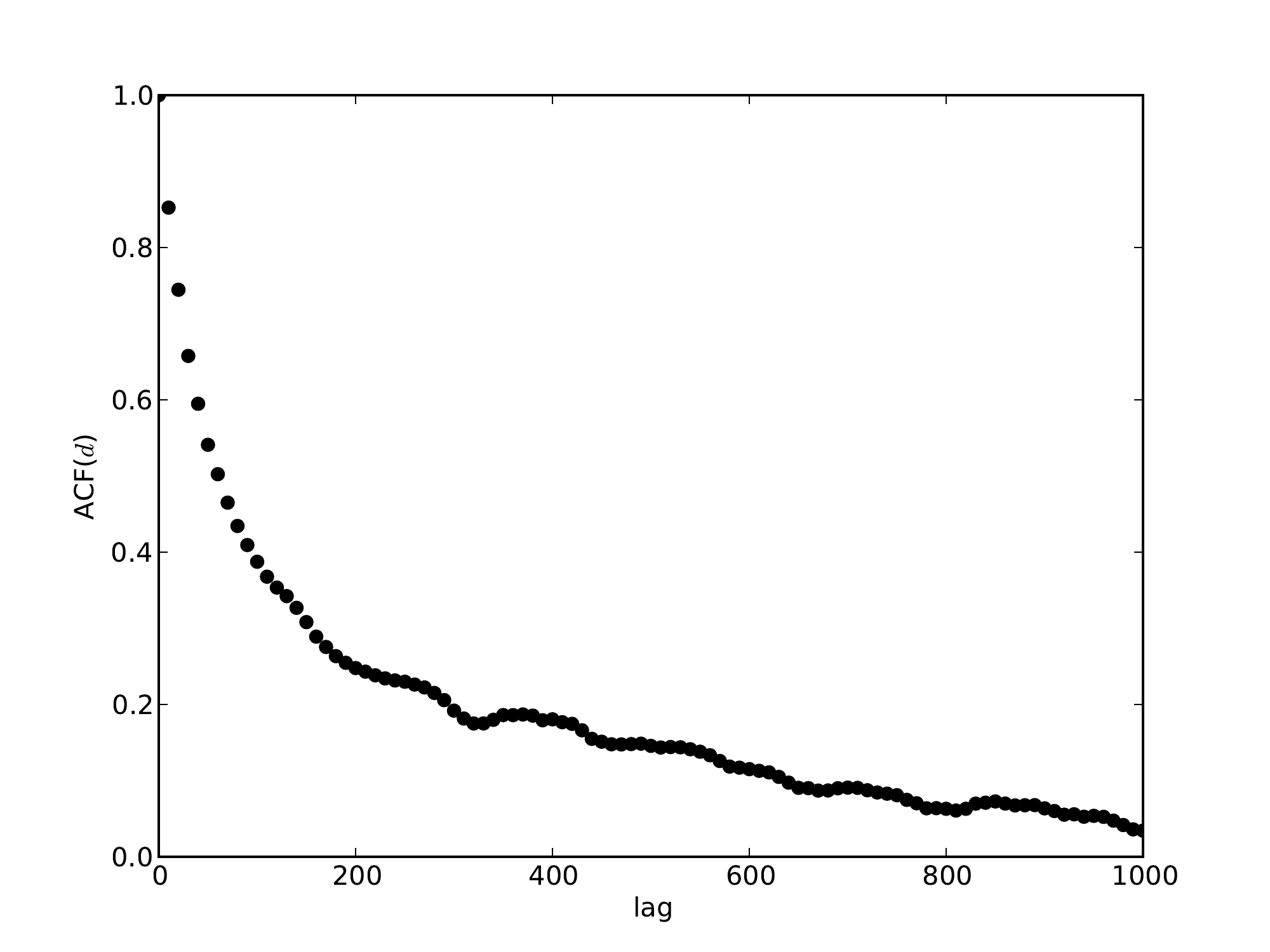}\caption{\selectlanguage{english}%
\label{fig:arfima_mcmc}\foreignlanguage{british}{Top row: one MCMC
run alternating between two minor modes (Left: trace of $k$, Right:
scatter-plot of $(k,d)$, with a jitter term added to $k$ as in Fig. \ref{fig:arfima}). Middle row: traces of $d$ for five out of the nine remaining MCMC
chains. Bottom row: adaptive MCMC; trace of $d$ (left) and ACF plot
(right), computed post burn-in. }\selectlanguage{british}%
}
\end{figure}

One chain seems to alternate between two local modes, one around $(k,d)=(20,10^{-3})$,
the other around $(k,d)=(30,10^{-3})$; see first row of Figure \ref{fig:arfima_mcmc}.
The nine other chains do not get trapped in these local modes; however
their mixing properties seem quite poor; see the MCMC traces for $d$
the middle row plot. Note how the darkest trace never seems to visit
the region $d>0.45$, which accounts for $20\%$ of the posterior
mass, according to our SMC runs.

These results are hardly satisfactory. We also experimented with $\bm{\Sigma}_{k}$
set to $\tau$ times the prior covariance matrix of $\bm{\theta}_{k}$,
but this did not give better results (not shown). Finally, we implemented
a (crudely) adaptive version of our random walk Metropolis sampler,
for the posterior conditional on $k=10$ (no reversible jump steps),
where, during a burn-in period of $5\times10^{4}$ iterations, the
covariance matrix of the random walk proposal is re-calibrated every
$10^{3}$ iterations, using past samples; specifically $\bm{\Sigma}_{k}$
is set to $(2.38^{2}/(k+1))\bm{S}_{t}$ \citep[see ][]{RobertsRosenthal:OptimalScalingMH},
where $\bm{S}_{t}$ is the empirical covariance matrix computed on
the $t$ first iterations. One sees that learning the correlation
structure of the posterior (conditional on $k$) significantly improves
the mixing of the chain, although the auto-correlation function (with
respect to $d$, and computed post burn-in) decays slowly, see bottom
row of \ref{fig:arfima_mcmc}. However, learning the posterior covariance
matrix, conditional on $k$, for each $k$ in a given range, would
of course be  wasteful.

\subsection{Consistency study\label{sub:Consistency-study}}

In this section, we show how both the approximate and the true posterior
evolve as the sample size grows. The goal is to assess both the asymptotic
properties of our FEXP semi-parametric procedure, and the effect of
the correction step on inference. Results are obtained from a single
SMC run.

\begin{figure}
\includegraphics[scale=0.3]{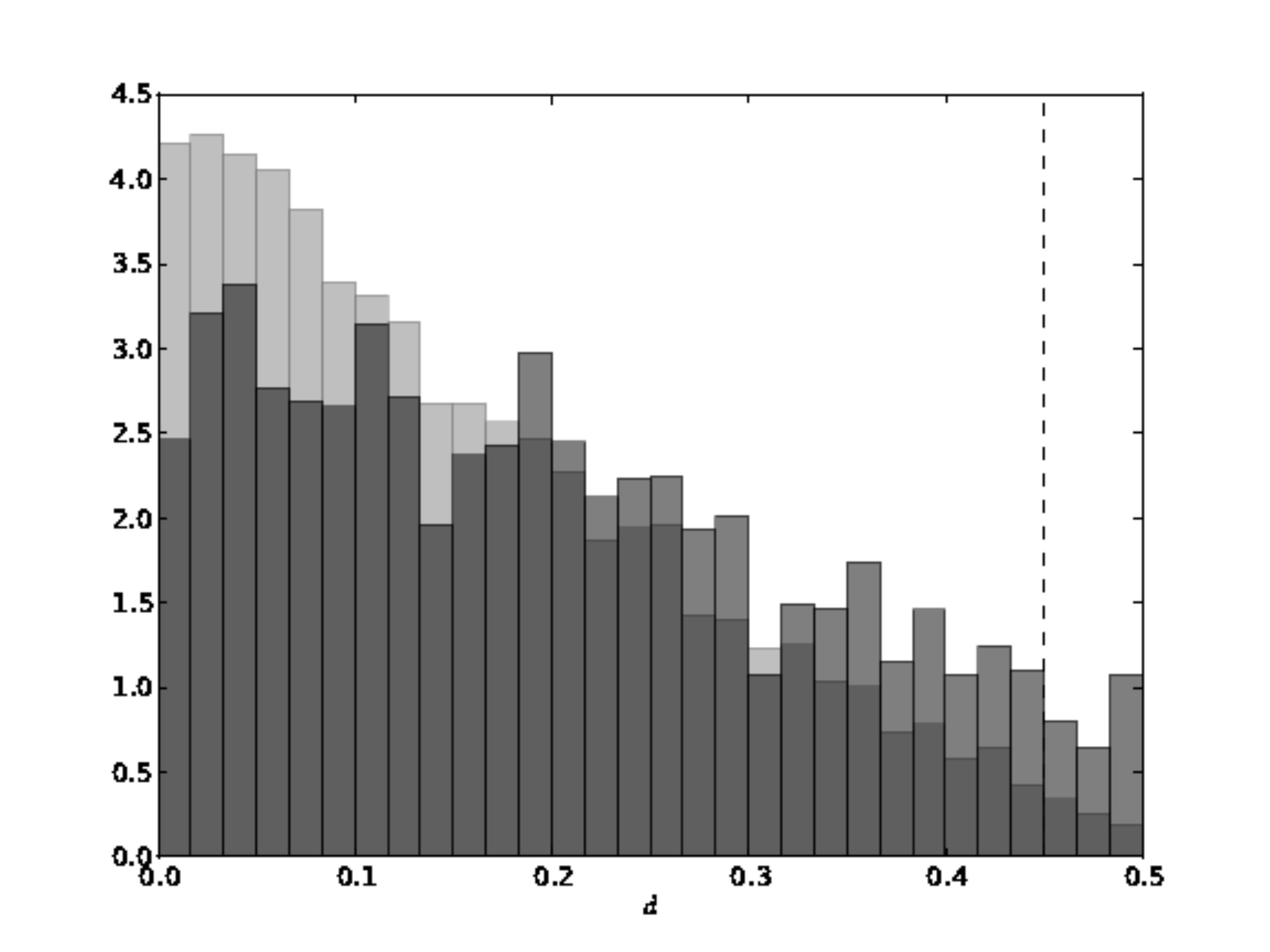}\includegraphics[scale=0.3]{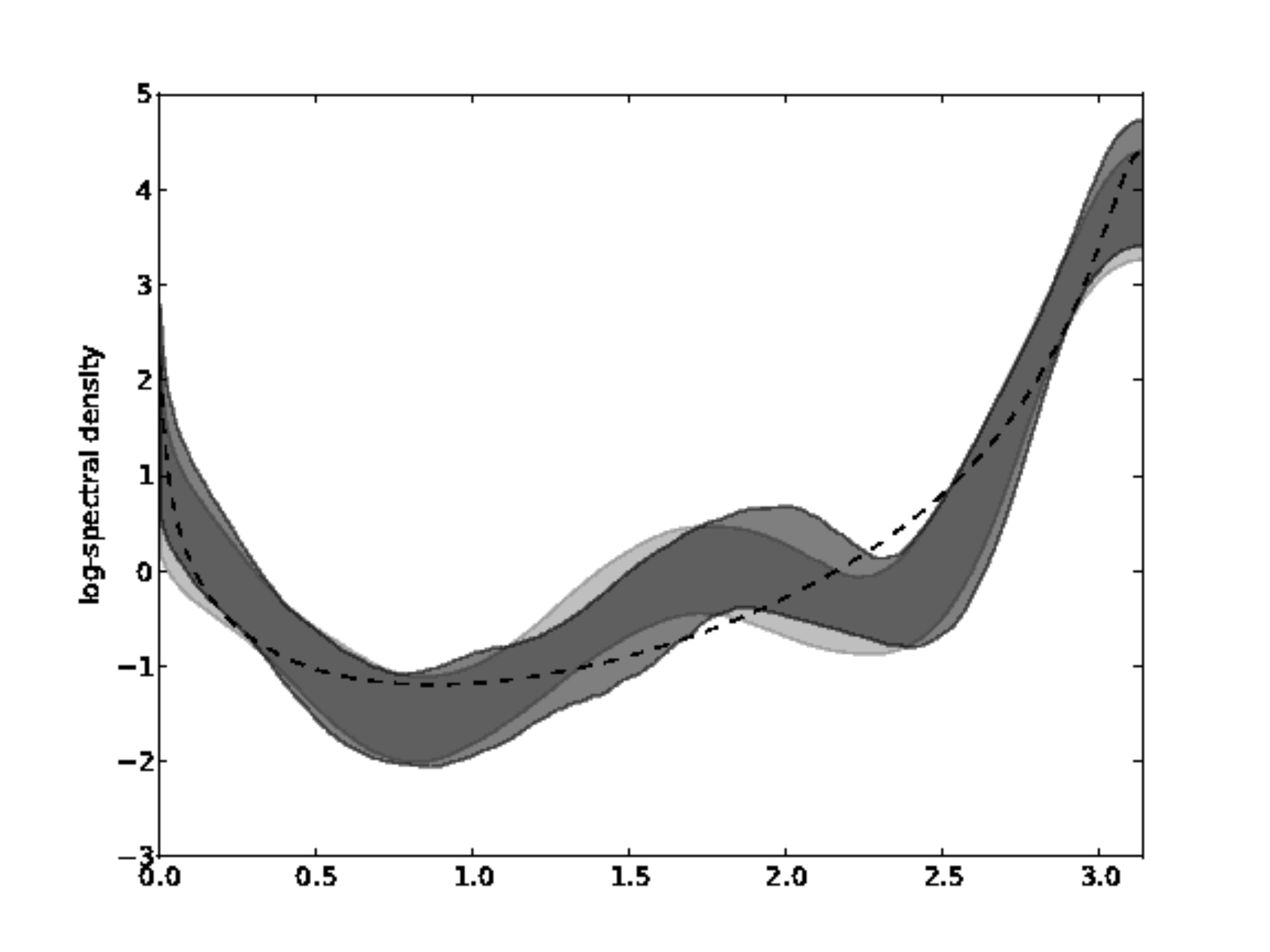}

\includegraphics[scale=0.3]{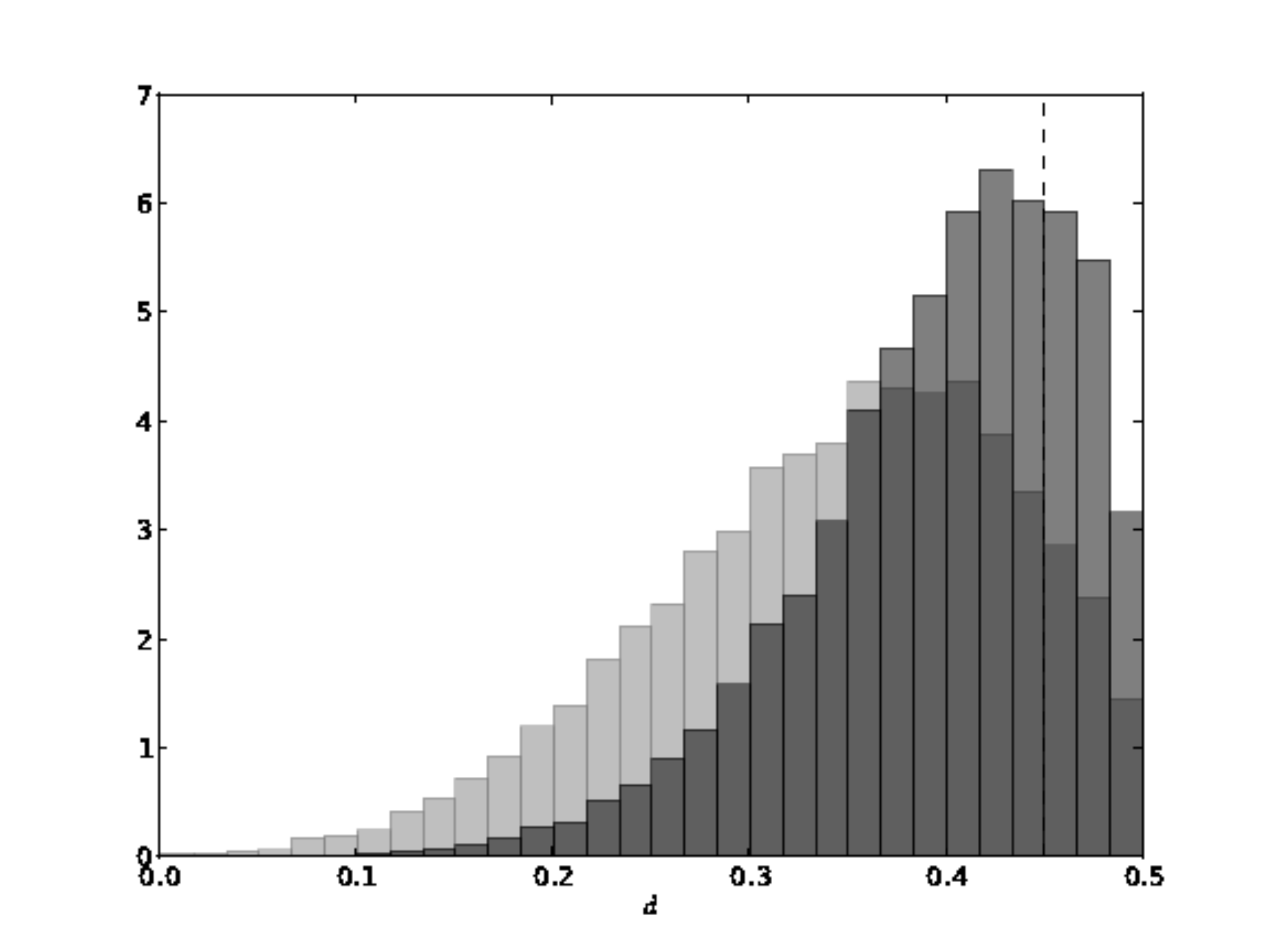}\includegraphics[scale=0.3]{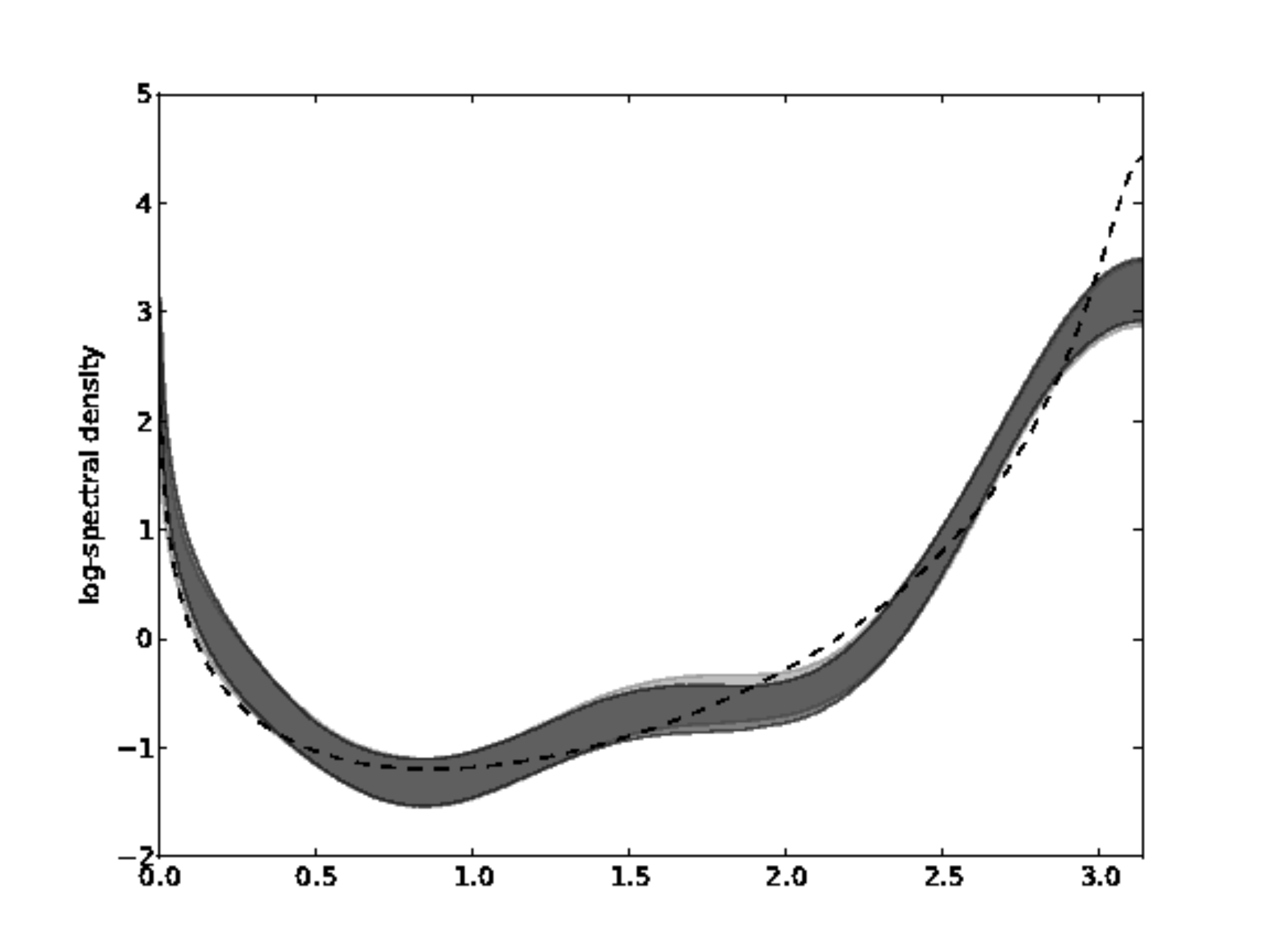}

\includegraphics[scale=0.3]{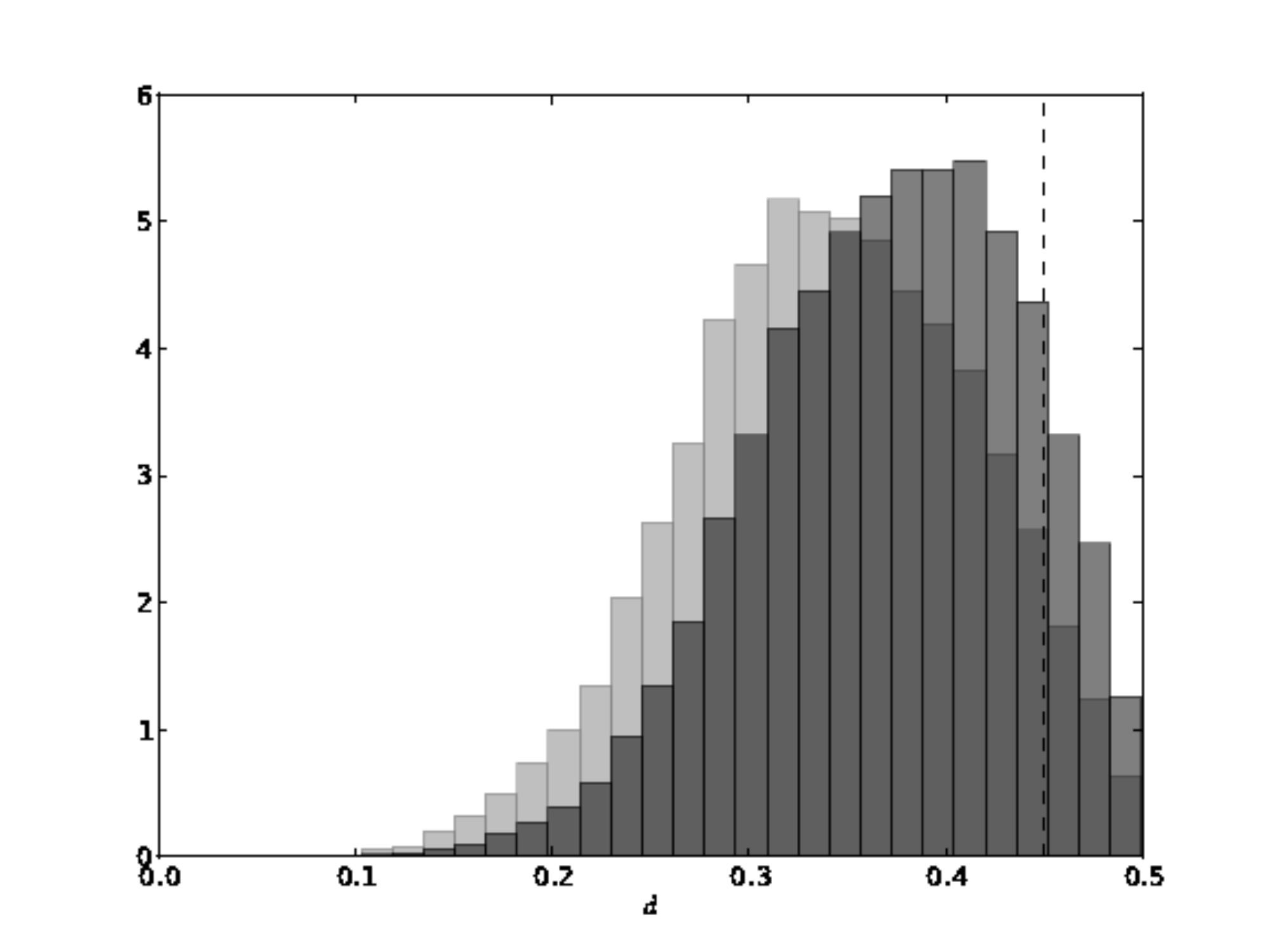}\includegraphics[scale=0.3]{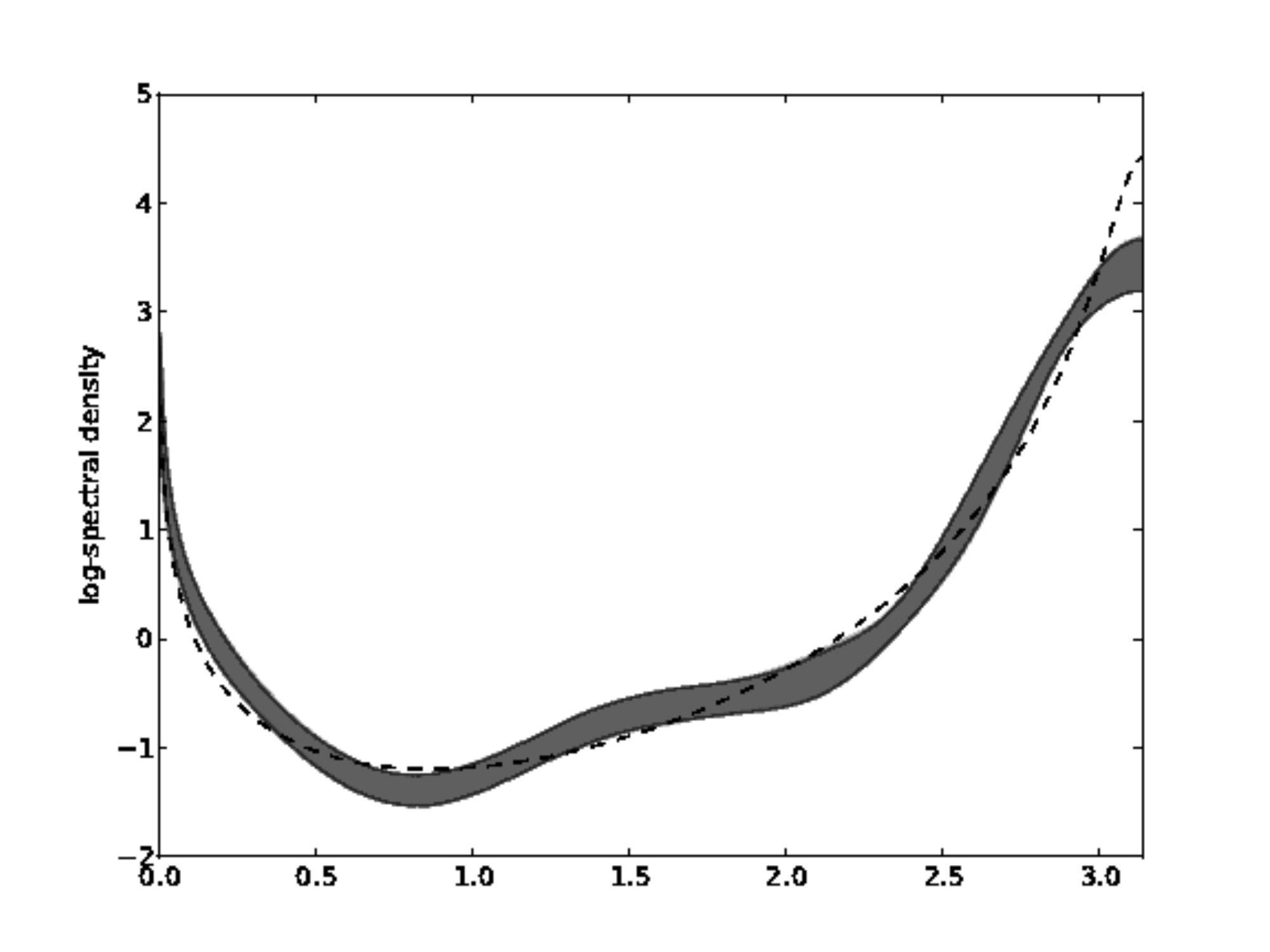}

\includegraphics[scale=0.3]{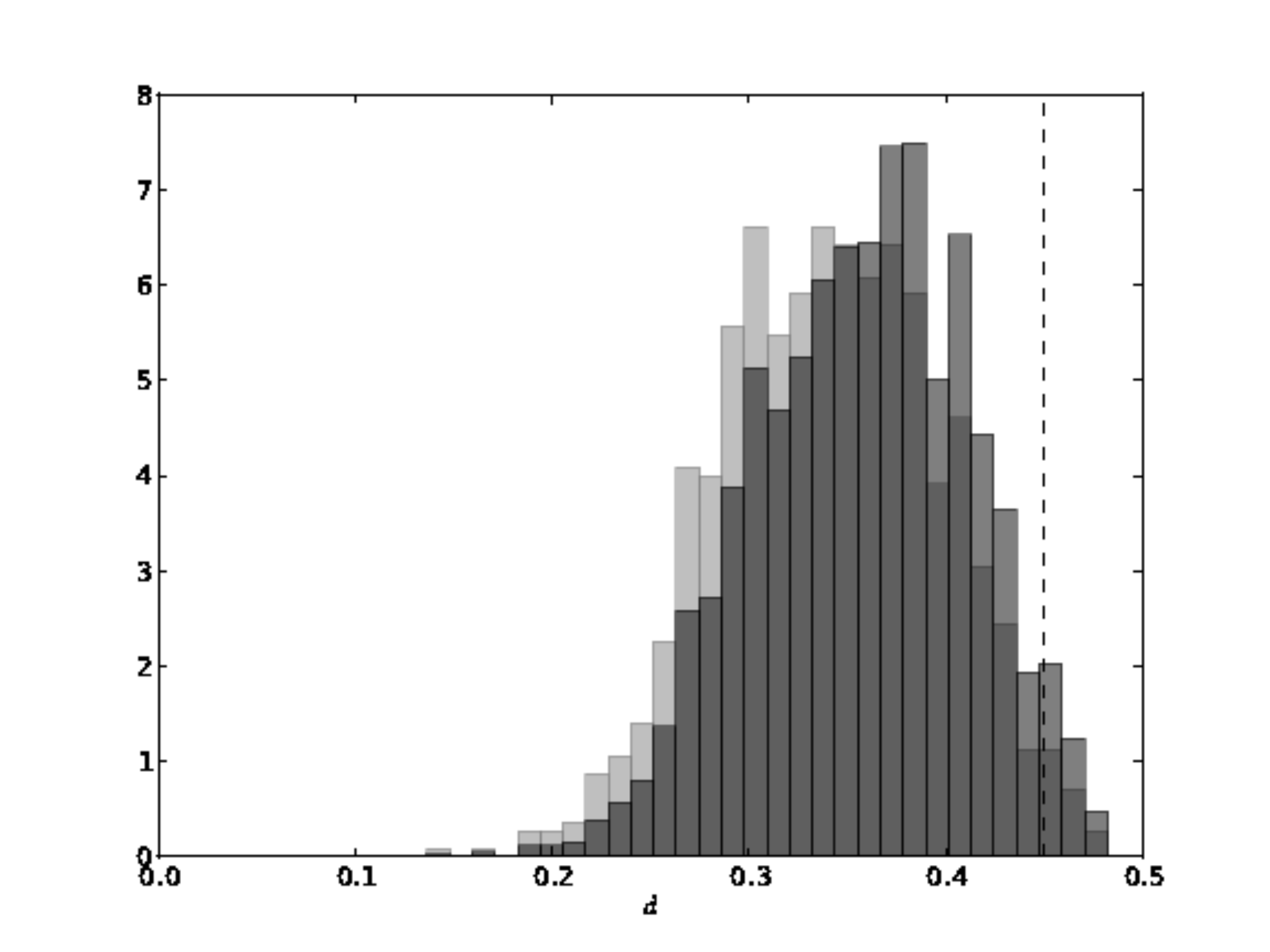}\includegraphics[scale=0.3]{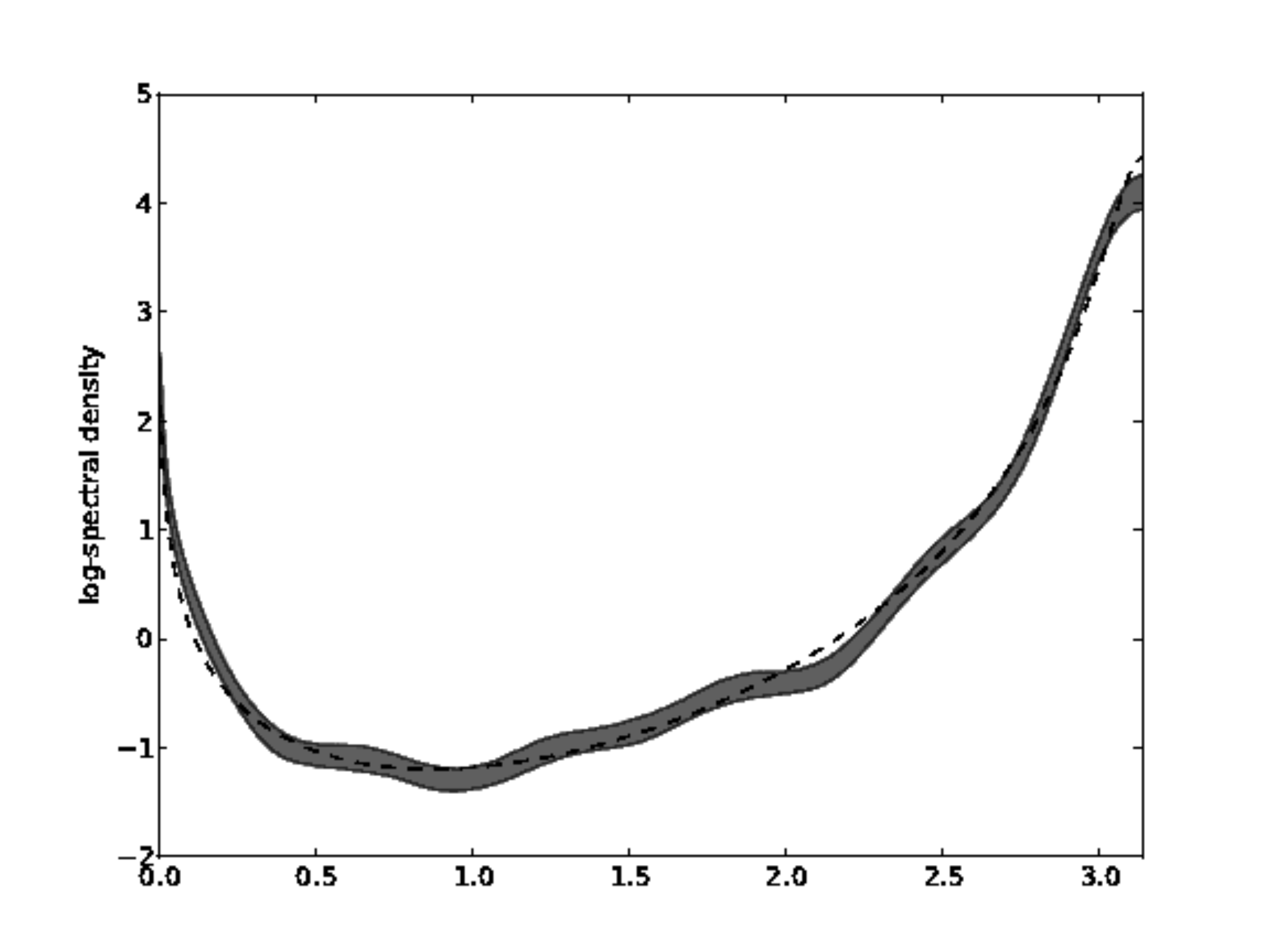}

\caption{\label{fig:Consistency}Consistency study for the ARFIMA data, Left:
marginal posterior of $d$, approximate (light grey histogram), or
exact (dark grey histogram); Right: $80\%$ confidence bands for the
log-spectral density (same color code), true spectral density (dashed
line). From top to bottom, sample size is $100$, $300$, $1000$,
$3000$.}
\end{figure}

\begin{figure}
\includegraphics[scale=0.3]{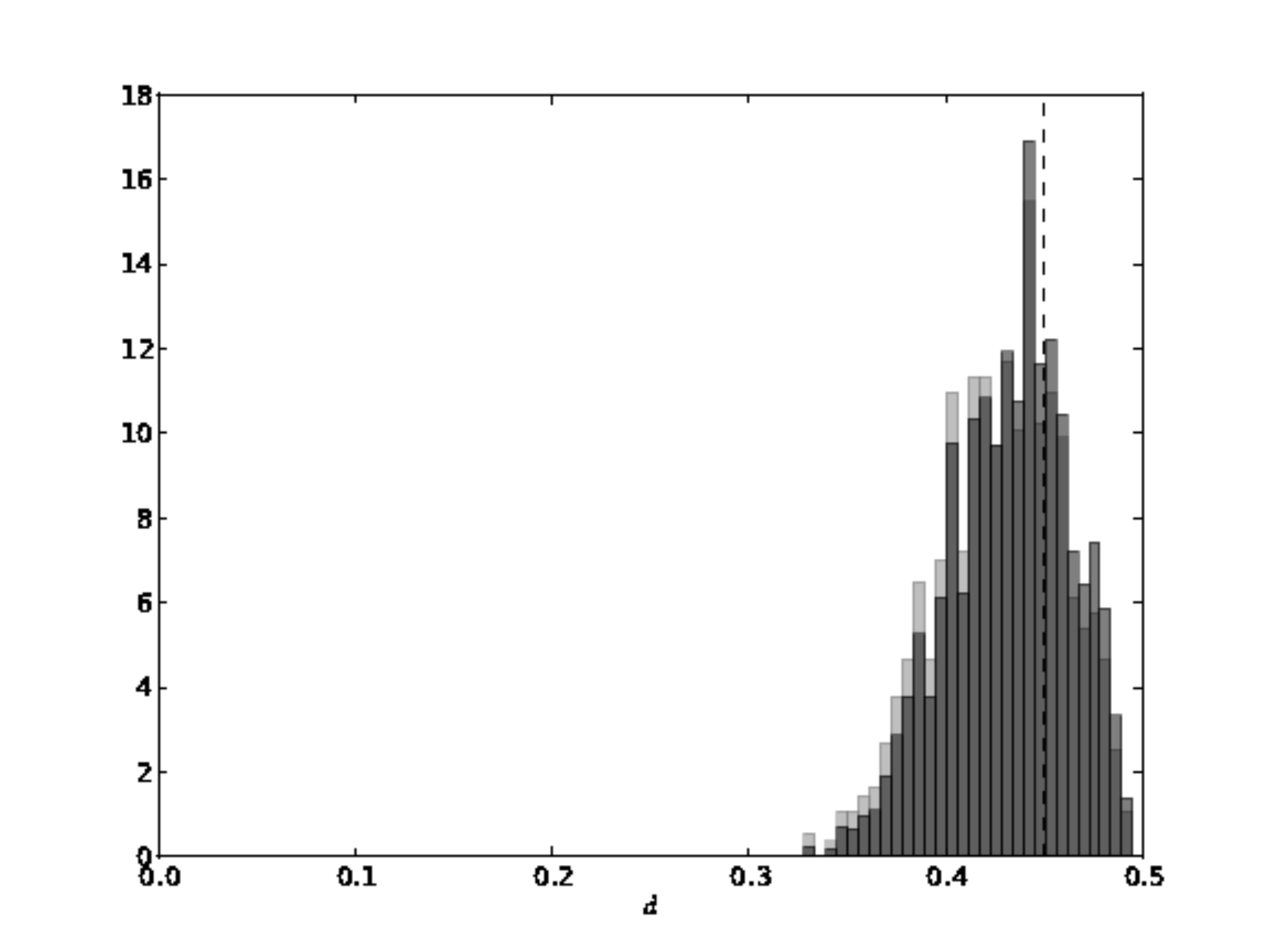}\includegraphics[scale=0.3]{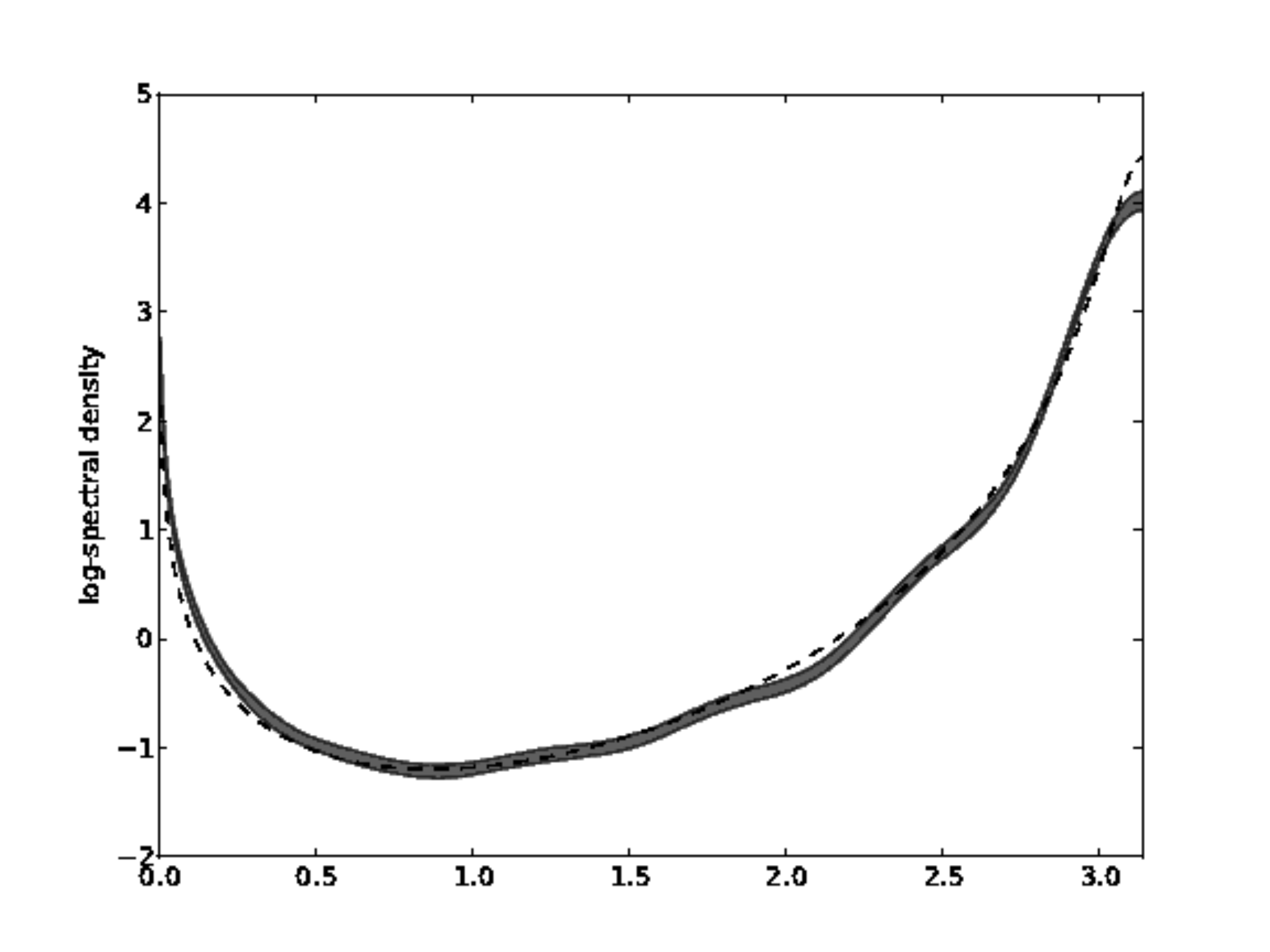}

\includegraphics[scale=0.3]{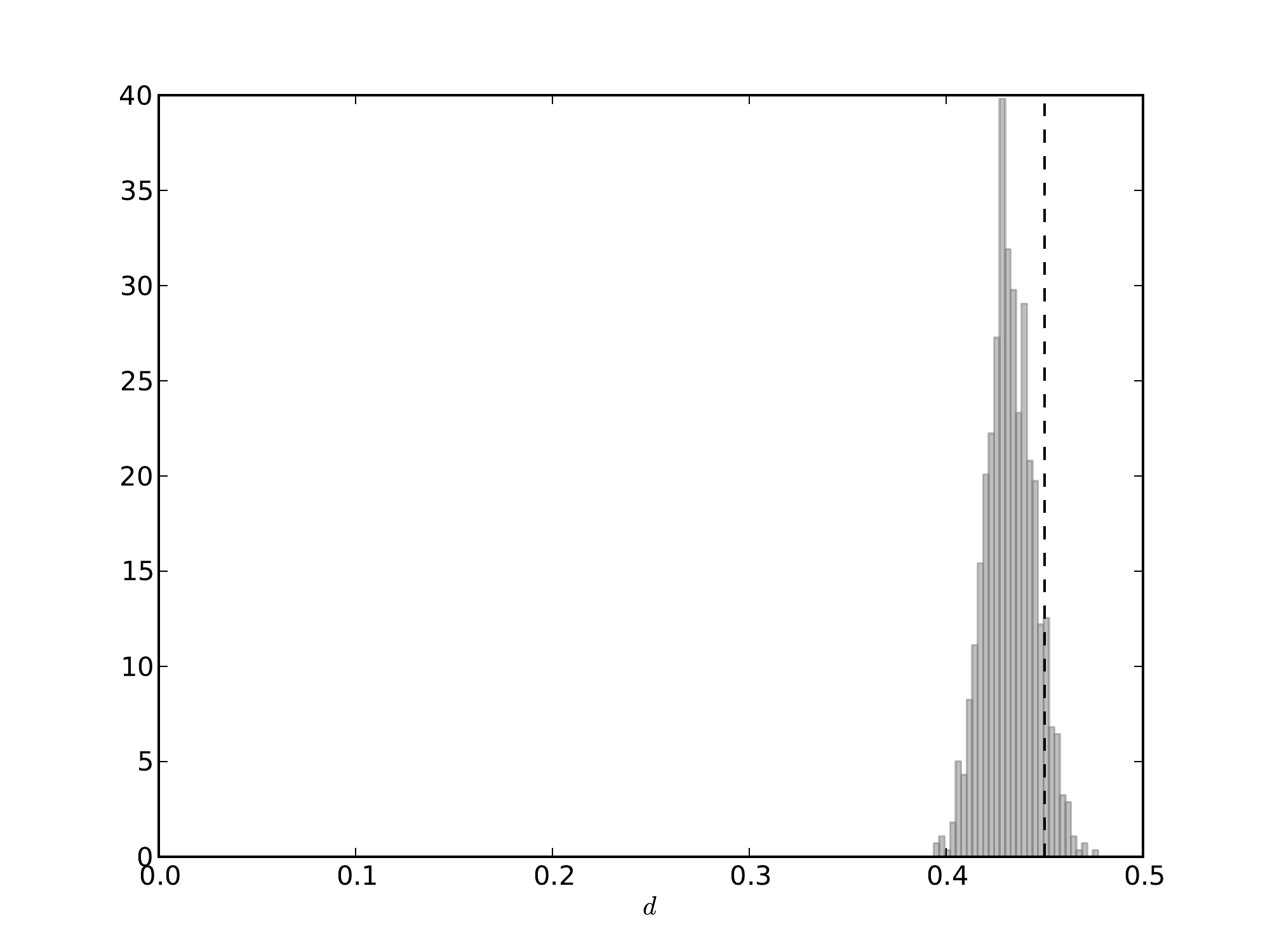}\includegraphics[scale=0.3]{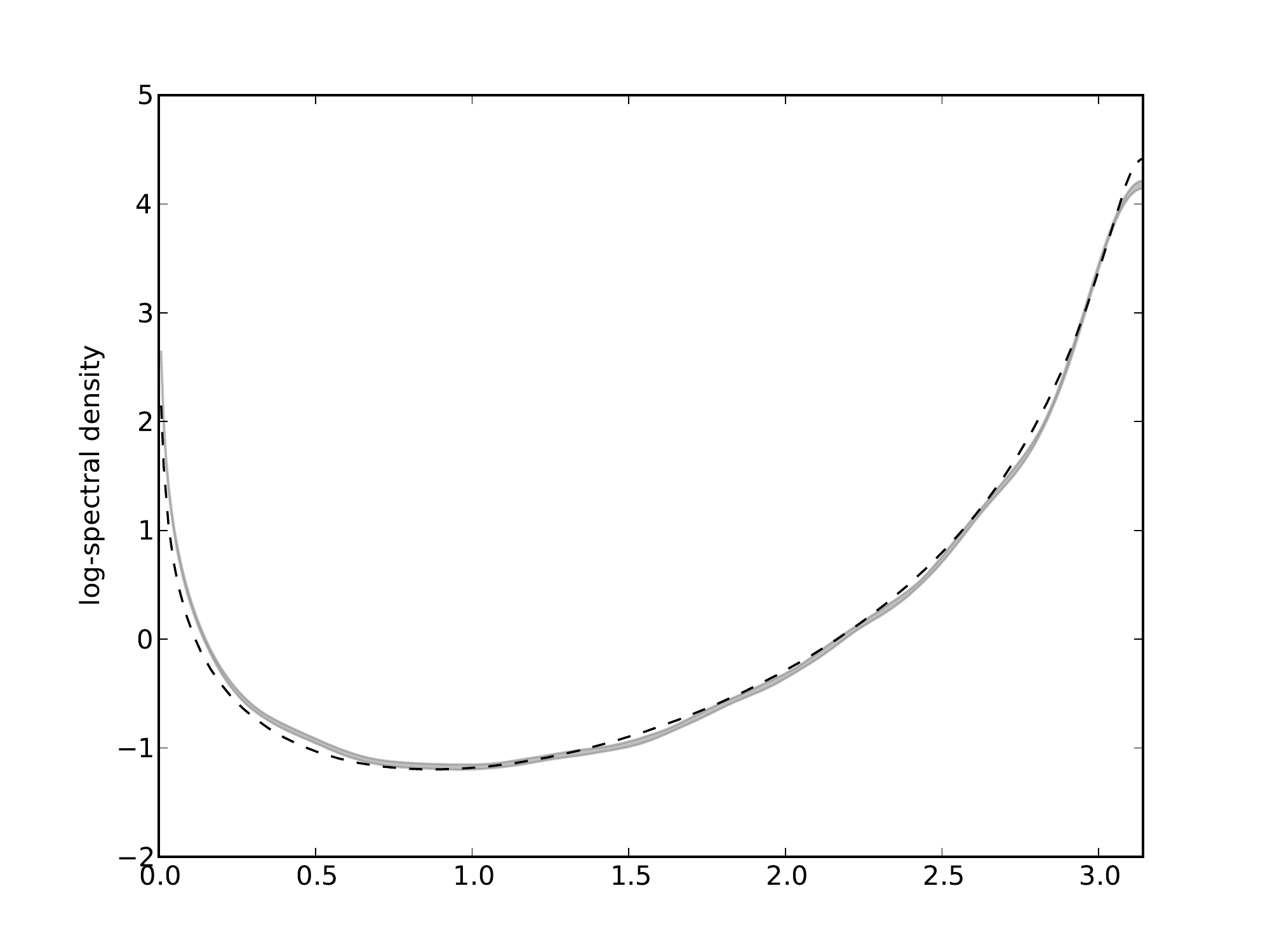}\caption{\label{fig:Consistency2}Consistency study for the ARFIMA data, same
caption as Fig. \ref{fig:Consistency}, sample size is $10^{4}$ (top)
and $10^{5}$ (bottom).}
\end{figure}

Figures \ref{fig:Consistency} and \ref{fig:Consistency2} summarise
posterior inference for the sample sizes, from top to bottom of Figure
\ref{fig:Consistency}, $n=100$, $300$, $1000$, $3000$, and, from
top to bottom of Figure \ref{fig:Consistency2}, $n=10^{4}$ and $n=10^{5}$.
On the left (resp. right) side, the marginal posterior distributions
of $d$ (resp. the $80\%$ confidence bands for the spectral density)
are compared; light grey is the approximate posterior, and dark grey
is the true posterior. (Transparency effects are used.) One sees that
the effect of the correction step becomes unnoticeable very quickly
as far as the estimation of the spectral density (away from $0$)
is concerned. The effect on the marginal posterior of $d$ takes a
bit more time to vanish, but is already negligible for $n$ around
$3000$; in fact, the effective sample size of the correction step
is already above $900$ (out of $1000$ particles) for $n=3000$.

The right panel of Figure \ref{fig:performanceSMC} compares the CPU
cost of the SMC sampler and the correction step. One sees that, unless
$n$ is of order of $10^{4}$ or more, the overall cost of the complete
procedure is dominated by that of the SMC sampler. For $n\gg10^{4}$,
the correction step becomes quickly too expensive, because of the
$O(n^{3})$ cost, and was not performed for $n=10^{5}$. (Which is
why the dark grey plots are missing in the bottom row of Figure \ref{fig:Consistency2}.)
Fortunately, one sees that for $n\gg10^{4}$ the correction step seems
to be superfluous.

\subsection{Prior sensitivity analysis\label{sub:prior-sensitivity-analysis}}

We have seen in Section \ref{sub:Settings} that the prior for the
FEXP coefficients $\xi_{j}$ was set to $N(0,100j^{-2\beta})$, with
$\beta=1$; this hyper-parameter is related to assumptions on the
smoothness of the true spectral density (the larger $\beta,$ the
smoother the true spectral density); see \citet{RousseauChopinLiseo2012}.
In this section, we consider briefly the effect of $\beta$ on posterior
inference.

\begin{figure}
\includegraphics[scale=0.3]{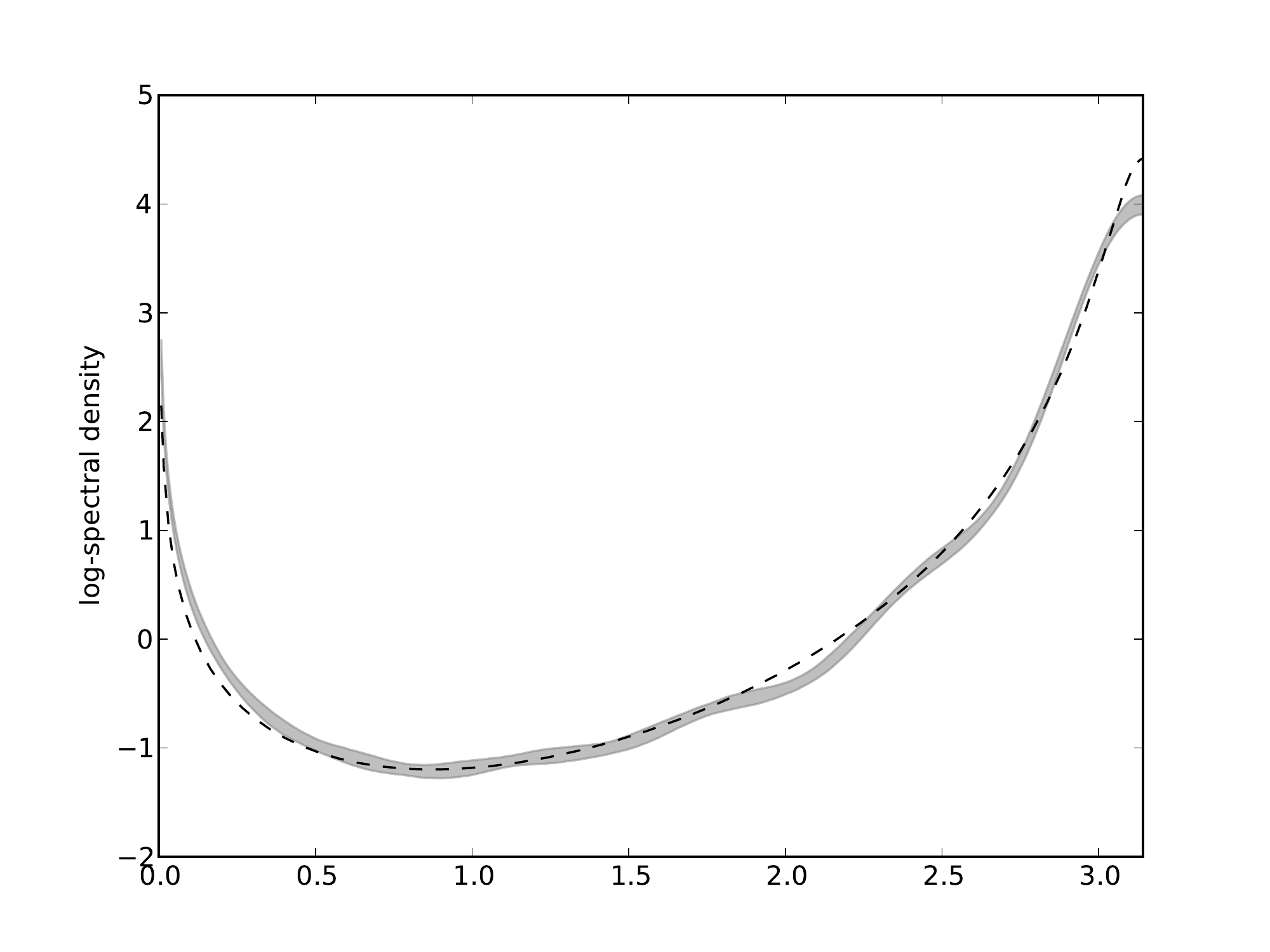}\includegraphics[scale=0.3]{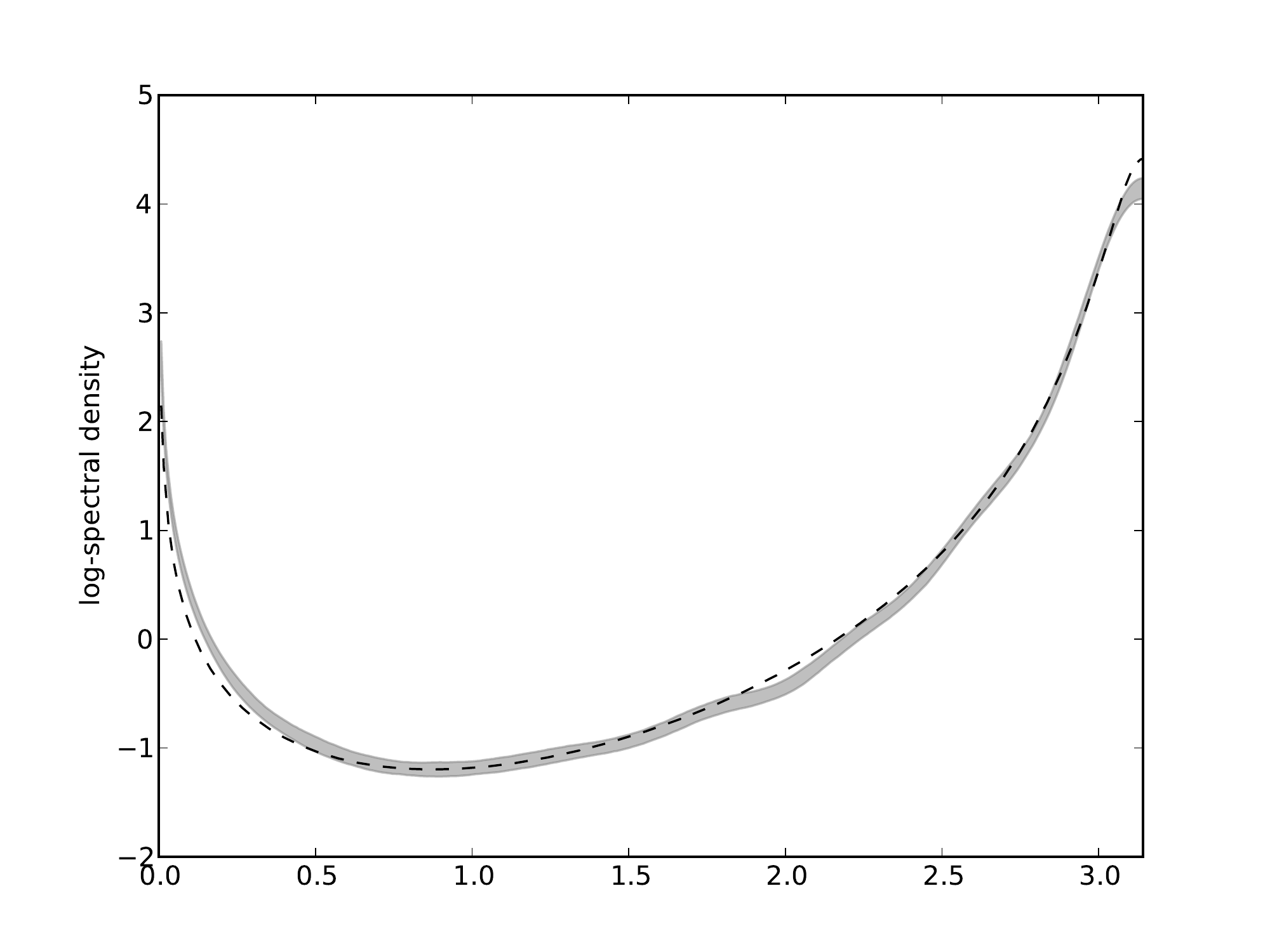}

\includegraphics[scale=0.3]{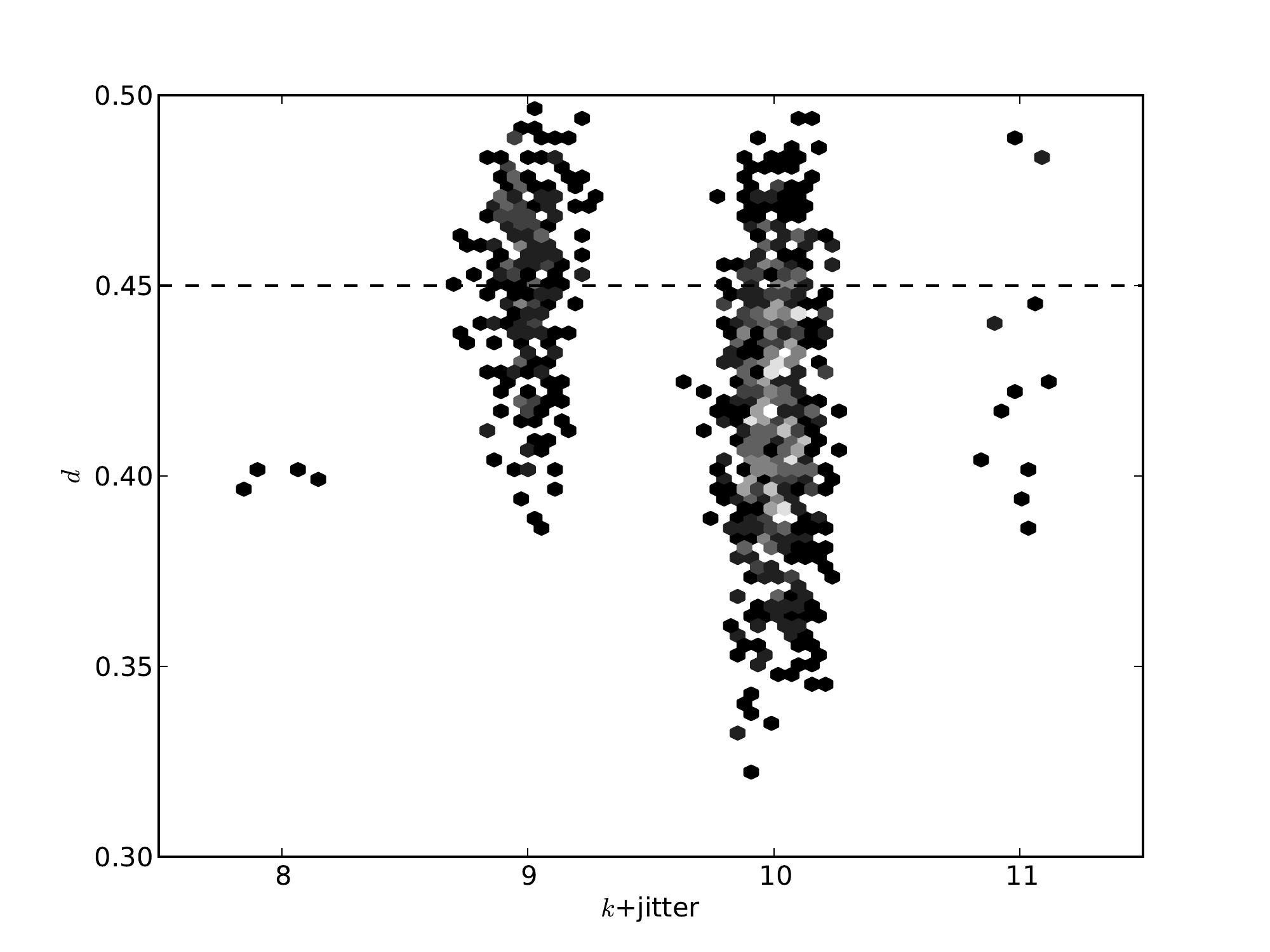}\includegraphics[scale=0.3]{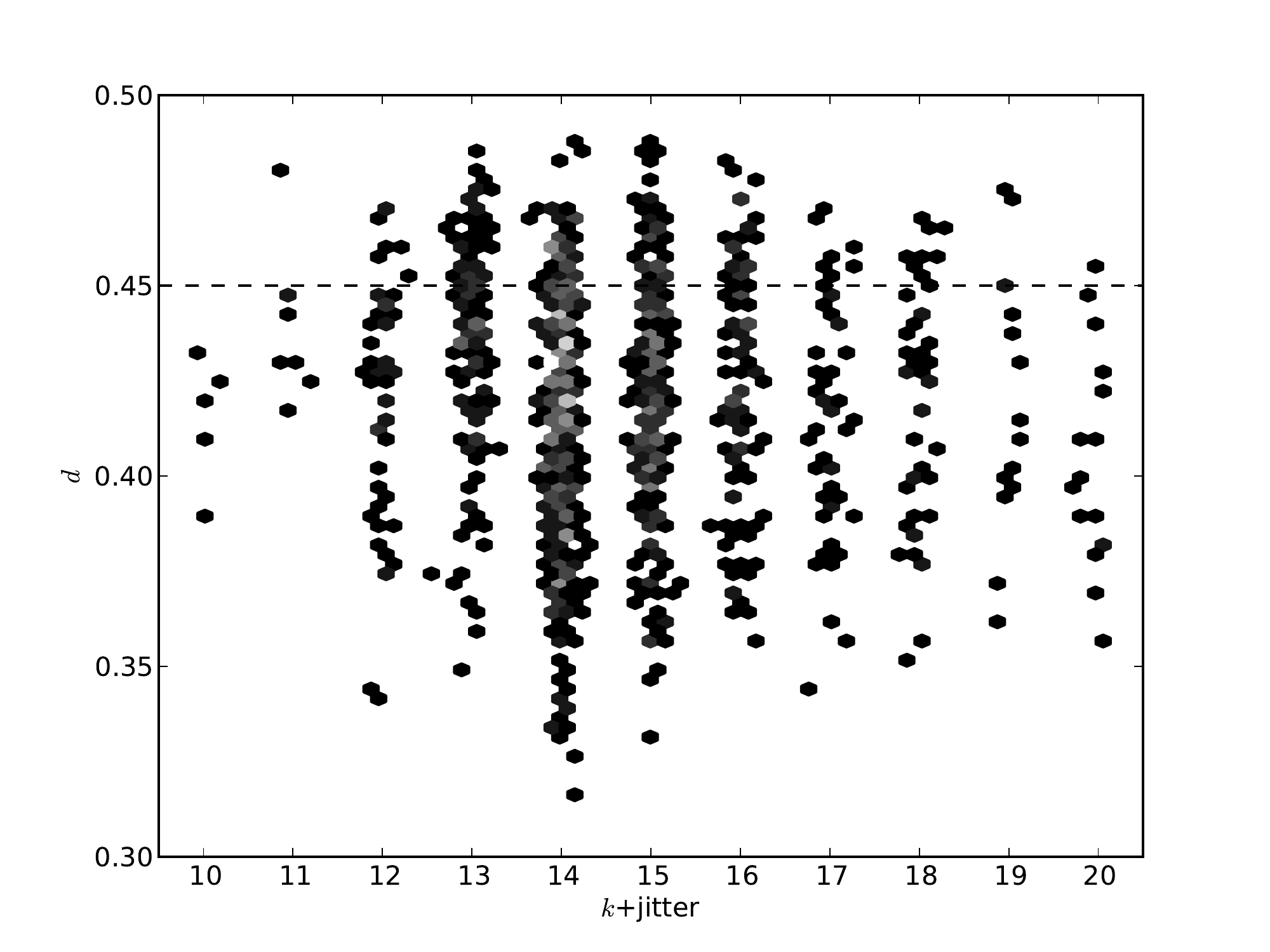}

\caption{\label{fig:PriorSens}Prior sensitivity analysis: $80\%$ posterior
confidence bands for log-spectral density (top), hexagonal bin plot
of $(k+\textrm{jitter},d)$, where jitter is $N(0,0.1^{2})$ noise
(bottom), for two different priors: $\beta=0$, i.e. $\xi_{j}\sim N(0,100)$
(left); and $\beta=2$, i.e. $\xi_{j}\sim N(0,100j^{-4})$ (right).}
\end{figure}

Figure \ref{fig:PriorSens} gives the same types of posterior plots
as in the previous sections, i.e. $80\%$ confidence bands for the
spectral density, and marginal posterior for the vector $(k,d)$,
for $\beta=0$ (Left side) and $\beta=2$ (right side). This must
be compared with the top right plot of Figure \ref{fig:Consistency2}
and the bottom right plot of \ref{fig:arfima}, for which $\beta=1$.
One sees that the choice of $\beta$ has a strong impact on the posterior
marginal distribution of $k$, but a rather moderate impact on either
$d$ or the spectral density itself, except maybe on the right edge
of the spectral density plots (for $\lambda$ close to $\pi$). Since
$k$ is essentially a nuisance parameter, one sees that the choice
of $\beta$ does not seem to be too critical for inferential purposes.
On the other hand, it is interesting to note the impact of $\beta$
on the computational difficulty to explore the posterior. We observe
that, for $\beta=2$, it is even more difficult for a MCMC sampler
to escape local modes such as those shown in the top row of Figure
\ref{fig:arfima_mcmc} (corresponding MCMC traces not shown).

\subsection{Real data study: Ethernet Traffic\label{sub:Real-data-study}}

In a preliminary study, we applied our methodology to the popular
Nile data set, but found the example not to be very challenging, either
from a computational point of view ($n=663$), or from an inferential
point of view (an ARFIMA$(0,d,0)$ model, or in other words, a fractional
Gaussian noise model, seems to fit the data well). More details may
be obtained from the first author. It is perhaps interesting to note 
that real datasets considered in the literature are rarely larger; 
see e.g.
the US GNP data of \citet{koop1997bayesian}, $n=172$, 
 the central England temperature data of \citet{pai1998bayesian}
$n=318$, the sunspot dataset of \citet{Choudhuri2004}, $n=288$,
and so on. This remark seems to give additional support to two
aspects of our work: (a) to perform exact Bayesian (rather than asymptotics-based)
inference; and (b) that the correction step, although $O(n^3)$, 
is typically negligible in practical examples, as shown in Fig. 
\ref{fig:performanceSMC}.

\begin{figure}
\includegraphics[scale=0.3]{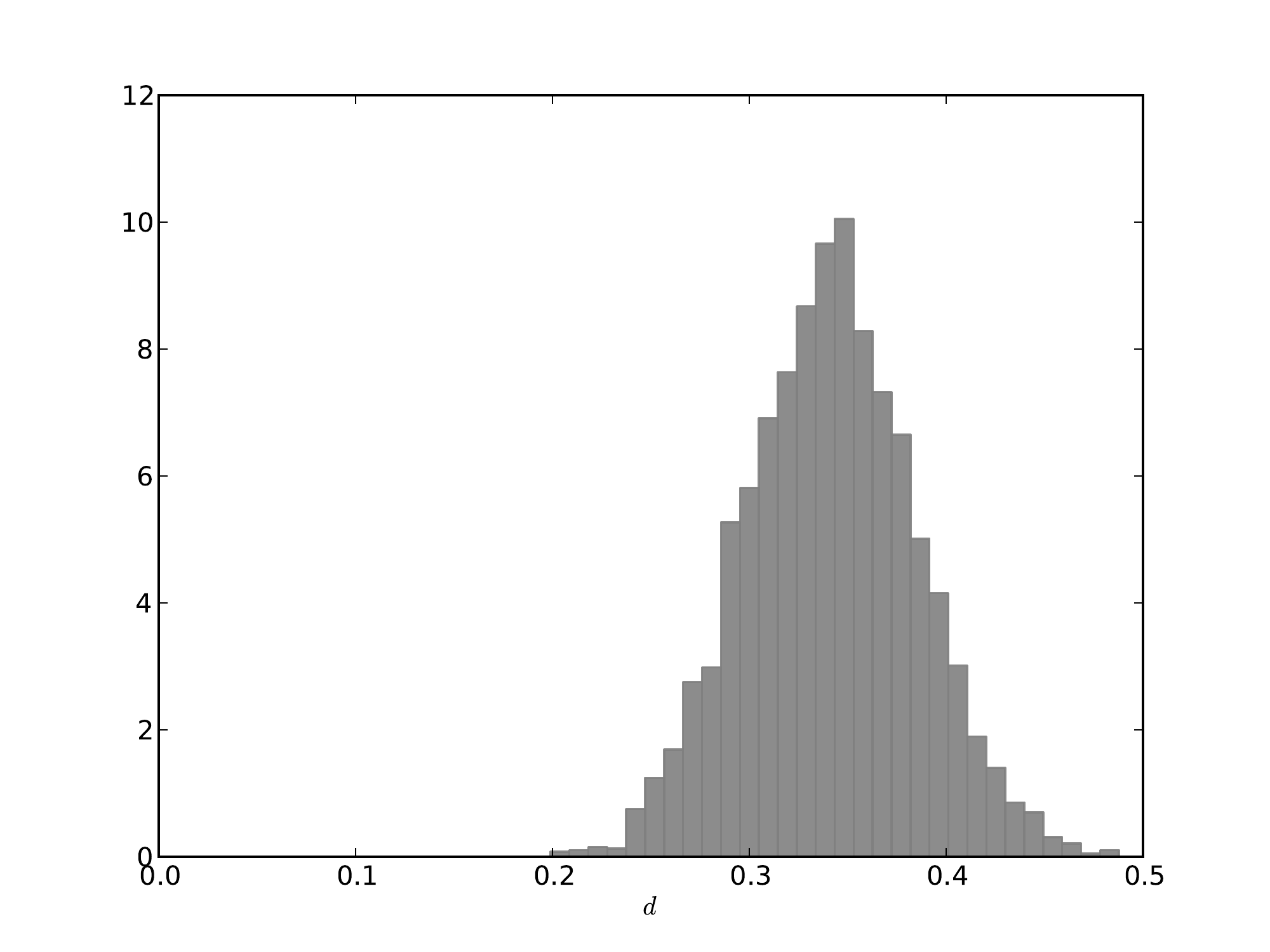}\includegraphics[scale=0.3]{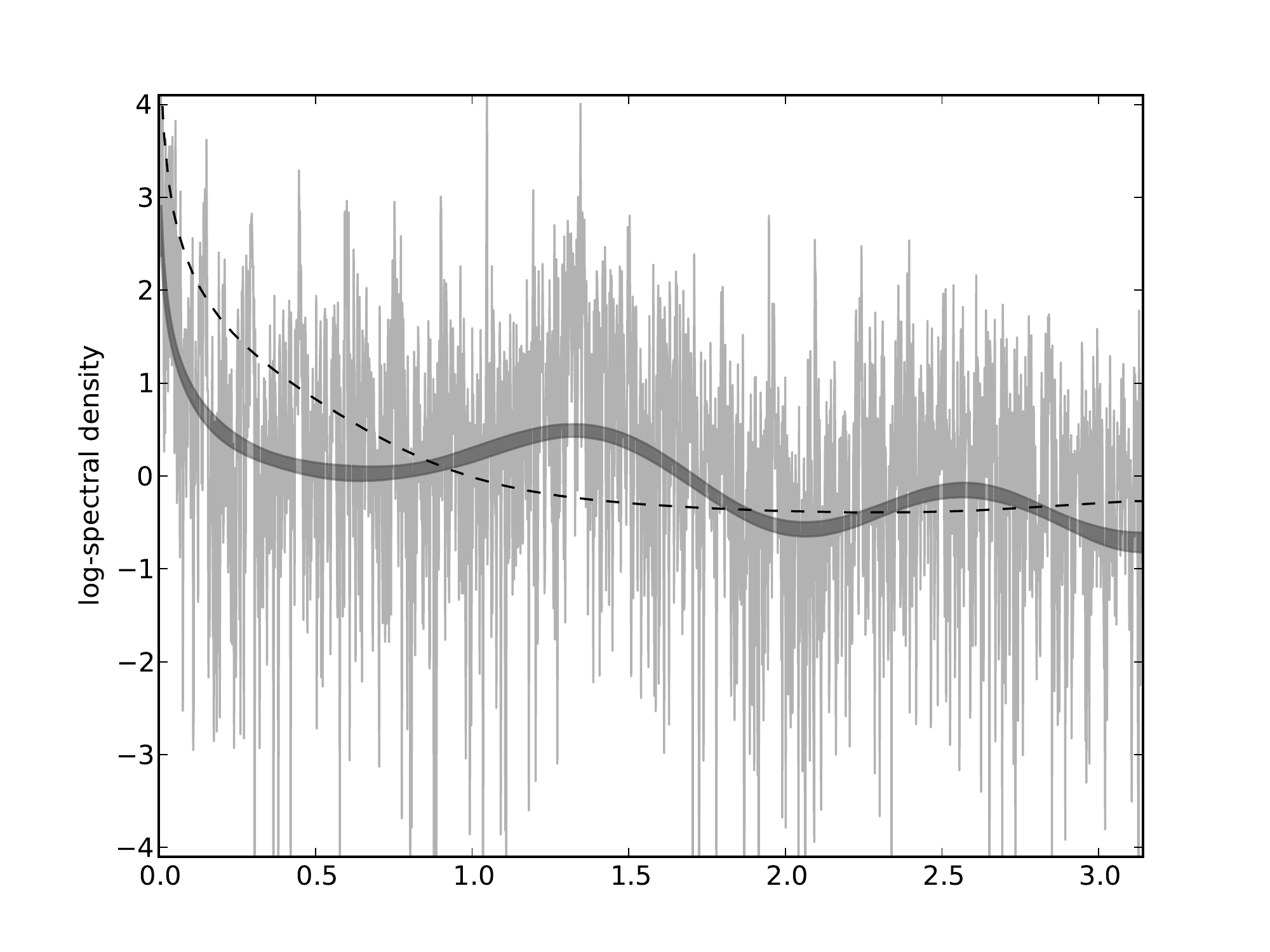}\caption{\label{fig:Ethernet}Ethernet data, Left: marginal posterior distribution
of $d$ from the FEXP semi-parametric model (as represented by a weighted
histogram obtained from the SMC sampler); Right: Bayesian 80\% confidence
band for the spectral density (dark grey), spectral density corresponding
to ML estimation of an ARFIMA model, with orders $p=q=3$ (dashed
line).}
\end{figure}

\texttt{In} this paper, we consider instead the Ethernet Traffic dataset
of \citet{leland1994self}, which can be found in the \texttt{longmemo}
\texttt{R} package. This is a time-series of length $n=4000$, which
records the number of packets passing through a particular network
per time unit. (For convenience, we divided the data values by 1000,
in order to use the same prior as for the simulated dataset.) The
right side of Fig. \ref{fig:Ethernet} plots the empirical spectrum
of this time series. Although the empirical spectrum is an asymptotically
biased estimator of the true spectral density under long-range dependence,
the bias is typically small for frequencies sufficiently far away
from $0$ \citep[e.g.][]{MoulinesSoulier2003}. Thus comparing the
empirical spectrum with a given estimator of the spectral density
provides at least some guidance on the performance of the said estimator.

Interestingly, we find these data to be quite challenging for frequentist
parametric procedures. The \texttt{FEXPest} command of the \texttt{longmemo
R} package, which computes \citet{Beran1993}'s estimator for a parametric
FEXP model, either returns the estimate $\hat{d}=0.52$ - although $d$ is supposed to be
in $(0,1/2)$ - or $\hat d=0.222$ or $\hat d=0.316$, according to a tuning parameter which
regulates the polynomial order selection.
The \texttt{fracdiff} command from the \texttt{fracdiff} package,
which does maximum likelihood estimation for ARFIMA$(p,q)$ models,
returns $\hat{d}=0.3$ for $p=q=3$ but the estimated spectral density,
plotted as a dashed line in the right panel of Figure \ref{fig:Ethernet}
does not seem to fit the spectrum of the data even for higher frequencies.
(According to the same procedure, the orders $p=q=3$ give the smallest
AIC among ARFIMA($p,p)$ models, with $p\leq10$).

In contrast, our Bayesian semi-parametric procedure seems to fit the
data rather well; see again the right side of Figure \ref{fig:Ethernet}.
We find strong evidence in favour of long-range dependence, as evidenced
by the marginal posterior distribution of $d$ in the left panel of
Figure \ref{fig:Ethernet}.

From a computational point of view, we mention briefly that we observe
the same things as in the previous section; that is, that the correction
step has a negligible impact on the results, and that the output SMC
sampler shows little variability when run several times.

\section{Conclusion}

Several straightforward extensions of this work could be considered. First, as said in the introduction, the approach we proposed could be adapted with little effort to other, parametric or semi-parametric, prior distributions for the spectral density $f$. 
Second, as pointed out by both referees, one may consider the problem of sequential estimation, that is, to compute sequentially the sequence of posterior distributions of the $t$ first data-points, where $t$ is incremented by one at each iteration. Of course, if a long-range dependent model is considered, then by construction the posterior distribution may not provide relevant information on $f$ if $t$ is too small, so perhaps sequential estimation should be started only after some initial sample of size $n_0$ has been processed. In that case, one could use a variant of the IBIS algorithm 
of \citep{Chopin:IBIS}, that is, a SMC sampler based on the sequence of posterior distributions based on $t$ data-points, $t=n_0,\ldots,n$, but with the likelihood replaced
by the approximation presented in Section \ref{sec:Approx}. This would lead to a $O(n^2\log(n))$ algorithm, where $n$ is the total number of observations, because computing the approximate likelihood at iteration $t$ would cost $O(t\log(t))$. (For the same reasons, using exact likelihoods would give a $O(n^4)$ algorithm, which would be far too expensive.) 
Thus, this approach would be useful only for sequential estimation, and not when one is interested only in the posterior over the complete data-set, as one may use instead in that case the SMC sampler described in this paper, which is $O(n\log(n))$.

\section*{Acknowledgements}

We are particularly grateful to both referees for their supportive comments. N. Chopin and J. Rousseau are supported by the ANR grant ``Bandhit\textquotedblright{} of the French Ministry of research.

\section*{Appendix: Proof of Theorem 1 }

The proof borrows several intermediate results from \citet{RousseauChopinLiseo2012}.
Regarding the prior distribution,  we make the following assumptions:
\begin{itemize} 
 \item the prior $p(\bm{\theta})$ decomposes as
$p(d) p(k) p_{k}(\bm{\xi}_{k})$,
where the support of $p(d)$ is $[0,1/2-t]$ for some $t>0$;
\item the prior for $K$, $p(k)$, is a Geometric distribution;
\item Conditional on $K=k$, the coefficients $\xi_j$'s in (\ref{eq:fexp_normalised})
have a Gaussian distribution
$$\xi_{j}(j+1)^{\alpha+1/2} \sim \mathcal{N}(0,1)$$
and they are independent except for the fact that the vector $\bm{\xi}_{k}$ must belong to
the set \[
\Xi_{k}(\alpha,L)=\{\bm{\xi}_{k}\in\mathbb{R}^{k+1};\sum_{j=0}^{k}(j+1)^{2\alpha+1}\xi_{j}
^{2}\leq L\}
\]
where $L$ is a large positive constant.
\end{itemize}

Regarding the true distribution of $\bm{x}$, which is denoted by $P_{o}^{n}$ 
from now on, we assume that $\bm{x}\sim N(0\times\bm{1},\bm{T}(f_{o}))$, where the
true spectral density $f_{o}$ admits an infinite FEXP representation, that is, a FEXP representation as defined by 
\eqref{eq:FEXPmodel}, but with parameters $d_{o}>0$, $k_{o}=+\infty$,
and $\xi_{0,o},\xi_{1,o},\ldots$ belonging to the Sobolev ball of
radius $L$, $\Xi(\beta,L)=\{(\xi_{0},\xi_{1},\ldots);\sum_{j=0}^{\infty}(j+1)^{2\beta+1}\xi_{j}^{2}\leq L\}$.
Throughout the proof we denote by $|A|_{Fr}=tr[A^{T}A]^{1/2}$ the
Froebinius norm of a matrix $A$ and by $\|A\|^{2}=\sup\{x^{T}A^{T}Ax;\|x\|=1\}$
its euclidian norm.

We now prove that, under the above conditions (on the prior distribution $p(\bm{\theta})$, and on the true distribution  $P_{o}^{n}$  of $\bm{x}$), 
and 
provided $0\leq d_{o}\leq1/2-t$, $\beta\geq\alpha>3/2$,
and for $L_{o}$ small enough (compared to $L$), one has: 
\[
\mathbb{E}^{\pi_{n}}\left[w^{2}(\bm{\theta})\right]=w_{o}^{2}\left\{ 1+o_{p}(1)\right\} ,\quad w_{o}^{2}=o_{p}(e^{u_{n}(\log n)^{3/2}})
\]
for any sequence $(u_{n})$ such that $u_{n}\rightarrow+\infty$,
and where $w$ (resp. $w_{o}$) is used as a short-hand for $w_{\mathrm{Corr}}$
(resp. $w_{\mathrm{Corr}}^{o}$) in the rest of the proof. One may
obtain the same type of results for $\mathbb{E}^{\pi_{n}}\left[w(\bm{\theta})\right]$
using the same calculations.

One has: 
\begin{eqnarray*}
\mathbb{E}^{\pi_{n}}\left[w^{2}(\bm{\theta})\right] & = & \frac{\int w^{2}(\bm{\theta})\exp\left\{ -\frac{1}{2}\bm{x}^{T}\bm{T}(1/4\pi^{2}f_{\bt})\bm{x}-l_{n,o}\right\} \left|\mathbf{T}(f_{\bm{\theta}})\right|^{-1/2}p(\bm{\theta})\, d\bm{\theta}}{\int\exp\left\{ -\frac{1}{2}\bm{x}^{T}\bm{T}(1/4\pi^{2}f_{\bt})\bm{x}-l_{n,o}\right\} \left|\mathbf{T}(f_{\bm{\theta}})\right|^{-1/2}p(\bm{\theta})\, d\bm{\theta}}\\
 & \eqdef & \frac{N_{n}}{D_{n}}
\end{eqnarray*}
where $l_{n,o}=-\left\{ \bm{x}^{T}\mathbf{T}(f_{o})^{-1}\bm{x}+\log\mathrm{det}\bm{T}(f_{o})\right\} /2$.

Let $\epsilon_{n}^{2}=(n/\log n)^{-\beta/(2\beta+1)}$. The idea of
the proof is to first show that 
\begin{equation}
\mathbb{E}^{\pi_{n}}\left[\frac{w^{2}(\bm{\theta})}{w_{o}^{2}}\right]=\mathbb{E}^{\pi_{n}}\left[\frac{w^{2}(\bm{\theta})}{w_{o}^{2}}\1_{S_{n,1}}(\bt)\1_{k\leq k_{n,1}}\right]+o_{p}(1)\label{step1}
\end{equation}
where $S_{n,1}=\{\bt;l(f_{o},f_{\bt})\leq\epsilon_{n}\}$, $k_{n,1}=k_{0}\epsilon_{n}^{-1/\alpha}$
for some $k_{0}>0$, and $l(f,f')$ is the $L^{2}$ distance between
spectral log-densities, $l(f,f')=\int_{-\pi}^{\pi}\left\{ \log\left(f/f'\right)\right\} ^{2}$.
Let $S_{n,1,k}=\left\{ \bm{\theta}_{k};\,\bt=(k,\bt_{k})\in S_{n,1}\right\} $.
In a second step, we show that 
\begin{equation}
\sup_{\bt\in S_{n,1,k}}\left|\log w(\bt)-\log w_{o}\right|=o_{p}(1)\label{step2}
\end{equation}
uniformly on $k\leq k_{n,1}$, which would then conclude the proof
of Theorem 1.

We now prove \eqref{step1}. As proven in \citet{RousseauChopinLiseo2012}
and \citet{kruijer:rousseau:11} (for the true posterior, but the
proof for the approximate posterior $\pi_{n}$ follows the same lines),
there exists $c>0$ such that 
\begin{equation}
P_{o}^{n}\left[D_{n}\geq e^{-cn\epsilon_{n}^{2}}\right]=o(1).\label{Dn}
\end{equation}
Indeed, let $k_{n}=\lfloor k_{0}(n/\log n)^{1/(2\beta+1)}\rfloor$
with $k_{0}>0$ and 
\[
\mathcal{B}_{n}=\left\{ \bm{\theta};\: k=k_{n},d_{o}\leq d\leq d_{o}+n^{-a},|\xi_{j}-\xi_{j,o}|\leq n^{-a},j=1,\cdots k_{n}\right\} ,
\]
then it is easy to see that, for some $c_{1}>0$, 
\[
\mathbb{P}^{p(\bt)}\left[\mathcal{B}_{n}\right]\geq e^{-c_{1}k_{n}\log n},
\]
where $\mathbb{P}^{p(\bt)}$ denotes the prior probability. As in
\citet{RousseauChopinLiseo2012} and \citet{kruijer:rousseau:11},
let $\Omega_{n}(\bt)=\{\bm{x};\tilde{l}_{n}(\bt)-l_{n,o}\geq-n\epsilon_{n}^{2}\}$,
with 
\[
\tilde{l}_{n}(\bt)=-\frac{1}{2}\bm{x}^{T}\bm{T}(1/4\pi^{2}f_{\bt})\bm{x}-\frac{1}{2}\log\mathrm{det}\left[\bm{T}(f_{\bt})\right]
\]
then we have that for all $\bt\in\mathcal{B}_{n}$, 
\begin{equation}
P_{o}^{n}\left[\Omega_{n}(\bt)^{c}\right]=o(1)\label{omegan}
\end{equation}
if $a$ is chosen large enough. Indeed let $A(\bt)=\T^{1/2}(f_{o})\T(1/4\pi^{2}f_{\bt})\T^{1/2}(f_{o})-\bm{I}_{n}$
and $B(\bt)=\T^{1/2}(f_{o})\T(f_{\bt})^{-1}\T^{1/2}(f_{o})-\bm{I}_{n}$.
Using Lemma 2.4 in the supplement of \citet{kruijer:rousseau:11},
for all $\bt\in\mathcal{B}_{n}$, we have that $\tr\left[\T^{-1/2}(f_{o})\T(f_{\bt})\T^{-1/2}(f_{o})\right]\leq2n$
for $n$ large enough so that all eigenvalues of $\T^{1/2}(f_{o})\T(f_{\bt})^{-1}\T^{1/2}(f_{o})$
are bounded from below by $1/2$ and 
\[
\log|\bm{I}_{n}+B(\bt)|=\tr\left[B(\bt)\right]-\frac{1}{2}\tr\left[\left((\bm{I}_{n}+\tau B(\bt))^{-1}B(\bt)\right)^{2}\right]\geq\tr\left[B(\bt)\right]-2\tr\left[B(\bt)^{2}\right].
\]
Moreover using Lemma 2.4 in the supplement of \citet{kruijer:rousseau:11},
we have that $\tr\left[B(\bt)\right]=\tr\left[A(\bt)\right]+O(n^{\epsilon})$
for all $\epsilon>0$ so that, since for $n$ large enough and $\bt\in\mathcal{B}_{n}$,
$2\tr\left[B(\bt)^{2}\right]+n^{\epsilon}\leq n\epsilon_{n}^{2}$,
\begin{eqnarray*}
P_{o}^{n}\left[\Omega_{n}(\bt)^{c}\right] & \leq & \mathbb{P}_{\bm{z}\sim\mathcal{N}(0,\bm{I}_{n})}\left[\bm{z}^{T}A(\bt)\bm{z}-\mathrm{tr}[A(\bt)]\geq n\epsilon_{n}^{2}\right].
\end{eqnarray*}
One also has: 
\[
|A(\bt)|_{Fr}^{2}=\frac{n}{2\pi}\int_{-\pi}^{\pi}\left(\frac{f_{o}(\lambda)}{f_{\bt}(\lambda)}-1\right)d\lambda+{\rm {Error},}
\]
where the ${\rm {Error}}$ is controlled by Lemma 2.6 in the supplement
of \citet{kruijer:rousseau:11}, since we have 
\[
|A(\bt)|_{Fr}^{2}=\tr\left[\left(\T(f_{\bt}^{-1})\T(f_{o}-f_{\bt})\right)^{2}\right]
\]
and \foreignlanguage{english}{$f_{o}=f_{\bt}b$ where (taking $\xi_{j}=0$
for $j>k$) 
\[
b(\lambda)=(2-2\cos\lambda)^{d-d_{o}}\exp\left\{ \sum_{j=0}^{\infty}(\xi_{o,j}-\xi_{j})\cos j\lambda\right\} 
\]
and we note that, for $\bt\in\mathcal{B}_{n}$, the function $\exp\left\{ \sum_{j=0}^{k}\xi_{j}\cos j\lambda\right\} $
is Lipschitz with constant $O(k_{n}^{(3/2-\be)_{+}})=O(1)$, for $a$
is large enough (see Lemma 3.1 in the supplement of \citet{kruijer:rousseau:11}).
Therefore we obtain 
\[
\mbox{Error}=O(\log n\|b-1\|_{\infty})=O(\log n),
\]
and $|A(\bt)|_{Fr}^{2}=O(n^{1-a}+\log n)=o(n^{\tau}n\epsilon_{n}^{2})$,
for some $\tau>0$, which combined with Lemma 1.3 in the supplement
of \citet{kruijer:rousseau:11}, leads to, for some $c>0$, 
\[
P_{o}^{n}\left[\Omega_{n}(\bt)^{c}\right]\leq e^{-cn^{\tau}}.
\]
We now turn to the second part of the proof. Let $S_{n,j,k}=\{\bm{\theta}_{k};l(f_{o},f_{(k,\bm{\theta}_{k})})\in((j-1)\epsilon_{n},j\epsilon_{n})\}$,
for $j=1,\cdots,J_{n}$ where $J_{n}=O(\epsilon_{n}^{-1})$, and let}

\begin{eqnarray*}
N_{n,j,k} & = & w_{o}\int_{S_{n,j,k}}w(\bt)^{2}\exp\left\{ \tilde{l}_{n}(\bt)-l_{o,n}\right\} p(\bt)\, d\bt\\
 & = & w_{o}\int_{S_{n,j,k}}\frac{\exp\left\{ l_{n}(\bt)-l_{o,n}\right\} }{w(\bt)}p(\bt)\, d\bt.
\end{eqnarray*}
We prove that uniformly in $J_{0}\leq j\leq J_{n}$ and $k\in\mathbb{N}$,
$N_{n,j,k}=o_{p}(e^{-cn\epsilon_{n}^{2}})$, for some $J_{0}>0$.
To do so we bound 
\[
P_{o}^{n}\left[\sup_{\bt\in S_{n,j,k}}\left|\log(w_{o}/w(\bt))\right|>v_{n,j,k}\right]
\]
for $j\geq2$ and for a properly chosen $v_{n,j,k}$. Let $k_{n,j}=k_{0}(j\epsilon_{n})^{-1/\alpha}$,
$J_{n,1}=J_{1}n^{-\alpha\epsilon}(\log n)^{\alpha}\epsilon_{n}^{-1}$,
with $J_{1}>0$. Define for $C>0$ (large enough) and $\tau>0$ (small
enough) 
\begin{equation}
\begin{split}v_{n,j,k} & =k^{3/2}n^{\epsilon}j\epsilon_{n},\quad\mbox{ if }k\leq k_{n,j},\quad j\geq1\\
v_{n,j,k} & =(C\log n(k\wedge n^{(1+\epsilon)/(\alpha-1/2)}k^{-1/(\alpha-1/2)})+k^{\tau})k_{n,j}^{-\alpha+1/2},\quad\mbox{ if }k>k_{n,j}.
\end{split}
\label{def:wnjk}
\end{equation}

Note that if $\epsilon$ is small, $Ck\log n=o(n^{\epsilon})$ for
some $k\geq k_{n,j}$ only if $(j\epsilon_{n})^{-1/\alpha}\lesssim n^{\epsilon}(\log n)^{-1}$,
i.e. if $j\gtrsim n^{-\alpha\epsilon}(\log n)^{\alpha}\epsilon_{n}^{-1}:=J_{n,1}$.
We write $-\log w(\bt)+\log(\xi_{o})=z'A(\bt)z/2$ with $\bm{z}=\T^{-1/2}(f_{o})\bm{x}\sim\mathcal{N}(0,I_{n})$,
and, under $P_{o}$, and we bound successively 
\[
\sup_{\bt\in S_{n,j,k}}\left(\bm{z}^{T}[A(\bt)-A(\txid)]\bm{z}^{T}+\tr[A(\bt)-A(\txid)]\right)
\]
and 
\[
\sup_{d}\left|\bm{z}^{T}[A_{o}-A(\bt_{d})]\bm{z}+\tr[A_{0}-A(\txid)]\right|
\]
with $\txid=(d,k,\bar{\bm{\xi}}_{d,k})$ and $\bar{\bm{\xi}}_{d,k}=\mbox{argmin}_{\bm{\xi}\in\R^{k+1}}l(f_{0},f_{d,k,\bm{\xi}})$,
see \citet{kruijer:rousseau:11}. We now study 
\[
\sup_{\bt\in S_{n,j,k}}\left(z'[A(\bt)-A(\txid)]z+\tr[A(\bt)-A(\txid)]\right).
\]
Note that $n^{-1}|A(\bt)-A(\txid)|_{Fr}^{2}=n^{-1}\tr\left[(A(\bt)-A(\txid))^{2}\right]$
converges towards 0, so we only need control the approximation error,
which we split into the approximation error of $\tr\left[(\T(f_{o})\T(f_{\bt}^{-1}-f_{\txid}^{-1}))^{2}\right]$
and of $\tr\left[(\T(f_{o})(\T(f_{\bt})^{-1}-\T(f_{\txid})^{-1})^{2})\right]$.
We now consider the first term. Note that 
\[
f_{\bt}^{-1}-f_{\txid}^{-1}=f_{\txid}^{-1}(\exp[\sum_{j=0}^{k}(\xi_{j}-(\bar{\xi}_{d,k})_{j})\cos(j\lambda)]-1):=f_{\txid}^{-1}b_{\bt}(\lambda)
\]
where 
\[
\sup_{\lambda\in[-\pi,\pi]}|b_{\bt}(\lambda)|\leq\sum_{j=0}^{k}|\xi_{j}-(\bar{\xi}_{d,k})_{j}|\leq\sqrt{k}\|\bm{\xi}-\bar{\bm{\xi}}_{d,k}\|\lesssim\sqrt{k}j\epsilon_{n}.
\]
Note that if $k\geq k_{n,j}$ then $k^{1/2-\alpha}\leq\sqrt{k}j\epsilon_{n}$
and we bound instead 
\[
\sup_{x}|b_{\bm{\theta}}(x)|\leq\sum_{j=0}^{k}|\xi_{j}-(\bar{\xi}_{d,k})_{j}|\leq\sqrt{k_{n,j}}\|\bm{\xi}-\bar{\bm{\xi}}_{d,k}\|+k_{n,j}^{-\alpha+1/2}.
\]
Lemma 2.1 of the supplement of \citet{kruijer:rousseau:11} implies
that the approximation error of the first term is bounded by $O((\sum_{j=1}^{k}|\xi_{j}-(\bar{\xi}_{d,k})_{j}|n^{\epsilon})^{2})=O((n^{\epsilon'}\sqrt{k}j\epsilon_{n})^{2})$
for all $\epsilon'>0$ and all $k\leq k_{n,j}$ and is bounded by
$O((k_{n,j}^{-\alpha+1/2}n^{\epsilon'})^{2})$ for all $k\geq k_{n,j}$
and all $\epsilon'>0$. From Lemma 2.4 of the supplement of \citet{kruijer:rousseau:11},
the same bounds apply to the second term. Finally we obtain that 
\[
\begin{split}|A(\bt)-A(\txid)|_{Fr} & =O(n^{\epsilon}\sqrt{k}j\epsilon_{n})\quad\mbox{ if }k\leq k_{n,j}\\
|A(\bt)-A(\txid)|_{Fr} & =O(n^{\epsilon}k_{n,j}^{-\alpha+1/2})\quad\mbox{ if }k>k_{n,j}.
\end{split}
\]
This implies that for all $\xi\in S_{n,j,k}$ :

$\bullet$ If $k\leq k_{n,j}$, since $v_{n,j,k}(\sqrt{k}j\epsilon_{n})^{-1}=kn^{\epsilon}$,
\[
\mathbb{P}\left(\bm{z}^{T}[A(\bt)-A(\txid)]\bm{z}+\tr[A(\bt)-A(\txid)]>v_{n,j,k}\right)\leq e^{-ckn^{\epsilon}},
\]

$\bullet$ If $k>k_{n,j}$, since $v_{n,j,k}k_{n,j}^{\alpha-1/2}\geq n^{\epsilon}$,
\[
Pr\left(\bm{z}^{T}[A(\bt)-A(\txid)]\bm{z}+\tr[A(\bt)-A(\txid)]>v_{n,j,k}\right)\leq e^{-cv_{n,j,k}k_{n,j}^{\alpha-1/2}}.
\]
Moreover a Taylor expansion of $A(\xi')$ around $A(\bt)$ implies
that $\forall\delta>0$ and for all $\xi,\xi'$ 
\begin{equation}
|\bm{z}^{T}[A(\bt)-A(\bt')]\bm{z}+\tr[A(\bt)-A(\bt')]|\lesssim(\bm{z}^{T}\bm{z}+n)n^{2(d-d_{o})_{+}+\delta}[\sum_{j=0}^{k}|\xi_{j}-\xi_{j}'|+|d-d'|].\label{bound:unif}
\end{equation}
Without loss of generality we can restrict ourselves on $\Omega_{n}=\{\bm{z}^{T}\bm{z}\leq2n\}$,
since $\mathbb{P}[\Omega_{n}^{c}]=o(e^{-cn})$ for some positive $c$.

$\bullet$ If $k\leq k_{n,j}$ and $j\leq J_{n,1}$, if $\|\bm{\xi}-\bm{\xi'}\|\leq n^{-1-\epsilon}k^{-1/2}$
and $|d-d'|\leq n^{-1-\epsilon}$, 
\[
|z'[A(\bt)-A(\bt')]z+tr[A(\bt)-A(\bt')]|=o(1),\quad\mbox{uniformly}.
\]
If $k\leq k_{n,j}$ and $j\geq J_{n,1}$, then 
\[
|z'[A(\bt)-A(\bt')]z+tr[A(\bt)-A(\bt')]|=o(n\epsilon_{n}^{2}j^{2}),\quad\mbox{uniformly}.
\]
Let $E_{n,j,k}$ be the covering number of $S_{n,j,k}$ by balls satisfying
the above constraint then 
\[
E_{n,j,k}\leq e^{C'k\log n}
\]

$\bullet$ If $k>k_{n,j}$. First if $k>k_{2}n\epsilon_{n}^{2}$,
for some $k_{2}>0$ possibly large, then uniformly over $\sum_{j=1}^{k}|\xi_{j}-\xi_{j}'|\leq n^{-1-\epsilon}k$
and $|d-d'|\leq n^{-1-\epsilon}$, 
\[
|\bm{z}^{T}[A(\bt)-A(\bt')]\bm{z}+\tr[A(\bt)-A(\bt')]|=o(k)
\]
The covering number of $S_{n,j,k}$ by the above constraints depends
on $k$. Define $K_{n}(k)$ such that $K_{n}(k)^{-(\alpha-1/2)}=n^{-1-\epsilon}k$,
i.e. $K_{n}(k)=n^{(1+\epsilon)/(\alpha-1/2)}k^{-1/(\alpha-1/2)}$.
Then 
\[
E_{n,j,k}\leq\exp(C\log n(k\wedge K_{n}(k)))
\]
Now if $k<k_{2}n\epsilon_{n}^{2}$, then uniformly over $\sum_{j=1}^{k}|\xi_{j}-\xi_{j}'|\leq n^{-1-\epsilon}(nj^{2}\epsilon_{n}^{2})$
and $|d-d'|\leq n^{-1-\epsilon}$, 
\[
|\bm{z}^{T}[A(\bt)-A(\bt')]\bm{z}+\tr[A(\bt)-A(\bt')]|=o(nj^{2}\epsilon_{n}^{2})
\]
and 
\[
E_{n,j,k}\leq\exp(Ck\log n).
\]

A simple chaining argument in $S_{n,j,k}$ implies that for all $j\leq J_{n,1}$
and all $k\leq k_{n,j}$, there exists $M>0$ such that 
\begin{equation}
\Pb\left[\Omega_{n}\cap\{\sup_{\bt\in S_{n,j,k}}\left(\bm{z}^{T}[A(\bt)-A(\txid)]\bm{z}+\tr[A(\bt)-A(\txid)]\right)>v_{n,j,k}+M\}\right]\leq e^{-ckn^{\epsilon}}\label{supAxi:1}
\end{equation}
and if $j\geq J_{n,1}$, for any $\delta>0$ and $n$ large enough
\begin{equation}
\Pb\left[\Omega_{n}\cap\{\sup_{\bt\in S_{n,j,k}}\left(\bm{z}^{T}[A(\bt)-A(\txid)]\bm{z}+\tr[A(\bt)-A(\txid)]\right)>v_{n,j,k}+\delta nj^{2}\epsilon_{n}^{2}\}\right]\leq e^{-ckn^{\epsilon}}.\label{supAxi:1b}
\end{equation}
Note also that for all $k\leq k_{n,j}$, $v_{n,j,k}=k^{3/2}n^{\epsilon}j\epsilon_{n}\leq n^{\epsilon}(j\epsilon_{n})^{1-3/(2\alpha)}$
so that whenever $\alpha>3/2$ and $j\leq J_{n,1}$, $v_{n,j,k}=o(1)$
and if $j\geq J_{n,1}$, $v_{n,j,k}=o(nj^{2}\epsilon_{n}^{2})$.

If $k>k_{n,j}$ and $k>k_{2}n\epsilon_{n}^{2}$, for all $j$, and
all $\delta>0$ 
\begin{equation}
\Pb\left[\sup_{\bt\in S_{n,j,k}}\left(\bm{z}^{T}[A(\bt)-A(\txid)]\bm{z}+\tr[A(\bt)-A(\txid)]\right)>v_{n,j,k}+\delta k\right]\leq e^{-ck^{\tau}},\label{supAxi:2}
\end{equation}
If $k<k_{2}n\epsilon_{n}^{2}$, for all $\delta>0$ and $n$ large
enough, 
\begin{equation}
\begin{split} & \Pb\left[\Omega_{n}\cap\{\sup_{\bt\in S_{n,j,k}}\left(\bm{z}^{T}[A(\bt)-A(\txid)]\bm{z}+\tr[A(\bt)-A(\txid)]\right)>v_{n,j,k}+\delta nj^{2}\epsilon_{n}^{2}\}\right]\\
 & \leq e^{-c(k\log n+n^{\epsilon})},\label{supAxi:3}
\end{split}
\end{equation}
and for all $k>k_{n,j}$, $v_{n,j,k}=o(k+nj^{2}\epsilon_{n}^{2})$.
We now study 
\[
\sup_{d}\left|\bm{z}^{T}[A(\bt_{o})-A(\txid)]\bm{z}+\tr[A(\bt_{o})-A(\txid)]\right|
\]
We have 
\[
\begin{split}|A(\bt_{o})-A(\txid)|_{Fr}^{2}=0+{\rm Error},\end{split}
\]
where Error is the approximation error of the trace by its limiting
integral. We bound separately the approximation error of $\tr\left[(\T(f_{o})\T(f_{o}^{-1}-f_{\txid}^{-1}))^{2}\right]$
and of $\tr\left[(\T(f_{o})(\T(f_{o})^{-1}-\T(f_{\txid})^{-1})^{2})\right]$.
We have, see \citet{kruijer:rousseau:11}, 
\[
f_{\txid}=f_{o}e^{(d-d_{o})H_{k}-\Delta_{d_{o},k}},\quad H_{k}(x)=\sum_{j>k}\eta_{j}\cos(jx),\quad\eta_{j}=2/j\quad\Delta_{d_{o},k}=\sum_{j>k}\theta_{o,j}\cos(jx)
\]
so that 
\[
f_{\txid}-f_{o}=f_{o}((d-d_{o})H_{k}-\Delta_{d_{o},k})+O(((d-d_{o})H_{k}-\Delta_{d_{o},k})^{2}x^{-2|d-d_{o}|})
\]
Therefore, from Lemma 2.6 in \citet{kruijer:rousseau:11} in the supplement,
\begin{equation}
\begin{split} & \mbox{error}\left(\tr\left[(\T(f_{o})\T(f_{o}^{-1}-f_{\txid}^{-1}))^{2}\right]\right)\\
 & =O((|d-d_{o}|+\|\Delta_{d_{o},k}\|_{\infty})^{2}\log n+(|d-d_{o}|+\|\Delta_{d_{o},k}\|_{\infty})l(f_{o},f_{\txid})^{1/2}(\log n)^{3}).
\end{split}
\end{equation}

Similarly using Lemma 2.4 in the supplement of \citet{kruijer:rousseau:11}
\[
\mbox{error}\left(\tr\left[(\T(f_{o})(\T(f_{o})^{-1}-\T(f_{\txid})^{-1})^{2})\right]\right)=O((|d-d_{o}|+\|\Delta_{d_{o},k}\|_{\infty})^{2}n^{\epsilon})
\]
for all $\epsilon>0$. Let $(k,d)$ be such that $\txid\in S_{n,j,k}$,
then 
\begin{equation}
|A(\xi_{o})-A(\txid)|_{Fr}=O((|d-d_{o}|+\|\Delta_{d_{o},k}\|_{\infty})n^{\epsilon})=O(n^{\epsilon}(j\epsilon_{n})^{(2\alpha-1)/(2\alpha)}),\quad\forall\epsilon>0\label{normAbd}
\end{equation}
see Lemma 3.1 in \citet{kruijer:rousseau:11}. We also have that for
all $|d-d'|\leq n^{-2}$, 
\[
|\bm{z}^{T}[A(\txid')-A(\txid)]\bm{z}-\tr[A(\txid')-A(\txid)]|\leq(\bm{z}^{T}\bm{z}+n)n^{-1}=O_{p}(1)
\]
uniformly, so that a simple chaining argument combined with \eqref{normAbd}
and Lemma 1.3 in the supplement of \citet{kruijer:rousseau:11} implies
that, for all $\epsilon>0$ 
\begin{equation}
\Pb\left(\sup_{d}|\bm{z}^{T}[A(\txid)-A(\bt_{o})]\bm{z}-\tr[A(\txid)-A(\bt_{o})]|>n^{\epsilon}(j\epsilon_{n})^{(2\alpha-1)/(2\alpha)}\right)\leq e^{-n^{\epsilon/2}}\label{supnormAbd}
\end{equation}
Hence, combining \eqref{supAxi:1}, \eqref{supAxi:2}, \eqref{supAxi:3}
and \eqref{supnormAbd} together with the fact that there exists $J_{0},C_{0}>0$
such that for all $j\geq J_{0}$, setting $S_{n,j}=\{\bt;(j-1)\epsilon_{n}\leq l(\bt_{o},\bt)\leq j\epsilon_{n}\}$,
\[
\int_{S_{n,j}}\exp\left\{ l_{n}(\bt)-l_{n}(\bt_{o})\right\} d\pi(\bt)\leq e^{-C_{0}j^2 n\epsilon_{n}^{2}},
\]
on a set having probability going to 1. Thus for all $\delta>0$,
there exists a set with $P_{o}$ probability going to 1 such that
\[
\begin{split} & \sum_{j=J_{0}}^{J_{n}}\sum_{k\in\N}p(k)N_{n,j,k}\\
 & \leq\sum_{j=1}^{J_{n}}\sum_{k=1}^{k_{n,1}}\sup_{(d,\bar{\bm{\xi}}_{d,k})\in S_{n,j,k}}\frac{w(\bt_{o})}{w(\txid)}\sup_{(d,\theta)\in S_{n,j,k}}\frac{w(\txid)}{w(\bt)}p(k)\int_{S_{n,j,k}}e^{l_{n}(\bt)-l_{n}(\bt_{o})}dp(\bt)\\
 & \leq\sum_{j=J_{0}}^{J_{n}}e^{n^{\epsilon}(j\epsilon_{n})^{(2\alpha-1)/(2\alpha)}}\sum_{k=1}^{k_{n,1}}e^{\delta(nj^{2}\epsilon_{n}^{2}+k)}p(k)\int_{S_{n,j,k}}e^{l_{n}(\bt)-l_{n}(\bt_{o})}dp(\bt)\\
 & \leq e^{-2cn\epsilon_{n}^{2}}
\end{split}
\]
Finally note that if there exists $\xi=(d,k,\theta)\in\cup_{j<J_{0}}S_{n,j,k}$,
then $\bd\in\cup_{j<J_{0}}S_{n,j,k}$ by definition of $\bd$. Replacing
$\epsilon_{n}$ by $J_{0}\epsilon_{n}$, with an abuse of notations,
we write $S_{n,1,k}:=\cup_{j\leq J_{0}}S_{n,j,k}$ for all $k$ and
we split the set $k$ into $k\leq k_{n,1}$ and $k>k_{n,1}$. 
\begin{multline*}
\mathbb{E}^{\pi_{n}}\left[w^{2}(\bt)\sum_{k=1}^{k_{n,1}}\1_{S_{n,1,k}}\right]\\
\leq w_{o}^{2}\sum_{k=1}^{k_{n,1}}\sup_{(d,\bar{\bm{\xi}}_{d,k})\in S_{n,1,k}}\frac{w^{2}(\txid)}{w_{o}^{2}}\sup_{(d,\bm{\xi})\in S_{n,1,k}}\frac{w^{2}(\bt)}{w^{2}(\txid)}p(k|x)\pi_{n}(S_{n,1,k}|k),
\end{multline*}
and 
\begin{multline*}
\mathbb{E}^{\pi_{n}}\left[w^{2}(\bt)\sum_{k=1}^{k_{n,1}}\1_{S_{n,1,k}}\right]\\
\geq w_{o}^{2}\sum_{k=1}^{k_{n,1}}\inf_{(d,\bar{\bm{\xi}}_{d,k})\in S_{n,1,k}}\frac{w^{2}(\txid)}{w_{o}^{2}}\inf_{(d,\theta)\in S_{n,1,k}}\frac{w^{2}(\bt)}{w^{2}(\txid)}p(k|x)\pi_{n}(S_{n,1,k}|k),
\end{multline*}
we have using \eqref{supnormAbd} associated with $j=1$, with probability
going to 1 uniformly in $(d,\bar{\bm{\xi}}_{d,k})\in\cup_{k<k_{n,1}}S_{n,1,k}$
$w^{2}(\txid)/w_{o}^{2}=1+o_{p}(1)$, and using \eqref{supAxi:1}
associated with $j=1$, $w^{2}(\bt)/w^{2}(\txid)=1+o_{p}(1)$ uniformly
in $(d,\theta)\in\cup_{k<k_{n,1}}S_{n,1,k}$. Let $k>k_{n,1}=k_{0}\epsilon_{n}^{-1/\alpha}\gtrsim(n/\log n)^{\frac{\beta/\alpha}{2\beta+1}}>>n\epsilon_{n}^{2}$,
then 
\begin{multline*}
\mathbb{E}^{\pi_{n}}\left[w^{2}(\bt)\sum_{k=k_{n,1}+1}^{\infty}\1_{S_{n,1,k}}\right]\\
\leq\frac{1}{w_{o}D_{n}}\sum_{k=k_{n,1}+1}^{\infty}p(k)\int_{S_{n,1,k}}\frac{w(\bt)^{-1}}{w(\txid)^{-1}}\frac{w(\txid)^{-1}}{w(\bt_{o})^{-1}}\exp\left\{ l_{n}(\bt)-l_{n}(\bt_{o})\right\} p(\bt)d\bt
\end{multline*}
and on the set or $x$ such that $D_{n}\geq e^{-cn\epsilon_{n}^{2}}$,
\begin{equation}
\begin{split} & \mathbb{E}^{\pi_{n}}\left[w^{2}(\bt)\sum_{k=k_{n,1}+1}^{\infty}\1_{S_{n,1,k}}\right]\\
 & \leq\frac{e^{cn\epsilon_{n}^{2}}}{w_{o}}\sum_{k=k_{n,1}+1}^{\infty}p(k)\sup_{(d,\theta)\in S_{n,1,k}}\frac{w(\bt)^{-1}}{w(\txid)^{-1}}\sup_{d:\txid\in S_{n,1,k}}\frac{w(\txid)^{-1}}{w(\bt_{o})^{-1}}\\
 & \qquad\times\int_{S_{n,1,k}}\exp\left\{ l_{n}(\bt)-l_{n}(\bt_{o})\right\} d\pi(d,\theta)\\
 & \leq\frac{e^{cn\epsilon_{n}^{2}}}{w_{o}}\sum_{k=k_{n,1}+1}^{\infty}p(k)e^{\epsilon k}\int_{S_{n,1,k}}\exp\left\{ l_{n}(\bt)-l_{n}(\bt_{o})\right\} p(\bt)d\bt,
\end{split}
\end{equation}
for all $\epsilon>0$, on a set of probability going to 1 using \eqref{supAxi:2}
and \eqref{supAxi:3}. A Markov inequality implies that 
\[
P_{o}^{n}\left[\sum_{k=k_{n,1}+1}^{\infty}p(k)e^{\epsilon k}\int_{S_{n,1,k}}\exp\left\{ l_{n}(\bt)-l_{n}(\bt_{o})\right\} d\pi(d,\theta)>u_{n}\sum_{k=k_{n,1}+1}^{\infty}p(k)e^{\epsilon k}\right]
\]
is $O(1/u_{n})$ for all $u_{n}$ going to infinity, so that with
probability going to 1, 
\begin{eqnarray*}
\mathbb{E}^{\pi_{n}}\left[w^{2}(\bt)\sum_{k=k_{n,1}+1}^{\infty}\1_{S_{n,1,k}}\right] & \leq & e^{-c_{1}k_{n,1}+cn\epsilon_{n}^{2}}=o(1)
\end{eqnarray*}
for some $c_{1},c>0$. We finally obtain that 
\[
\mathbb{E}^{\pi_{n}}\left[w^{2}(\bt)\right]=w^{2}(\bt_{o})(1+o_{p}(1))+w^{-1}(\bt_{o})e^{-Cn\epsilon_{n}^{2}}o_{p}(1),
\]
for some $C>0$. Simple computations show that for all $\epsilon>0$,
\[
P_{o}^{n}\left[\left|\log w(\bt_{o})\right|>n^{\epsilon}\right]=o(1),
\]
we can also make precise the upper bound on $\left|\log w(\bt_{o})\right|$
by controlling better $|A(\bt_{o})|$, leading to a term of order
$(\log n)^{3}$, which terminates the proof.

 \bibliographystyle{apalike}
\bibliography{complete}

\end{document}